\documentclass[12pt,a4paper]{article}
\pdfoutput=1
\usepackage[utf8]{inputenc}
\usepackage[T1]{fontenc}
\usepackage{style}
\usepackage[english]{babel}
\usepackage{url}
\usepackage{footnote}

\usepackage{amsmath}
\usepackage{graphicx} 
\usepackage{verbatim} 

\newcommand{\Mpc}{\text{Mpc}}

\newcommand{\fid}{{\rm fid}}

\renewcommand{\L}{\mathcal{L}}

\renewcommand{\l}{\lambda}
\renewcommand{\d}{\partial}

\newcommand{\be}{\begin{equation}}
\newcommand{\ee}{\end{equation}}
\newcommand{\beqa}{\begin{eqnarray}}
\newcommand{\eeqa}{\end{eqnarray}}
\newcommand{\bsm}{\begin{smallmatrix}}
\newcommand{\esm}{\end{smallmatrix}}

\newcommand\m{\mu}

\newcommand\G{\mathcal{G}}

\renewcommand\r{\rho}

\renewcommand\l{\lambda}

\renewcommand\k{{\bf k}}
\newcommand\q{{\bf q}}

\newcommand{\e}{\eta}

\def\e{{\rm e}}
\def\d{\partial}
\newcommand{\bseq}{\begin{subequations}}
\newcommand{\eseq}{\end{subequations}}

\renewcommand{\ln}{\mathop{\rm ln}\nolimits}

\newcommand{\F}{\mathcal{F}}
\renewcommand{\L}{\Lambda}

\renewcommand{\k}{{\bf k}}

\newcommand{\z}{{\bf z}}

\renewcommand{\d}{\partial}

\newcommand{\lin}{\mathrm{lin}}

\newcommand{\tot}{{\rm tot}}
\newcommand{\obs}{{\rm obs}}

\newcommand{\true}{{\rm true}}

\def\l{\left(}
\def\r{\right)}
\usepackage[textwidth=3cm,textsize=footnotesize]{todonotes}

\definecolor{darkgreen}{RGB}{0,120,0}

\title{Non-linear perturbation
theory extension of the Boltzmann code \texttt{CLASS} 
}

\author[a,b]{Anton Chudaykin\footnote{\texttt{chudy@ms2.inr.ac.ru}}}
\author[c,a]{Mikhail M. Ivanov\footnote{\texttt{mi1271@nyu.edu}}}
\author[d,e]{Oliver H.\,E. Philcox\footnote{\texttt{ohep2@cantab.ac.uk}}}
\author[f]{Marko Simonovi\'c\footnote{\texttt{marko.simonovic@cern.ch}}}
\affiliation[a]{Institute for Nuclear Research of the
Russian Academy of Sciences, \\ 
\normalsize \it  60th October Anniversary Prospect, 7a, 117312
Moscow, Russia}
\affiliation[b]{Moscow Institute of Physics and Technology,\\
	Institutsky lane 9, Dolgoprudny, Moscow region, 141700, Russia}
\affiliation[c]{Center for Cosmology and Particle Physics, Department of Physics,
New York University,\\
New York, NY 10003, USA}
\affiliation[d]{Department of Astrophysical Sciences, Princeton University,\\ Princeton, NJ 08540, USA}
\affiliation[e]{Department of Applied Mathematics and Theoretical Physics, University of Cambridge,\\ Cambridge CB3 0WA, UK}
\affiliation[f]{Theoretical Physics Department, CERN,\\
1 Esplanade des Particules, Geneva 23, CH-1211, Switzerland}

\abstract{We present a new open-source code that calculates one-loop power auto- and cross-power spectra for matter fields and biased tracers in real and redshift space. 
These spectra incorporate all ingredients required for a direct application to data:
non-linear bias, redshift-space distortions,
infra-red resummation, counterterms, and the Alcock-Paczynski effect.
Our code is based on the Boltzmann solver \texttt{CLASS} and inherits its advantage:
user friendliness,
ease of modification, 
high speed, and simple interface with other software. 
We present detailed descriptions of the theoretical model, the code structure,
approximations, and accuracy tests. 
A typical end-to-end run for one cosmology takes $\sim0.3$~seconds, which is 
sufficient for Markov Chain Monte Carlo parameter extraction.
As an example, we apply the code to data from the Baryon Oscillation Spectroscopic Survey (BOSS) 
and infer cosmological parameters from 
the shape of the galaxy power spectrum.
Our code and custom-built BOSS likelihoods are available at
\href{https://github.com/Michalychforever/CLASS-PT}{
\textcolor{blue}{https://github.com/Michalychforever/CLASS-PT}}
and \href{https://github.com/Michalychforever/lss_montepython}{
\textcolor{blue}{https://github.com/Michalychforever/lss\_montepython}}
}

\begin{document}

\begin{flushright}
	INR-TH-2020-016 \\
	CERN-TH-2020-062
\end{flushright}

\maketitle
\flushbottom

\section{Introduction}

Observations of temperature and polarization fluctuations in the cosmic microwave 
background (CMB) are one of the main pillars 
of the $\L$CDM 
model (see \cite{Aghanim:2018eyx} and references therein).
The most important tools connecting CMB data and cosmological parameters are Boltzmann codes,
which allow one to compute various observables in a given cosmological model. 
Building upon years of development starting with \texttt{CMBFast} \cite{Seljak:1996is},
the two most popular and independently designed
Boltzmann solvers that have emerged 
are \texttt{CAMB} \cite{Lewis:2002ah} and \texttt{CLASS}
\cite{Blas:2011rf}. Both 
are very efficient and accurate, allowing for fast and robust extraction of CMB likelihoods.
These two codes and their various extensions (see Refs.~\cite{Bellini:2017avd,Brinckmann:2018cvx} for some reviews) 
have been widely used in the cosmology community.

Another source of cosmological information 
that is becoming increasingly important is the large-scale structure 
(LSS) clustering of galaxies in the late universe. 
This clustering is measured in redshift surveys such as the Baryon Oscillation Spectroscopic Survey (BOSS)~\cite{Alam:2016hwk}. 
Next generation surveys like Euclid~\cite{Laureijs:2011gra,Amendola:2016saw} and DESI~\cite{Aghamousa:2016zmz} will
map a significant volume of the universe across a wide range of redshifts.  
In order to prepare for these future surveys and 
eventually harvest cosmological information encoded in the LSS data
as efficiently as possible, it is imperative to build
simple and robust 
extensions of the standard Boltzmann codes
that can reevaluate LSS likelihoods as one scans over different 
cosmologies.\footnote{The commonly used Boltzmann codes do have nonlinear modules
featuring fitting formulas like HALOFIT \cite{Takahashi:2012em}. 
However, the application of these modules to galaxy clustering is quite limited for several reasons.
For instance, these formulas do not accurately capture the behavior of the matter power spectrum 
on mildly-nonlinear scales, in particular the non-linear evolution of the BAO wiggles.
Also, they were calibrated only on a small grid of cosmological parameters,
which does not cover many beyond-$\L$CDM extensions.}
With this work we present one such tool, a modified \texttt{CLASS}
code---\texttt{CLASS-PT}---that embodies an end-to-end calculation of various power spectra using the 
state-of-the-art perturbation theory models that incorporate
all ingredients required for a direct application to data.

We provide a \texttt{Jupyter} notebook\footnote{\texttt{CLASS-PT/notebooks/nonlinear\_pt.ipynb}} that generates the spectra for galaxies and matter in real and redshift space.
Additionally, we share a \texttt{Mathematica} 
notebook\footnote{\texttt{CLASS-PT/read\_tables.nb}} that reads the spectra 
from output tables produced by \texttt{CLASS-PT} if it is run using a \texttt{.ini} file.
Additionally,  
we publicly release custom-built BOSS galaxy power spectrum likelihoods written for the Markov Chain Monte Carlo sampler 
\texttt{Montepython}~\cite{Audren:2012wb,Brinckmann:2018cvx}, 
which can be used for various cosmological analyses. It is worth stressing that 
\texttt{CLASS-PT} has been already applied to the analysis of the BOSS data~\cite{Ivanov:2019pdj,Ivanov:2019hqk,Philcox:2020vvt}, 
and used in a blind cosmology challenge based on a large-volume numerical simulation~\cite{Nishimichi:2020tvu}.
Moreover, in Ref.~\cite{Chudaykin:2019ock} it was used 
to assess the accuracy of the neutrino mass and cosmological parameter measurements  
with a future Euclid-like galaxy survey.

Many important developments have led 
to~\texttt{CLASS-PT}, both in theoretical modeling
and practical implementation of perturbation theory calculations. We postpone details
of the relevant theoretical results for the next section. Here we only briefly review the history 
of some numerical methods and publicly available software dedicated to perturbation theory (PT).
To the best of our knowledge, this begins
with \texttt{COPTER} \cite{Carlson:2009it}, 
which was designed to compute the 1-loop and 2-loop matter power spectra 
in Standard Perturbation Theory (SPT) and its extensions. 
The first code to compute statistics beyond the power spectrum was \texttt{Zelca} \cite{Tassev:2013zua},
designed to evaluate the matter power spectrum and bispectrum in the Zel'dovich
approximation.
Later on, \texttt{FnFast} \cite{Bertolini:2015fya}
was developed to compute the one-loop power spectrum,
bispectrum, and trispectrum in SPT and in the Effective Field Theory of Large-scale Structure (EFTofLSS).
In all these codes perturbation theory loop integrals were evaluated by direct numerical integration.
Recently, it has been realized that the computation of the one-loop power spectrum integrals
can be significantly optimized by using their relatively simple structure in position space.
These new methods are based on the Fast Fourier Transform (FFT) \cite{Schmittfull:2016jsw,McEwen:2016fjn} and
they were implemented in \texttt{FAST-PT} \cite{McEwen:2016fjn,Fang:2016wcf}, a \texttt{python} code
that evaluates the one-loop Eulerian perturbation theory power spectra for matter 
and biased tracers in real and redshift space.
The FFT approach leads to a significant  
boost in the performance over the direct numerical integration, opening the door 
to the use of
complete perturbation theory templates in a realistic Monte Carlo Markov Chain (MCMC) analysis. 

While \texttt{CLASS-PT} is built on some of these results, it also brings several novelties. 
In particular, it uses an FFT method that is very different from the original proposals of 
Refs.~\cite{Schmittfull:2016jsw,McEwen:2016fjn}. 
This approach was put forward in Ref.~\cite{Simonovic:2017mhp} (see Section~\ref{sec:code} for more details). 
Another major difference with respect to previous Eulerian PT codes is that \texttt{CLASS-PT}
properly describes the nonlinear evolution of the BAO wiggles, implemented via the so-called 
infrared (IR) resummation scheme. This is particularly 
important for redshift surveys where the correct shape of the BAO wiggles 
is crucial for reliable cosmological constraints. 

Finally, we note that, 
whilst our paper was in the final stages of preparation, 
a new code \texttt{PyBird} has appeared~\cite{DAmico:2020kxu}, which is based on the same perturbation theory 
model as \texttt{CLASS-PT}, but with different implementation of some of the key ingredients. 
The two codes agree within the designed precision when evaluated for the same cosmological parameters. 

This paper is structured as follows. Section~\ref{sec:model} discusses the main theoretical 
ingredients and presents the corresponding formulae used in \texttt{CLASS-PT}, before we review the structure of the code in Section~\ref{sec:code}. 
In Section~\ref{sec:approx}
we discuss in detail the technical implementation of the non-linear model 
and test various approximations.
A busy reader who is only interested in the final results, can skip directly to Sections~\ref{sec:plots} and~\ref{sec:boss}.
Section~\ref{sec:plots} contains some examples and important 
caveats that must be kept in mind when using our code.
As an illustration, in Section~\ref{sec:boss} we apply \texttt{CLASS-PT} to 
the cosmological analysis of the BOSS galaxy clustering data.
We release our BOSS likelihoods along with the code.
In Section~\ref{sec:concl} we draw conclusions. 
Two short appendices contain some useful additional information: explicit 
expressions for the FFTLog redshift-space master integrals (Appendix~\ref{app:rsdfft})
and a quick installation manual (Appendix~\ref{app:manual}). Appendix~\ref{app:bias} discusses the pirors for the nuisance parameters used in our 
BOSS analysis.

\section{The Power Spectrum Model}
\label{sec:model}

In this Section we describe the theoretical model used in \texttt{CLASS-PT}. 
We start with a brief summary of 
theoretical developments that have led to a complete and consistent description of large-scale clustering. A reader familiar with these 
results may wish to
skip this Section.  
We will give details of all relevant ingredients 
needed for the description of the nonlinear power spectrum: the clustering of matter and biased tracers in 
real space, IR-resummation, 
the effects of redshift space and 
Alcock-Paczynski (AP) distortions. 

\subsection{Brief Overview of Perturbation Theory}

Since Yakov Zel'dovich proposed a first model for nonlinear gravitational clustering of 
cosmological fluctuations in 1970~\cite{1970A&A.....5...84Z},
there have been 
numerous attempts to build a consistent theoretical description 
of large-scale structure in the mildly-nonlinear regime.
Historically, the most popular approach was SPT (\cite{Scoccimarro:1995if,Scoccimarro:1996se,Scoccimarro:1997st}, for a review see~\cite{Bernardeau:2001qr}), 
where dark matter is treated as a pressureless perfect fluid and the nonlinear equations of motion 
are solved perturbatively in Eulerian space. The major problem of SPT is that higher
order perturbative corrections to the power spectrum do 
not lead to significant 
improvements on mildly-nonlinear scales~\cite{Crocce:2005xy,Blas:2013aba}. This apparent breakdown of 
perturbation theory led to attempts to partially resum the diagrammatic expansion in order to 
improve convergence properties~\cite{Crocce:2005xy,Crocce:2005xz}. However, such resummation schemes
were insufficient, as 
can be seen from a simple example of the one-dimensional universe~\cite{McQuinn:2015tva}.
In that case, the whole standard perturbation theory expansion can be explicitly and exactly resummed, 
but it does not lead 
to any notable improvement compared to the linear theory prediction.

Efforts to resolve this problem have 
led to the 
development of the Effective Field Theory of Large-Scale Structure~\cite{Baumann:2010tm}.
The key insight was that the ideal fluid approximation is inconsistent even on large scales, and that 
the true equations of motion are those of an imperfect fluid with various contributions to the effective stress-tensor. 
Starting from the Boltzmann equation (which is the true description of the dynamics for dark matter particles)
and focusing on the dynamics of the long-wavelength fluctuations (averaging over the short modes), 
one can show that the imperfect fluid terms naturally arise and can be organized in a perturbative derivative expansion.   
Whilst the form of these terms is dictated by symmetry, their amplitudes are unknown free parameters which have to be measured from the data.
Since these free parameters---the counterterms--- 
capture the effects of the poorly-known short-scale physics, 
including them in the
power spectrum 
significantly improves the performance of the theory~\cite{Carrasco:2012cv,Carrasco:2013mua}. 
The realization that the LSS theory must include unknown free parameters has
finally resolved the long-standing problem of the consistent description of matter clustering on mildly non-linear scales.
Another major advantage of the EFT approach is that it provides
reliable estimates of theoretical errors, allowing 
theoretical uncertainties to be included in the 
total error budget and guaranteeing unbiased inference of cosmological parameters~\cite{Baldauf:2016sjb}.

Another problem of Eulerian perturbation theory concerns
the long-wavelength displacement of 
dark matter particles, which can be very large in our universe. Whilst the effects of these bulk flows are locally unobservable due to the
Equivalence Principle~\cite{Peloso:2013zw,Kehagias:2013yd,Creminelli:2013mca}, they still affect 
features in the power spectrum, such as the BAO wiggles.  
It is well-known that treating them perturbatively leads to significant errors in the 
description of the BAO peak~\cite{Eisenstein:2006nj,Crocce:2007dt,Sugiyama:2013gza}, 
even though their effect on the broadband part of the correlation function (or the power spectrum) remains under perturbative control. 
Since the dominant dynamical effect of the bulk flows is a simple translation produced by the linear theory displacements, 
there exists
a relatively straightforward way to take them into account non-perturbatively~\cite{Senatore:2014via,Baldauf:2015xfa,Vlah:2015zda,Blas:2015qsi,Blas:2016sfa,Senatore:2017pbn,Ivanov:2018gjr,Lewandowski:2018ywf}. 
In other words, large contributions from these displacements at different orders in perturbation 
theory can be rigorously resummed. For this reason, this procedure is referred to as infrared (IR) resummation.
It 
allows one to take advantage of simplicity of the Eulerian description,
while keeping the impact of large displacements exact and hence significantly improving 
predictions for the shape of the BAO wiggles.

To make connection to observations, two additional ingredients are necessary;  the first being
the nonlinear description of biased tracers.
Following the
first attempts to build such a description in terms of a local-in-density
bias expansion, 
an important milestone was the realization 
that various tidal and higher derivative bias operators must also be included~\cite{McDonald:2009dh,Assassi:2014fva}. Furthermore, 
since the formation of biased tracers is nonlocal in time~\cite{Senatore:2014eva}, the 
expansion has to include additional terms
that cannot be expressed in terms of local operators involving only two derivatives of gravitational and 
velocity potentials~\cite{Senatore:2014eva,Lewandowski:2014rca,Mirbabayi:2014zca}. The perturbative bias model, (at least up to third order, which is needed for the one-loop 
power spectrum), is now well established and tested against various numerical simulations (for a review see~\cite{Desjacques:2016bnm}).
The second important ingredient is the treatment of redshift space distortions (RSD). 
The 
standard perturbation theory kernels in the presence of RSD have been
known for a long time~\cite{Bernardeau:2001qr}, however, a consistent calculation of 
the one-loop power spectrum in redshift space requires
additional counterterms related to the velocity field~\cite{Senatore:2014vja,Perko:2016puo}. 
Below, we discuss these contributions in detail. 

While \texttt{CLASS-PT} is entirely based on Eulerian perturbation theory, 
it is worth emphasizing that similar progress has been made in Lagrangian Perturbation 
Theory (LPT) as well~\cite{Matsubara:2007wj,Porto:2013qua,Carlson:2012bu,Vlah:2015sea,Vlah:2016bcl,Modi:2017wds}. 
An advantage of the LPT is that IR resummation is automatically incorporated at all orders, but this comes at the cost of larger
computational complexity.
Nevertheless, 
when all relevant biases and counterterms are included, 
the two approaches are fully consistent~\cite{Vlah:2016bcl,Modi:2017wds,Schmittfull:2018yuk}.
If both theories are well-defined, this is, in a sense, equivalent, since
the Lagrangian and Eulerian schemes are just two different ways of solving the 
exact same equations of motion.

\subsection{Dark Matter Power Spectrum}
To begin, let us consider the model for
the matter power 
spectrum in real space. On very large scales (or early times) the dark matter fluctuations evolve linearly.
Thus, to a very good approximation, their power spectrum is given by
\be
P_{\rm lin}(z,k) = D^2(z)P_{\rm lin}(k) \, ,
\ee
where $D(z)$ is the linear growth factor and $P_{\rm lin}(k)$ is the linear power spectrum 
at redshift zero. In the mildly-nonlinear regime one can calculate perturbative corrections to 
this simple result, the first of which
is the so-called one-loop contribution. 
For dark matter in real space, 
this is the 
sum of two terms;
\be
\label{eq:1ldm}
P_{\text{1-loop}}(z,k) = P_{\text{1-loop,\,SPT}}(z,k)+P_{\rm ctr}(z,k)\,,
\ee
where $P_{\text{1-loop,\,SPT}}(z,k)$ is the SPT contribution~\cite{Bernardeau:2001qr}
and $P_{\rm ctr}(z,k)$ is the counterterm needed for the consistency 
of the one-loop result~\cite{Baumann:2010tm,Carrasco:2012cv}.
The explicit expression for the counterterm at this order in perturbation theory is given by
\be
\label{eq:pdm_count}
P_{\rm ctr}(z,k)=- 2c_s^2(z)k^2 P_{\text{lin}}(z,k) \,, 
\ee
where $c_s^2(z)$ is an effective parameter (sometimes refereed to as the effective sound speed), whose amplitude and time dependence 
are not known \textit{a priori}. Thus, $c_s^2(z)$ must be treated as a nuisance 
parameter in data analysis. The SPT one-loop term 
can be written as a sum of two well-known pieces,
\be
\label{eq:pdm1loop}
P_{\text{1-loop,\,SPT}}(z,k) = D^4(z)\left(P_{13}(k) +  P_{22}(k)\right) \,,
\ee
each of which is given by a particular convolution integral,
\be 
\label{eq:pdm1l}
\begin{split}
& P_{22}(k) = 2\int_\q \,F_2^2(\q,\k-\q) P_{\rm lin}(q)P_{\rm lin}(|\k-\q|)\,,\\
& P_{13}(k) = 6P_{\rm lin}(k)\int_\q\,F_3(\k,-\q,\q) P_{\rm lin}(q)\, .
\end{split}
\ee
Here, and throughout the rest of the paper, we use the notation $\int_\q\equiv\int\frac{d^3q}{(2\pi)^3}$. 
The convolution kernels~$F_2$ and~$F_3$ are the usual perturbation theory 
kernels~\cite{Bernardeau:2001qr,Blas:2015qsi}.
Equations~\eqref{eq:pdm_count} and~\eqref{eq:pdm1l} give a complete description of the 
one-loop power spectrum of dark matter in real space. This model has been exhaustively
tested against N-body simulations and found to predict the
nonlinear matter power spectrum at mildly-nonlinear scales quite well, see e.g. Ref.~\cite{Baldauf:2015aha}.

Strictly speaking, our
Eq.~\eqref{eq:pdm1loop} is 
correct only in the
Einstein-de Sitter (EdS) universe, where the momentum and time-dependences of the 
loop integrals factorize. 
However, even in 
a more general case the common 
practice is to retain
the EdS perturbation theory kernels but replace the 
the growth factor in EdS\footnote{Note that in the EdS cosmology the growth factor is identical to the scale factor, $D=a$. }
with the linear growth factor computed in the true cosmology.
We will use this approximation throughout the paper and we also implement it in the 
\texttt{CLASS-PT} code.\footnote{In principle, one can still do the 
exact calculation using the appropriate Green's functions. In this case the momentum integrals have similar form 
to those
in EdS, but the time integrals have to be evaluated numerically.} We make this choice for two reasons.
First, this approximation is quite accurate.  Indeed, 
the residual difference with respect to the full 
calculation is so small that it is irrelevant even for future galaxy surveys~\cite{Pietroni:2008jx,Fasiello:2016qpn,delaBella:2017qjy}.
Second, it allows the nonlinear corrections to be easily calculated for {\em{any}} time by simply 
rescaling the result at redshift zero.\footnote{This is not true if IR-resummation is included.} In other words, one can
rewrite Eq.~\eqref{eq:pdm1loop} as
\be
P_{\rm 1-loop,\, SPT}(z, k) =  P_{13}(z, k) +  P_{22}(z, k)\,,
\ee
where $P_{13}(z,k)$ and $P_{22}(z,k)$ are obtained from Eq.~\eqref{eq:pdm1l}
by performing
the loop integrals with the linear power spectrum
evaluated at the redshift of interest, $P_{\rm lin}(z,k)$.

\subsection{Power Spectrum of Biased Tracers}
To calculate the one-loop power spectrum of biased tracers, we have
to include all possible operators up to third order in the bias expansion:
\be 
\label{eq:bias_expansion}
\delta_{g} =b_1\delta +\epsilon+\frac{b_2}{2}\delta^2+b_{\mathcal{G}_2}\mathcal{G}_2
+ \frac{b_3}{6}\delta^3 + b_{\delta\mathcal G_2} \delta\mathcal G_2 + b_{\mathcal G_3}\mathcal G_3 +  b_{\Gamma_3}\Gamma_3 +R^2_*\d^2\delta \,.
\ee
Here we have defined the Galileon operator
\be 
\G_2 (\Phi_g)\equiv (\d_i \d_j \Phi_g)^2 - (\d_i^2\Phi_g)^2\,,
\ee
where $\Phi_g$ is gravitational potential. The only cubic operator 
that gives a nontrivial contribution to the one-loop power spectrum can be written as  
\be 
\Gamma_3\equiv \G_2 (\Phi_g)-\G_2 (\Phi_v)\,,
\ee
where $\Phi_v$ is velocity potential.\footnote{The two potentials $\Phi_g$ and 
$\Phi_v$ are the same in linear theory, but 
differ at higher orders in perturbation theory.}
For the definition of $\mathcal G_3$ and relations of our operators to other equivalent choices 
of basis, see~\cite{Desjacques:2016bnm}.
The term $\epsilon$ denotes the stochastic contribution which is uncorrelated with the 
large-scale density field. In the simplest approximation, we may treat the power spectrum of~$\epsilon$ as Poissonian, and thus constant.
In practice, 
it is more complicated and 
has scale-dependent corrections.
Finally, the last term in Eq.~\eqref{eq:bias_expansion} is the higher derivative bias which
we keep for consistency and completeness. In general, $b_1, b_2, b_{\mathcal{G}_2}, b_3, b_{\delta\mathcal G_2}, b_{\mathcal G_3}, b_{\Gamma_3}$ and $R^2_*$
are free parameters.

Using the particular bias expansion given above, the one-loop auto-power spectrum of the bias tracers takes the following form~\cite{Senatore:2014eva,Assassi:2014fva,Desjacques:2016bnm},
\be
\begin{split}
\label{Pg}
P_{\rm gg}(z,k)=&~b^2_1(z) (P_{\text{lin}}(z,k)+P_{\text{1-loop,\,SPT}}(z,k))+b_1(z) b_2(z) \mathcal{I}_{\delta^2}(z,k)\\
&+2b_1(z)b_{\mathcal{G}_2} (z) \mathcal{I}_{\mathcal{G}_2}(z,k)+b_1(z)\left(2b_{\mathcal{G}_2}(z) + \frac{4}{5}b_{\Gamma_3}(z)\right)\mathcal{F}_{\mathcal{G}_2}(z,k)\\
&+\frac{1}{4}b^2_2(z)\mathcal{I}_{\delta^2\delta^2}(z,k)+b^2_{\mathcal{G}_2}(z)\mathcal{I}_{\mathcal{G}_2\mathcal{G}_2}(z,k)+b_2(z)b_{\mathcal{G}_2}(z) \mathcal{I}_{\delta^2\mathcal{G}_2}(z,k)\\
& +P_{\nabla^2 \delta}(z,k)+P_{\epsilon\epsilon}(z,k)\,,
\end{split}
\ee
where $P_{\epsilon\epsilon}(z,k)$ is the power spectrum of the stochastic component. 
This uses 
the following definitions~\cite{Assassi:2014fva}:
\bseq 
\begin{align}
\label{eq:ireals}
&{\cal I}_{\delta^2}(z,k) \equiv 2\int_\q F_2(\q,\k-\q)P_{\text{lin}}(z,|\k-\q|)P_{\text{lin}}(z,q)\,,\\
&{\cal I}_{\mathcal{G}_2}(z,k) \equiv 2\int_\q \sigma^2(\q,\k-\q)F_2(\q,\k-\q)P_{\text{lin}}(z,|\k-\q|)P_{\text{lin}}(z,q)\,,\\
&{\cal F}_{\mathcal{G}_2}(z,k) \equiv 4P_{\text{lin}}(z,k)\int_\q \sigma^2(\q,\k-\q)F_2(\k,-\q)P_{\text{lin}}(z,q)\,,\\
&{\cal I}_{\delta^2\delta^2}(z,k) \equiv 2\int_\q P_{\text{lin}}(z,|\k-\q|)P_{\text{lin}}(z,q) - 2\int_\q P^2_{\text{lin}}(z,q)\,,\\
&{\cal I}_{\mathcal{G}_2\mathcal{G}_2}(z,k) \equiv 2\int_\q \sigma^4(\q,\k-\q)P_{\text{lin}}(z,|\k-\q|)P_{\text{lin}}(z,q) \,,\\
&{\cal I}_{\delta^2\mathcal{G}_2}(z,k) \equiv 2\int_\q \sigma^2(\q,\k-\q)P_{\text{lin}}(z,|\k-\q|)P_{\text{lin}}(z,q) \,,\\
& P_{\nabla^2 \delta}(z,k)  \equiv  -2b_1(z) \big( R_*^2(z) + c_s^2(z) b_1(z) \big) k^2 P_{\lin}(z,k)\,,
\end{align}
\eseq 
where $\sigma^2(\k_1,\k_2) \equiv (\k_1\cdot \k_2)^2/(k_1^2k_2^2)-1$. 

Three important comments are in order here.
First, we define ${\cal I}_{\delta^2\delta^2}(z,k)$ by subtracting 
the low-$k$ constant contribution, to ensure it has an
$\mathcal O(k^2)$ behavior 
on large scales. 
The constant contribution is reabsorbed in the stochastic power 
spectrum since it is perfectly degenerate with the shot noise.
Second, the dark matter counterterm is combined with the higher derivative bias 
since they are perfectly degenerate for the galaxy power spectrum. 
Third, the contributions from operators 
$\delta^3,~\delta\mathcal G_2,~\mathcal G_3$
disappeared after renormalization.
This is the reason why $b_3, b_{\delta\mathcal G_2}, b_{\mathcal G_3}$ are absent 
in Eq.~\eqref{Pg}.

Using the same bias model we can also calculate the galaxy-matter cross-spectrum
which is of relevance, for instance, for lensing surveys.  It has the following form~\cite{Assassi:2014fva}:
\be 
\begin{split}
P_{\rm gm}(z,k)=&~b_1(z) (P_{\text{lin}}(z,k)+P_{\text{1-loop,\,SPT}}(z,k))+\frac{1}{2}b_2(z)\mathcal{I}_{\delta^2}(z,k)\\
&+\left(b_{\mathcal{G}_2}(z) + \frac{2}{5}b_{\Gamma_3}(z)\right)\mathcal{F}_{\mathcal{G}_2}(z,k)\\
&+b_{\mathcal{G}_2}(z) \mathcal{I}_{\mathcal{G}_2}(z,k) - \big(R_*^2(z) + 2c_s^2(z)b_1(z) \big)k^2 P_{\lin}(z,k)\,.
\end{split}
\ee
Note that the matter counterterm and the higher-derivative bias enter
the cross-spectrum and the the auto-spectrum
in different combinations. In principle, 
This allows one 
to break the degeneracy 
between them using the galaxy-lensing observations.

\subsection{Power Spectrum of Biased Tracers in Redshift Space}

The radial positions of galaxies in a survey are assigned using their redshifts, which 
are contaminated by the peculiar velocity field.
This gives rise to the so-called redshift-space distortions RSD,
which allow one to probe the velocity field along the line-of-sight direction $\hat\z$.
We will work within the flat-sky plane-parallel approximation, where the redshift-space mapping
can be fully characterized by the cosine of the angle 
between the line-of-sight $\hat\z$ and the wavevector of a given Fourier mode $\k$, $\mu\equiv (\hat\z\cdot \k)/k$. In this setup, the expression for the one-loop redshift-space power spectrum reads 
(see Refs.~\cite{Senatore:2014vja,Perko:2016puo}):
\be
\label{eq:mastz}
\begin{split}
P_{\rm gg,RSD}(z,k,\mu)= & Z^2_1(\k)
P_{\text{lin}}
(z,k)+ 2\int_{\q}Z^2_2(\q,\k-\q)
P_{\text{lin}}(z,|\k-\q|)
P_{\text{lin}}(z,q)\\
& + 6Z_1(\k)P_{\text{lin}}(z,k)\int_{\q}Z_3(\q,-\q,\k)P_{\text{lin}}(z,q)\\
& + P_{\text{ctr,RSD}}(z,k,\mu)
 + P_{\epsilon\epsilon,\text{RSD}}(z,k,\mu)\,,
\end{split}
\ee
where the redshift-space kernels are given by
\bseq 
\begin{align}
&Z_1(\k)  = b_1+f\mu^2\,,\\
&Z_2(\k_1,\k_2)  =\frac{b_2}{2}+b_{\mathcal{G}_2}\left(\frac{(\k_1\cdot \k_2)^2}{k_1^2k_2^2}-1\right)
+b_1 F_2(\k_1,\k_2)+f\mu^2 G_2(\k_1,\k_2)\notag\\
&\qquad\qquad\quad~~+\frac{f\mu k}{2}\left(\frac{\mu_1}{k_1}(b_1+f\mu_2^2)+
\frac{\mu_2}{k_2}(b_1+f\mu_1^2)
\right)
\,,\\
&Z_3(\k_1,\k_2,\k_3)  =2b_{\Gamma_3}\left[\frac{(\k_1\cdot
     (\k_2+\k_3))^2}{k_1^2(\k_2+\k_3)^2}-1\right]
\big[F_2(\k_2,\k_3)-G_2(\k_2,\k_3)\big]
\notag
\\  
&\quad
+b_1 F_3(\k_1,\k_2,\k_3)+f\mu^2 G_3(\k_1,\k_2,\k_3)+\frac{(f\m k)^2}{2}(b_1+f \mu_1^2)\frac{\m_2}{k_2}\frac{\m_3}{k_3}
\notag
\\
&\quad
+f\mu k\frac{\mu_3}{k_3}\left[b_1 F_2(\k_1,\k_2) + f \mu^2_{12} G_2(\k_1,\k_2)\right]
+f\m k (b_1+f \mu^2_1)\frac{\m_{23}}{k_{23}}G_2(\k_2,\k_3)
\notag
\\
&\quad+b_2 F_2(\k_1,\k_2)+2b_{\mathcal{G}_2}\left[\frac{(\k_1\cdot (\k_2+\k_3))^2}{k_1^2(\k_2+\k_3)^2}-1\right]F_2(\k_2,\k_3)
+\frac{b_2f\mu k}{2}\frac{\mu_1}{k_1}
\notag
\\
&\quad+b_{\mathcal{G}_2}f\mu k\frac{\m_1}{k_1}\left[\frac{(\k_2\cdot \k_3)^2}{k_2^2k_3^2}-1\right]
\,,
\end{align} 
\eseq
where $\k=\k_1+\k_2+\k_3$ and $G_n$ are 
the velocity divergence kernels \cite{Bernardeau:2001qr}. 
Note that $Z_3(\k_1,\k_2,\k_3)$ contains only bias parameters that give nontrivial contributions
to the redshift-space one-loop power spectrum and that it must be symmetrized over its 
momentum arguments when used in Eq.~\eqref{eq:mastz}. Furthermore, we have omitted the time dependence of $f\equiv {d \log D}/{d\log a}$ and biases for clarity.

Let us discuss the structure of the last two terms in Eq.~\eqref{eq:mastz} in some detail. 
The leading counterterm contributions in redshift space can be seen as a simple 
generalization of the dark matter sound speed \cite{Senatore:2014vja,Lewandowski:2015ziq},
\be
\label{eq:pctrrsd}
\begin{split}
P_{{\rm ctr, RSD,} \nabla^2\delta} (z,k,\mu) = & -2\tilde c_0(z) k^2 P_{\text{lin}}(z,k) \\
& -2\tilde c_2(z) f(z) \mu^2 k^2 P_{\text{lin}}(z,k)
-2\tilde c_4(z) f^2(z)\mu^4 k^2 P_{\text{lin}}(z,k)\,,
\end{split}
\ee
where $\tilde c_0(z)$, $\tilde c_2(z)$ and $\tilde c_4(z)$ are quantities 
that are generically expected to have similar value to the real-space dark matter 
sound speed in units of $[\text{Mpc}/h]^2$. 
However, due the presence of fingers-of-God \cite{Jackson:2008yv}
these counterterms can be more significant for some tracers than na\"ively expected.
Since the fingers-of-God are induced by the higher-derivative terms in the non-linear RSD
mapping, one may include an additional counterterm proportional to $k^4\mu^4 P_{\rm lin}(z,k)$
as a proxy of the higher-order contributions,
\be 
P_{{\rm ctr,RSD},\,\nabla_\z^4 \delta} (z,k,\mu) =- \tilde{c}(z)f^4(z) \mu^4  k^4 (b_1(z)+f(z)\mu^2)^2P_{\rm lin}(z,k)\,,
\ee
where we have inserted the linear Kaiser factor $(b_1(z)+f(z)\mu^2)^2$ \cite{Kaiser:1987qv} for convenience.
Whilst, we leave the systematic derivation of all corrections of this order for future work, we
stress 
that addition of this term can be important in order to 
fit the data or results from N-body simulations, if fingers-of~God effects are large~\cite{Ivanov:2019pdj,Nishimichi:2020tvu}. 
The full counterterm contribution
is then given by
\be
P_{\rm ctr,RSD}(z,k,\mu)=  P_{{\rm ctr,RSD,} \nabla^2\delta}(z,k,\mu) + P_{{\rm ctr,RSD},\,\nabla_\z^4 \delta}(z,k,\mu) \,,
\ee
and 
depends on four free functions of time~$\tilde c_0(z),\tilde c_2(z),\tilde c_4(z)$ 
and $\tilde{c}(z)$.

Finally, the stochastic power spectrum in redshift space has the following structure at next-to-leading order in derivative expansion:
\be
P_{\epsilon\epsilon,{\rm RSD}}(z,k,\mu)=P_{\rm shot}(z)+ a_0(z) k^2 + a_2(z) \mu^2 k^2 \,,
\ee
where $P_{\rm shot}$ describes a constant shot noise and the additional two terms are scale-dependent
shot noise contributions for the monopole and the quadrupole. Note that the amplitude of the shot noise
and the two coefficients $a_0$ and $a_2$ are functions of time only, while the $k$ and $\mu$ dependence 
of the stochastic power spectrum is very simple.  
It is worth mentioning that the pair-counting Poissonian contribution $1/\bar n$ is often subtracted from the power spectrum estimator. 
It is however important to keep the residual constant $P_{\rm shot}$ in the model in order to capture deviations from the Poissonian prediction, which are expected on general grounds.

Whilst all the terms presented above should be kept in a data analysis for consistency, 
some 
contributions are quite degenerate at the level of galaxy power spectra.
For instance, the $P_{{\rm ctr,RSD},\,\nabla_\z^4 \delta}$ counterterm is 
very 
degenerate with the $a_2\mu^2k^2$ stochastic contribution, given the slope of the linear power spectrum 
on mildly nonlinear scales. 
Therefore, as far as galaxy clustering is concerned and depending on the required precision, one can opt to 
keep only one of the two terms. 
For example,
the recent re-analyses of BOSS data kept only the higher derivative counterterm 
~\citep{Ivanov:2019pdj},
whilst the analysis of Ref.~\cite{DAmico:2019fhj} 
includes only
the $a_2 \mu^2 k^2$ contribution. 
As 
can be seen from these two papers,
the particular choice does not impact the inference of cosmological parameters.
Moreover, the authors of Ref.~\cite{Schmittfull:2018yuk},
have shown that the $a_0$ contribution can be neglected on scales with $k\lesssim 0.3$ $h/$Mpc. 
Given these reasons, we will neglect the $a_0$ and $a_2$ 
terms henceforth.

Thus far we have presented the perturbation theory model for the redshift-space 
power spectrum, keeping the full $k$ and $\mu$ dependence. However, it is generally more convenient to summarize 
the full angular information in a few multipoles, using the relation
\be
P_{\rm gg,RSD}(z,k,\mu) =\sum_{\ell \; {\rm even}} {\cal L}_\ell(\mu) P_{\ell}(z,k)\,,
\ee
where ${\cal L}_\ell(\mu)$ are Legendre polynomials. The galaxy power spectrum multipoles are thus 
\be
\label{eq:mult}
P_{\ell}(z,k) \equiv \frac{2\ell+1}{2}\int_{-1}^{1}d\mu \, {\cal L}_\ell(\mu) 
P_{\rm gg,RSD} (z,k,\mu)\,.
\ee 
In both this paper and our code we will focus on the monopole ($\ell=0$),
quadrupole ($\ell=2$) and hexadecapole ($\ell=4$), since they contain 
the bulk of cosmological information.
(Recall that these are the only moments that appear at zeroth (linear) order of perturbation theory.) 
To compute these at next-to-leading order, we will take into account all terms induced by
the one-loop corrections up to $\mathcal O(\mu^8)$.

The final expression for the 
galaxy power spectrum multipoles
follows from Eq.~\eqref{eq:mastz} and can be written analogously to Eq.~\eqref{Pg};
\begin{subequations}
\label{Pell}
\begin{align}
\notag
 P_{0}(z,k)= & ~(P^{\text{lin}}_{0,\theta \theta}(z,k)+ 
P^{\text{1-loop, SPT}}_{0,\theta \theta}(z,k))
 +b_1(z) (P^{\text{lin}}_{0,\theta \delta}(z,k)
 + P^{\text{1-loop, SPT}}_{0,\theta \delta}(z,k)) \\
 &   \notag
 +b_1^2(z)(P^{\text{lin}}_{0,\delta\delta}(z,k)
  +P^{\text{1-loop, SPT}}_{0,\delta\delta}(z,k)) +0.25b_2^2(z)\mathcal{I}_{\delta^2\delta^2}(z,k)\\
  \notag
 &+b_1(z)b_2(z)\mathcal{I}_{0,\delta \delta^2}(z,k)
+b_2(z)\mathcal{I}_{0,\theta \delta^2}(z,k)
+b_1(z)b_{\mathcal{G}_2}(z)\mathcal{I}_{0,\delta \mathcal{G}_2}(z,k)
 \\
 \notag
& +b_{\mathcal{G}_2}(z)\mathcal{I}_{0,\theta \mathcal{G}_2}(z,k)
+b_2(z)b_{\mathcal{G}_2}(z)\mathcal{I}_{\delta^2\mathcal{G}_2}(z,k)
+b_{\mathcal{G}_2}^2(z)\mathcal{I}_{\mathcal{G}_2\mathcal{G}_2}(z,k) \\
& \notag +(2b_{\mathcal{G}_2}(z)+0.8b_{\Gamma_3}(z))
(b_1(z)\mathcal{F}_{0,\delta \mathcal{G}_2}(k)+\mathcal{F}_{0,\theta \mathcal{G}_2}(z,k))\\
&+c_0(z) P_{0,\nabla^2\delta}(z,k)+ \tilde{c}(z) P_{0,\nabla^4_\z\delta}(z,k)+
P_{\text{shot}}(z)\,,\\
\notag
P_{2}(z,k)= &~
(P^{\text{lin}}_{2,\theta \theta}(z,k)
+P^{\text{1-loop, SPT}}_{2,\theta \theta}(z,k))
+b_1(z)(P^{\text{lin}}_{2,\theta \delta}(z,k)
+P^{\text{1-loop, SPT}}_{2,\theta \delta}(z,k))\\
&\notag 
+b_1^2(z)P^{\text{1-loop, SPT}}_{2,\delta \delta}(z,k)
+b_1(z)b_2(z)\mathcal{I}_{2,\delta \delta^2}(z,k)
+b_2(z)\mathcal{I}_{2,\theta \delta^2}(z,k)
\\
\notag
& 
+b_1(z)b_{\mathcal{G}_2}(z)\mathcal{I}_{2,\delta \mathcal{G}_2}(z,k)
+b_{\mathcal{G}_2}(z)\mathcal{I}_{2,\theta \mathcal{G}_2}(z,k)
+(2b_{\mathcal{G}_2}(z)+0.8b_{\Gamma_3}(z))\mathcal{F}_{2,\theta\mathcal{G}_2}(z,k)
\\
&
+c_2(z) P_{2,\nabla^2\delta}(z,k)
+ \tilde{c}(z) P_{2,\nabla^4_\z\delta}(z,k)\,,
\\
\notag 
P_{4}(z,k)= & ~
(P^{\text{lin}}_{4,\theta \theta}(z,k)+
P^{\text{1-loop, SPT}}_{4,\theta \theta}(z,k))
+b_1(z)P^{\text{1-loop, SPT}}_{4,\theta \delta}(z,k)+b_1^2(z)P^{\text{1-loop, SPT}}_{4,\delta\delta}(z,k)\\
&+b_2(z)\mathcal{I}_{4,\theta \delta^2}(z,k)+b_{\mathcal{G}_2}(z)\mathcal{I}_{4,\theta \mathcal{G}_2}(z,k)+
c_4(z) P_{4,\nabla^2\delta}(z,k)+ \tilde{c}(z) P_{4,\nabla^4_\z\delta}(z,k)\,,
\end{align}
\end{subequations}
where $P_{\delta \delta}$, $P_{\theta \delta}$, $P_{\theta \theta }$ are 
the auto- and cross-spectra of the density field $\delta$ and the velocity divergence field $\theta$.
The different contributions ${\cal I}_{\ell,n}$ and ${\cal F}_{\ell,n}$ 
are redshift-space generalizations of the real-space bias loop integrals 
\eqref{eq:ireals}. Note that 
the basis of counterterms has been changed to have a single free coefficient
for each multipole moment, and
the new contributions are defined as
\be
\begin{split}
 P_{\ell,\nabla^2\delta}(z,k)\equiv 
 \frac{2\ell+1}{2}\int_{-1}^1d\mu\,\mathcal{L}_\ell(\mu) \mu^{\ell} f^{\ell/2} k^2 P_{\rm lin}(k)\,.
\end{split} 
\ee
The mapping between the old and new coefficients is given by\footnote{This mapping, strictly speaking, is not exact
when IR resummation and the AP effect are present, 
but we have checked that the residual difference is smaller both than our baseline accuracy of $0.1\%$
and the size of the two-loop corrections.} 
\be
\begin{split}
 c_0 & \equiv  \tilde{c}_0+\frac{f}{3}\tilde{c}_2 +\frac{f^2}{5} \tilde{c}_4 \,,\quad c_2 \equiv  \tilde{c}_2 + \frac{6 f}{7} \tilde{c}_4 \,,\quad 
 c_4 \equiv \tilde{c}_4\,.
\end{split}
\ee

\subsection{IR Resummation}
\label{sec:irresth}

As previously
discussed, IR resummation is imperative to 
properly describe the spread of the BAO peak. 
In this Section we present our implementation of this effect, using
two closely related, but distinct, 
approaches in the real and redshift space cases. Since the large bulk flows affect only the BAO wiggles, 
the common starting point is to split the linear power spectrum into 
the smooth $P_{\text{nw}}$ and wiggly component $P_\text{w}$;
\be
P_{\text{lin}}(k)= P_{\text{nw}}(k)+P_\text{w}(k)\,.
\ee
The details of the algorithm used to perform this splitting 
is given in Section~\ref{sec:approx}.

In real space we follow the approach presented in Refs.~\cite{Blas:2016sfa}, which was developed 
in the context of time-sliced Perturbation Theory (TSPT)~\cite{Blas:2015qsi}.
Following
the wiggly-smooth decomposition one computes the damping factor\footnote{Note the additional factors of $2\pi$
compared to Refs.~~\cite{Blas:2016sfa,Ivanov:2018gjr}; these are a result of using a different Fourier transform 
convention.}
\be
\label{eq:Sigma}
\Sigma^2(z)\equiv\frac{1}{6\pi^2}\int_0^{k_S}dq\, P_{\text{nw}}(z,q)\left[1-j_0\l\frac{q}{k_{osc}}\r+2j_2\l\frac{q}{k_{osc}}\r\right]\,,
\ee
where $k_{osc}$ is the wavenumber corresponding to the BAO wavelength~$\ell_{\rm BAO}\sim 110\,h/$Mpc, $j_n(x)$ are spherical Bessel functions of order~$n$,
and~$k_S$ is the scale separating the long and short modes. We use the value $k_S=0.2~h$/Mpc as advocated in Ref.~\cite{Blas:2016sfa}, even though 
any other choice in the physically relevant range $(0.05 -
0.1)~h$/Mpc produces a very similar result. 
When we perform the one-loop calculation, the residual dependence of the final result on $k_S$ 
is comparable to the two-loop wiggly contribution and hence should be treated as a small theoretical error.
Once the damping factor $\Sigma^2(z)$ is obtained, one computes the tree-level IR-resummed dark matter 
power spectrum as
\be
\label{eq:rLO}
P_{\rm mm,\, LO}(z,k)=P_{\text{nw}}(z,k)  + \e^{-k^2\Sigma^2(z)} P_{\text{w}}(z,k)\,.
\ee
The various one-loop IR-resummed 
power spectra for matter (XY=mm), galaxy (XY=gg), and the matter-galaxy cross spectrum (XY=gm) 
can be obtained from the usual one-loop integrals evaluated using $P_{\rm mm,\, LO}(z,k)$
as an input instead of the linear power spectrum. Schematically, we can write 
\be
\label{eq:irres1loop}
\begin{split}
P_{\rm XY} = P_{\rm tree,\, XY}[P_{\rm mm,\,LO}] + P_{\rm 1-loop,\,XY}[P_{\rm mm,\, LO}]\,,
\end{split} 
\ee
where the various spectra $P_{\rm tree,\, XY}$ are given by 
\be
\begin{split}
& P_{\rm tree,\,mm} = P_{\text{nw}}(z,k)  + \e^{-k^2\Sigma^2(z)} P_{\text{w}}(z,k)(1+k^2\Sigma^2(z))\,,\\
& P_{\rm tree,\,gm} = b_1 P_{\rm tree,\,mm}\,,\quad P_{\rm tree,\,gg} = b^2_1 P_{\rm tree,\,mm}\,.
\end{split}
\ee
Note that the additional term $k^2\Sigma^2(z)\e^{-k^2\Sigma^2(z)} P_{\text{w}}(z,k)$ prevents double-counting of the bulk flow contributions
that are contained in the one-loop expression.

Let us now focus on the redshift-space 
power spectrum of galaxies.
IR resummation becomes more complicated in this case, since
the tree-level IR resummed matter power spectrum
picks up non-trivial angular dependence from the anisotropic damping factor~\cite{Ivanov:2018gjr}, 
\be
\label{Plo}
P_{\text{mm, LO}}(z, k,\mu) \equiv  (b_1(z)+f(z)\mu^2)^2 \left(P_{\text{nw}}(z, k)+\e^{-k^2\Sigma^2_{\rm tot}(z, \mu)}P_\text{w}(z, k)\right)\,,
\ee
where we have introduced the new damping function, which depends on
the logarithmic growth factor, $f(z)$;
\be
\label{eq:sigmatot}
\Sigma^2_\tot(z,\mu)=(1+f(z)\mu^2(2+f(z)))\Sigma^2(z)+f^2(z)\mu^2(\mu^2-1)\delta\Sigma^2(z)\,.
\ee
This is a function of
the real-space damping \eqref{eq:Sigma} and on a new contribution,
\be
\begin{split}
&\delta\Sigma^2(z)\equiv \frac{1}{2\pi^2}\int_0^{k_S}dq\,P_{\text{nw}}(z,q)j_2\l\frac{q}{k_{osc}}\r \,.
\end{split}
\ee
Due to the anisotropy of the BAO damping, the one-loop calculation
strictly requires 
computation of anisotropic loop integrals, which
in contrast to the real space case, 
cannot be reduced to one-dimension. 
However, these can be simplified 
by splitting the one-loop contribution itself into a smooth 
and wiggly part. More precisely, one first computes the usual redshift-space one-loop integrals with a smooth 
part only. 
Second, one evaluates the same integrals with one insertion of the unsuppressed wiggly power spectrum 
and applies the direction-dependent damping factor \eqref{eq:sigmatot} to the output, giving~\cite{Baldauf:2015xfa}
\be
\label{eq:irres1loopfact}
\begin{split}
	& P_{\rm gg}(z, k,\mu) =   (b_1(z) + f(z)\mu^2)^2\left(  P_{\text{nw}}(z, k) 
+ \e^{-k^2\Sigma^2_{\text{tot}}(z,\mu)}
P_{\text{w}}(z,k)(1+k^2\Sigma^2_{\rm tot}(z,\mu))
	\right) \\
	& + P_{\text{gg, nw, RSD, 1-loop}}(z,k,\mu) +
\e^{-k^2\Sigma^2_{\text{tot}}(z,\mu)} P_{\text{gg, w, RSD, 1-loop}}(z,k,\mu)\,.
\end{split}
\ee
Here $P_{\text{...1-loop}}[P_\lin]$ are treated as functionals of the input linear power spectrum;
\be 
\begin{split}
& P_{\text{gg, nw, RSD},\,\text{1-loop}}(z, k,\mu)\equiv P_{\text{gg, RSD, 1-loop}}[P_{\text{nw}}] \,,\\
& P_{\text{gg, w, RSD},\,\text{1-loop}}(z,k,\mu)\equiv P_{\text{gg, RSD, 1-loop}}[P_{\text{nw}}+P_{\text{w}}]
-P_{\text{gg, RSD, 1-loop}}[P_{\text{nw}}] \,.
\end{split}
\ee
For simplicity we have neglected the one-loop contributions 
obtained from two insertions of the wiggly power spectrum (since these 
scale as $P_{\rm w}^2$).
Once the two contributions $P_{\rm gg, w}$ and $P_{\rm gg, nw}$ are summed, 
the eventual IR-resummed anisotropic power spectrum 
can be used to compute the multipoles 
in Eq.~\eqref{eq:mult}.

It is important to stress that our implementation of IR resummation at one loop order contains four potential sources of error:
\begin{itemize}
	\item Imperfectness of the wiggly-non-wiggly decomposition;
	\item Dependence of the damping factor on the separation cutoff;
	\item Inaccuracy of the factorization prescription;
	\item One-loop corrections of $\mathcal{O}(P_{\rm w}^2)$ from two insertions of $P_{\rm w}$.
\end{itemize}
In Refs.~\cite{Blas:2016sfa,Ivanov:2018gjr} is was shown that these effects are smaller than the two-loop 
contribution. 
Furthermore, it can be shown that these errors 
can be consistently subtracted 
and shifted to the next order
at any given order of perturbation theory.
This will be
additionally discussed 
in Section~\ref{sec:approx}.

\subsection{Alcock-Paczynski Effect} 

The observed galaxy distribution is a function of angles and redshifts. 
However, it is more convenient to switch 
to the geodesic distances between galaxies and consider the ``de-projected'' 3-dimensional 
power spectrum instead of the 2-dimensional angular power spectrum (at redshift $z$)~\cite{1978ApJ...221....1D}. 
In practice, this change of coordinates is realized by 
means of 
assuming 
some trial fiducial cosmology~\cite{Matsubara:1996nf,Ballinger:1996cd,Heinesen:2018hnh,Heinesen:2019phg}. 
Importantly, if the trial cosmology differs
from the correct one, the reconstructed 3D power 
spectrum appears distorted,
which known as the Alcock-Paczynski effect~\cite{Alcock:1979mp}.
These effects
are routinely used to constrain cosmological parameters from galaxy surveys, 
see e.g.~\cite{Alam:2016hwk}.

One does not need, of course, to assume a wrong cosmology 
to generate 
Alcock-Paczynski distortions.
If the fiducial cosmology is correct, there are no distortions in the data, but they are present 
in the theoretical templates that are fitted to these data. 
After all, the Alcock-Paczynski coordinate conversion is only a technical tool to extract 
the distance information that is encoded in the angle- and redshift-dependence of the galaxy distribution.
Mathematically, it does not change the information content of the galaxy power spectrum.

To account for the Alcock-Paczynski effect one has to compute the observable galaxy power spectrum 
using the following formula:
\be
P_\obs(z,k_\obs,\mu_\obs) = P_{\rm gg}(z,k_\true[k_\obs,\mu_\obs],\mu_\true[\mu_{\obs}]) \cdot \frac{D^2_{A,\fid}(z)H_\true(z)}{D^2_{A,\true}(z)H_\fid(z)},
\ee
where $k_{\text{true}}$ and $\mu_{\text{true}}$ are the values that one would obtain in the true cosmology (and those used to evaluate the theory model), 
whereas $k_{\text{obs}}$ and $\mu_{\text{obs}}$ refer to quantities 
in the
fiducial cosmology that was used
to build galaxy catalogs. 
The relation between the true and observed wavenumbers 
and angles is given by (suppressing the explicit time dependences)
\be
\begin{split}\label{AP_k_mu}
	k^2_\true&=k^2_\obs\left[\l\frac{H_\true}{H_\fid}\r^2\mu_\obs^2+\l\frac{D_{A,\fid}}{D_{A,\true}}\r^2(1-\mu_\obs^2)\right]\,,\\
	\mu^2_\true&=\l\frac{H_\true}{H_\fid}\r^2\mu^2_\obs\left[\l\frac{H_\true}{H_\fid}\r^2\mu_\obs^2+\l\frac{D_{A,\fid}}{D_{A,\true}}\r^2(1-\mu_\obs^2)\right]^{-1}\,.
\end{split}
\ee
These formulas realize the map $(k_{\text{true}},\mu_{\text{true}})\to (k_{\text{obs}},\mu_{\text{obs}})$, which 
is used in our code.

During the likelihood analysis one samples cosmological parameters in an attempt to find the 
true vales $H_\true$ and $D_{A,\true}$ given the fiducial $H_\fid$ and $D_{A,\fid}$ 
used to 
create catalogs. Including the AP effect, the final galaxy multipoles
are given by
\be 
\label{eq:APPl}
P_{\ell,\text{AP}}(z,k) = \frac{2\ell+1}{2} \int_{-1}^1 d\mu_{\text{obs}}\, P_{\text{obs}}(z,k_{\text{obs}},\mu_{\text{obs}}) \cdot 
\mathcal{L}_\ell(\mu_{\text{obs}})\,.
\ee

Note that the AP effect and IR resummation lead to the leakage of some 
bias contributions to higher order multipoles. 
For instance, in the absence of these effects the term ${\cal I}_{\delta^2\delta^2}$
only contributes to the monopole moment, whilst including 
the AP effect produces some non-trivial angle-dependence and 
generates contributions into higher multipole moments.
\texttt{CLASS-PT} explicitly computes these contributions, 
though we
drop them in the \texttt{python} wrapper 
\texttt{classy} for memory optimization reasons, since they are found to be highly negligible. 
The plots with these contributions can be found in the \texttt{Mathematica} notebook
in the code web folder.

\subsection{Tree-level IR-resummed Bispectrum }

The tree-level IR-resummed bispectrum in real space can be easily obtained form our code as well. This is easily formed by
taking the usual expression for the tree-level
matter bispectrum and replacing $P_{\rm lin}(z,k)$ with 
the leading order IR-resummed spectrum given in~\eqref{eq:rLO}. Note that this replacement is the exact result in
real space. In redshift space, one should 
use the anisotropic expression~\eqref{Plo}   
and consistently average over the angular variables that include the AP effect;a
procedure that will be implemented in future versions of~\texttt{CLASS-PT}.

\section{Structure of the Code}
\label{sec:code}

Our code is executed as a module, \texttt{nonlinear\_pt.c}, in the standard \texttt{CLASS} code \texttt{v2.6.3.}, which was the latest CLASS version when the work on the code started. The new module 
is implemented
as a clone of the \texttt{nonlinear.c} module that evaluates HALOFT. 
A work cycle of our modifications 
can be schematically represented by 
a sequence of the following three steps:
\begin{enumerate}
\item  The function \texttt{nonlinear\_pt\_pk\_l()} takes the linear transfer functions from the module \texttt{perturbations.c} and convolves them 
with the primordial power spectrum from \texttt{primordial.c} to get the linear matter power spectra at redshifts specified by the user;
\item For each required redshift, various non-linear power spectra are evaluated 
by the function \texttt{nonlinear\_pt\_loop()}, which uses 
FFTLog with pre-computed cosmology-independent matrices $M$ (see Eq.~\eqref{eq:matricesM});
\item These spectra are passed to subsequent modules similarly to the non-linear spectra 
computed by HALOFIT in \texttt{nonlinear.c}\,.
\item Alternatively, there is an option to use external linear $P(k)$ instead of the one computed directly 
by \texttt{CLASS}.
\end{enumerate}

The most important ingredients that made our FFTLog calculation 
possible are the CLASS realization of FFT 
developed in \texttt{CLASS-matter}\footnote{\href{https://github.com/lesgourg/class_public/tree/class_matter}{
\textcolor{blue}{https://github.com/lesgourg/class\_public/tree/class\_matter
}}
} (see Ref.~\cite{Schoneberg:2018fis}),
and the fast matrix multiplication algorithms included 
in the
open-source \texttt{C} library \texttt{OpenBLAS}\footnote{\href{https://github.com/xianyi/OpenBLAS}{
\textcolor{blue}{https://github.com/xianyi/OpenBLAS
}}
}.
We stress that \texttt{OpenBLAS} is the only external library used in our code.
It is free and its installation is fast and straightforward. 
A detailed installation manual for \texttt{CLASS-PT} is given in Appendix~\ref{app:manual}.

Crucially, our alterations do not 
alter the way \texttt{CLASS} works. 
The module is written in \texttt{C} and it is wrapped as a \texttt{python}
library \texttt{classy}. Compared to the usual \texttt{classy}, just 
one function is modified, \texttt{pk(k,z)}, and several new functions added.
Examples of working sessions of our code are given in the \texttt{Jupyter}
notebook available at the code webpage. 

Several
important flags 
regulate our non-linear module:
 
\begin{itemize}
	\item \texttt{non-linear = PT}: If set, this flag executes the non-linear module. The syntax here is analogous to
	the one used to execute the HALOFIT module, ``\texttt{non-linear = Halofit}''.
	\item \texttt{z\_pk=0,0.61,1100}: Redshifts for which the non-linear corrections should be computed.
	\item \texttt{IR resummation = Yes}: Decide whether IR resummation is performed.
	\item \texttt{Bias tracers = Yes}: Decide whether the loop integrals for biased tracers are computed.
	\item \texttt{RSD = Yes}: Decide whether the redshift-space loop integrals are computed.
    \item \texttt{AP = Yes}: If \textit{both} ``\texttt{RSD}'' and ``\texttt{IR resummation}'' are switched on, this activates calculation of
    the AP effect. 
    \item \texttt{Omfid=0.31}: Pass the fiducial value of $\Omega_{m}$ that was used to create the catalogs
    with the AP effect. \textit{One must never scan over this parameter in an MCMC analysis}. The value of $\Omega_{m,\,{\rm fid}}$ must be always fixed to the one used in the survey data production.
     \item \texttt{FFTLog mode=Normal}: Depending on a particular situation, the user can either run the code with high precision settings (which is a default choice), or in the fast mode, which is slightly less accurate but much faster. 
     This regime can be activated 
     by the flag ``\texttt{FFTLog mode=FAST}'' If the code is run in the default regime, the flag ``\texttt{FFTLog mode}''
     does not need to be specified.
      \item \texttt{output format=Normal}: This flag speficies the size of the wavenumber grid used to compute and store the power spectra. 
      By default, our module uses the standard \texttt{CLASS} array of wavenumber. 
      However, if one computes both the CMB power spectra $C_\ell$ and the non-linear power spectra $P(k)$ in one \texttt{CLASS} call,
      one is encouraged to use the flag ``\texttt{output format=FAST}.'' 
      In this case the non-linear power spectra are stored on a reduced wavenumber grid, 
      which leads to a notable gain in speed.
      \item \texttt{cb=Yes}. By default, \texttt{CLASS-PT} uses the
      linear power spectrum of the cold dark matter and baryon (``cb'') fluid 
      as an input for the non-linear calculations. This is motivated by
      the evidence that galaxies trace the cb-fluid and not the total matter density that 
      includes massive neutrinos. If the user is willing to compute the non-linear corrections to the total 
      matter density field, they should use the flag ``\texttt{cb=No}.''
      \item \texttt{External Pk = file\_pk.dat}. 
      This option should be activated if the user wants to perform nonlinear calculations with 
      an external linear matter power spectrum (tabulated in e.g.~\texttt{`file\_pk.dat'}).
\end{itemize}

A concrete understanding of 
our module's architecture 
is not essential for its running. 
Moreover, it will likely change in the future to match the most recent official version of \texttt{CLASS}.
We warn the users that some functions that exist in the current version of  
\texttt{CLASS-PT} are redundant
and will be optimized in the future. 
The current stable version of the code is available at 
\href{https://github.com/Michalychforever/CLASS-PT}{
\textcolor{blue}{https://github.com/Michalychforever/CLASS-PT}}, where modifications will be commented upon.

\section{Technical Implementation and Approximations}
\label{sec:approx}

Numerical algorithms and approximations 
are essential elements of our code.
In this Section, we describe 
some of these technical details including the FFTLog algorithm used to evaluate loop integrals, 
the wiggly-non-wiggly decomposition, as well as their application to the
evaluation of the IR-resummed redshift-space galaxy power spectrum multipole moments.

\subsection{Basics of the FFTLog Method}

The one-loop perturbation theory integrals
involve convolution kernels that reduce to simple multiplications in position space.
This inspired methods that evaluate these by using
the
Fast Fourier Transform (FFT) to switch between Fourier and position space to evaluate these integrals~\cite{Schmittfull:2016jsw,McEwen:2016fjn}.
Alternatively, FFTs (with uniform binning in $\log k$) can be used instead 
as a tool to decompose 
the linear power spectrum into complex power laws~\cite{Hamilton:1999uv}. 
Whilst the loop integrals are then {\em not} deconvolved, 
they have a simple
analytical solution for the power-law universes~\cite{Simonovic:2017mhp}. We refer to this particular
approach as the FFTLog method. 
Importantly, it 
can be extended to the one-loop bispectrum
and the two-loop power spectrum~\cite{Simonovic:2017mhp}, which cannot be written as 
not simple convolution integrals.\footnote{For
some related results regarding the two-loop power spectrum see also~\cite{Schmittfull:2016yqx,Slepian:2018vds}.}
Keeping in mind these statistics as our eventual goal, 
we choose FFTLog for the 
evaluation of loop integrals in \texttt{CLASS-PT}.
Another advantage of this algorithm is that it is very easy to implement, since it boils down to simple 
multiplications of the cosmology-independent matrices
with the cosmology-dependent vectors, which can be easily obtained 
from FFTs of the linear matter power spectrum. 

The discrete approximation to the linear power spectrum
in a finite momentum interval $[k_{\rm min},k_{\rm max}]$, denoted as~$\bar P(z,k)$,
can be written as
\be
\bar P_{\rm lin}(z,k)=\sum_{m=-N/2}^{m=N/2}c_m k^{\nu +i\eta_m}\,,
\ee 
where the Fourier coefficients $c_m$ and exponents $\eta_m$ are given by
\be
\begin{split}
c_m =  \frac{1}{N}\sum_{j=0}^{N-1}P_{\rm lin}(z,k_j)k_j^{-\nu}k_{\rm min}^{-i\eta_m} e^{-2\pi i m j/N}\,,
\quad \eta_m = \frac{2\pi m}{\ln(k_{\rm max}/k_{\rm min})}\,.
\end{split} 
\ee
The parameter $\nu$ is sometimes refered to as~``bias''. 
In principle, $\nu$ can be an arbitrary real number, however, the convergence 
properties of convolution integrals on different scales 
vary 
depending on the value $\nu$. Thus, the freedom to choose $\nu$ can be used to 
boost the efficiency of numerical evaluation. 

As mentioned above, the perturbation theory loop integrals over each power-law function $k^{\nu +i\eta_m}$ can be done analytically, which
allows one to reduce the evaluation of the whole loop integral to a matrix multiplication.
Crucially, the elements of this matrix are cosmology-independent and can be pre-computed and saved as a table. 
All the cosmology-dependence resides in the coefficients $c_m$,
whose evaluation takes very little time by virtue of the FFT algorithm.

In order to find analytical solutions for the loop integrals with power-law power spectra, 
the integration has to be performed over the whole momentum range, i.e.~for $q\in [0,\infty]$.
This implies that perturbation theory loop integrals are evaluated with the same integration boundaries.
One may be worried about this in the context of perturbation theory, since we are integrating 
over the small scales where the perturbative description breaks down. However, as 
already emphasized, 
the purpose of counterterms in the EFT approach is precisely to absorb all small scale dependence of the loop integrals.
In this way, it is guaranteed that the final results are independent of the exact 
short-distance behavior of the power spectrum.\footnote{Alternatively, one may introduce a 
UV cutoff $\Lambda$ by simply padding the power spectrum with zeros for all wavenumbers 
$k\geq \Lambda$. In this case the EFT counterterms  
absorb the cutoff dependence of the loops and ensure that the final result
for the one-loop power spectrum does not depend on $\Lambda$. 
Thus, even though we use $\Lambda = \infty$, this choice 
is irrelevant for the cosmological constraints and 
can only affect the amplitudes of the counterterms.
}

\subsubsection{FFTLog in Redshift Space}

In real space all one-loop integrals can be expressed in terms of the single `master integral'
\be
\label{eq:Imast}
\int_\q \frac{1}{q^{2\nu_1}|\k-\q|^{2\nu_2}} = k^{3-2\nu_{12}}{\sf I}(\nu_1,\nu_2)\,,
\ee
where $\nu_{12}\equiv \nu_1+\nu_2$ and 
\be
{\sf I}(\nu_1,\nu_2) \equiv \frac{1}{8\pi^{3/2}}\frac{\Gamma\left(\frac{3}{2}-\nu_1\right)
\Gamma\left(\frac{3}{2}-\nu_2\right)
\Gamma\left(\nu_{12}-\frac{3}{2}\right)
}{\Gamma(\nu_1)\Gamma(\nu_2)\Gamma(3-\nu_{12})}\,.
\ee
for Gamma function $\Gamma$. However, in redshift space the loop integrals become more complicated due to the 
anisotropy introduced by the line-of-sight direction $\hat\z$. 
One can find up to four loop momenta multiplying $\hat\z$ in the one-loop integrands.
To evaluate these integrals, we generalize 
Eq.~\eqref{eq:Imast} as follows:
\be
\label{eq:intAs}
\begin{split}
& \int_{\q} \frac{q^i}{q^{2\nu_1}|\k-\q|^{2\nu_2}} = k^{3-2\nu_{12}} \cdot A_1 k^i \;,\\
&\int_{\q} \frac{q^iq^j}{q^{2\nu_1}|\k-\q|^{2\nu_2}} = k^{3-2\nu_{12}} \cdot (k^2 A_2 \mathcal O_{2a}^{ij} + B_2 \mathcal O_{2b}^{ij} ) \;,\\
& \int_{\q} \frac{q^iq^jq^l}{q^{2\nu_1}|\k-\q|^{2\nu_2}} = k^{3-2\nu_{12}} \cdot (k^2 A_3 \mathcal O_{3a}^{ijl} + B_3 \mathcal O_{3b}^{ijl} ) \;,\\
& \int_{\q} \frac{q^iq^jq^lq^m}{q^{2\nu_1}|\k-\q|^{2\nu_2}} = k^{3-2\nu_{12}} \cdot (k^4 A_4 \mathcal O_{4a}^{ijlm} + k^2 B_4 \mathcal O_{4b}^{ijlm} +C_4 \mathcal O_{4c}^{ijlm} ) \;,
\end{split}
\ee
where $A_n,B_n$ and $C_n$ are some functions of $\nu_1$ and $\nu_2$, and 
we have introduced the following operators:
\be
\begin{split}
& \mathcal O_{2a}^{ij} = \delta^{ij} \;, \qquad \mathcal O_{2b}^{ij} = k^ik^j \;,\\
& \mathcal O_{3a}^{ijl} = \frac13(\delta^{ij}k^l+{\rm 2\; perms.}) \;, \qquad \mathcal O_{3b}^{ijl} = k^ik^jk^l \;,\\
& \mathcal O_{4a}^{ijlm} = \frac13(\delta^{ij}\delta^{lm}+{\rm 2\; perms.}) \;, \qquad \mathcal O_{4b}^{ijlm} = \frac16(\delta^{ij} k^l k^m+{\rm 5\; perms.}) \;\\
& \mathcal O_{4c}^{ijlm} = k^ik^jk^lk^m \; .
\end{split}
\ee
By contracting the left hand sides of the integrals \eqref{eq:intAs} with 
different powers of $\q$ and $\k$, one can reduce these integrals to the form \eqref{eq:Imast}. 
The resulting formulas are 
a set of simple algebraic equations 
that can be solved to find the functions $A_n,B_n$ and $C_n$. The explicit solutions can be found in Appendix~\ref{app:rsdfft}. 
Plugging these expressions into \eqref{eq:intAs}, it is straightforward to obtain the following redshift-space master integrals:
\be
\begin{split}
& \int_{\q} \frac{(\hat{\z}\cdot \q)}{q^{2\nu_1}|\k-\q|^{2\nu_2}} = k^{3-2\nu_{12}} \cdot k \mu \; A_1 \;,\\
& \int_{\q} \frac{(\hat{\z}\cdot \q)^2}{q^{2\nu_1}|\k-\q|^{2\nu_2}} = k^{3-2\nu_{12}} \cdot k^2 (A_2 + \mu^2 B_2) \,,\\
& \int_{\q} \frac{(\hat{\z}\cdot \q)^3}{q^{2\nu_1}|\k-\q|^{2\nu_2}} = k^{3-2\nu_{12}} \cdot k^3\mu (A_3+\mu^2 B_3) \,,\\
& \int_{\q} \frac{(\hat{\z}\cdot \q)^4}{q^{2\nu_1}|\k-\q|^{2\nu_2}} = k^{3-2\nu_{12}} \cdot k^4 (A_4+\mu^2 B_4 + \mu^4 C_4) \;.
\end{split}
\ee

With these formulas in 
hand one can compute the one-loop redshift-space integrals
in the discrete FFTLog representation just like in the real-space case~\cite{Simonovic:2017mhp}. 
Crucially, the dependence on $\mu$ is given by simple polynomials, e.g.~the one-loop matter 
power spectrum takes the following form;
\be
\label{eq:P1lzs}
P_{\rm 1-loop,RSD}(z, k,\mu)=(1+f(z)\mu^2)\sum_{n=0}^3 P^{(n)}_{13}(z,k)\mu^{2n}+\sum_{n=0}^4 P^{(n)}_{22}(z,k)\mu^{2n}\,,
\ee
and each $P^{(n)}$ can be computed via FFTLog in full analogy with the real-space case
\be
\label{eq:matricesM}
\begin{split}
& P_{22}^{(n)}=k^3 D^4(z) \sum_{m_1,m_2}c_{m_1}k^{-2\nu_1}M_{22}^{(n)}(\nu_1,\nu_2)c_{m_2} k^{-2\nu_2}\,,\\
& P_{13}^{(n)}=k^3  P_{\rm lin}(z, k) D^2(z) \sum_{m_1}c_{m_1}k^{-2\nu_1} M^{(n)}_{13}(\nu_1)\,,\\
\end{split} 
\ee
where $M_{22}^{(0)},~M_{13}^{(0)}$
are the standard real-space matrices~\cite{Simonovic:2017mhp} and 
$M_{22}^{(n)},~M_{13}^{(n)}$ with $n>0$, are their redshift-space generalizations.
The explicit expressions for these matrices are quite cumbersome and thus not quoted here. 
They can be found in the main body of the code.

Since the $\mu$-dependence of basic perturbation theory one-loop integrals is known explicitly,
one can easily do the Legendre integrals analytically 
at the level of the FFTLog matrices. 
This allows one to obtain master matrices $M_{22,\,\ell}$ and  $M_{13,\,\ell}$.
Using these matrices each multipole can be computed with only two matrix multiplications, just like in the real-space case. This is not the case when IR resummation and the Alcock-Paczynski effect
are present. To account for them, we evaluate each integral entering \eqref{eq:P1lzs}
separately, combine them into the full $P(k,\mu)$
and then do the $\mu$-integrals numerically. This procedure will be discussed in more detail shortly. 

\subsubsection{Practical Realization}

To carry out the FFTLog, we first create a 
grid with $N_{\rm FFTLog}=256$ (default mode) and $N_{\rm FFTLog}=128$ (fast mode) harmonics spanning the range 
\[
[5\cdot 10^{-5},100]~h/{\rm Mpc}\,,
\]
and 
use two different values of the FFTLog ``bias'' exponent $\nu$ 
for the matter 
and bias tracer loop integrals (see~\cite{Simonovic:2017mhp} for details)
\be
\begin{split}
& \nu =-0.3\quad \text{(matter)}\,,\quad \nu=-1.6\quad \text{(biased tracers)} \,.
\end{split}
\ee
It is important to stress that the choice $\nu = -0.3$ leads to poor convergence for the matter one-loop integrals at small scales, $k>1~h$/Mpc.
To alleviate this issue, we apply an exponential cutoff for these high $k$'s, which
is justified because the 
one-loop predictions are not valid on these 
scales at the redshifts relevant for current and future galaxy surveys. 
If necessary, 
one can always choose a different value of the bias 
for which the FFTLog calculation will be better convergent for large wavenumbers. 

\subsubsection{Accuracy Tests}

\begin{figure}[h!]
\begin{center}
\includegraphics[width=0.49\textwidth]{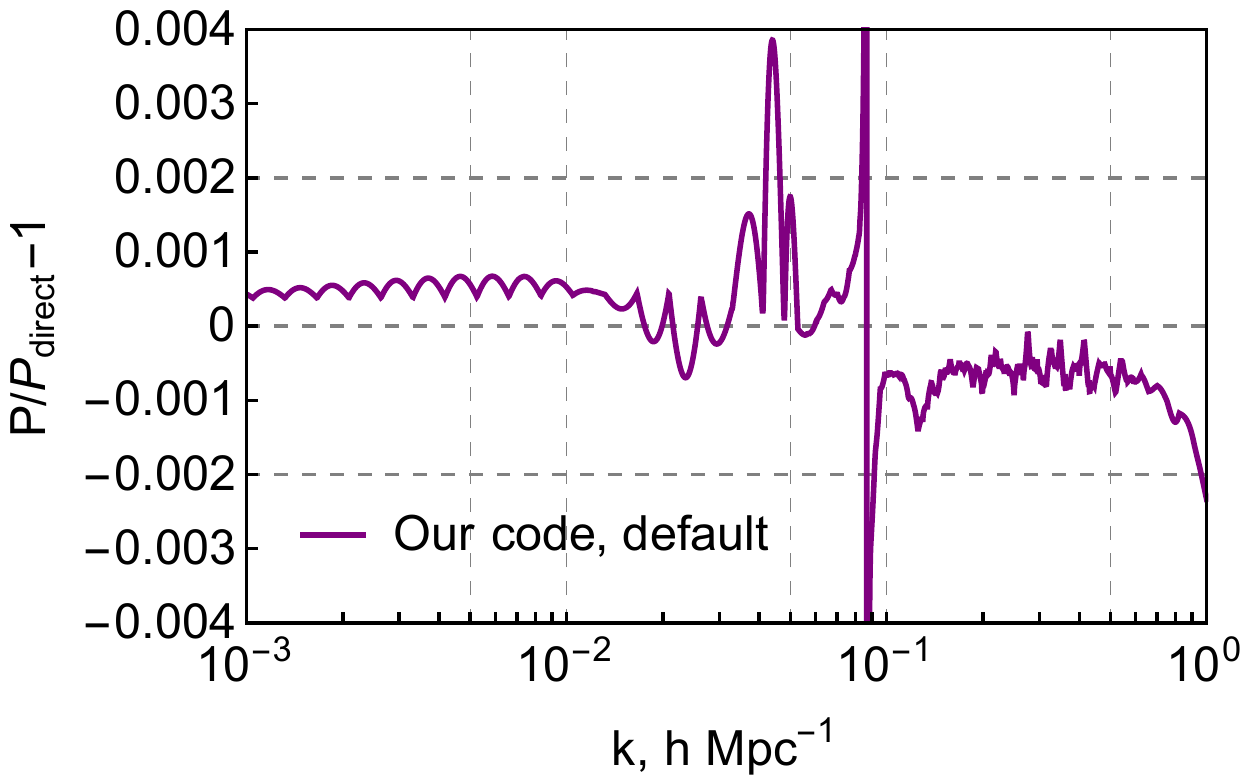}
\includegraphics[width=0.49\textwidth]{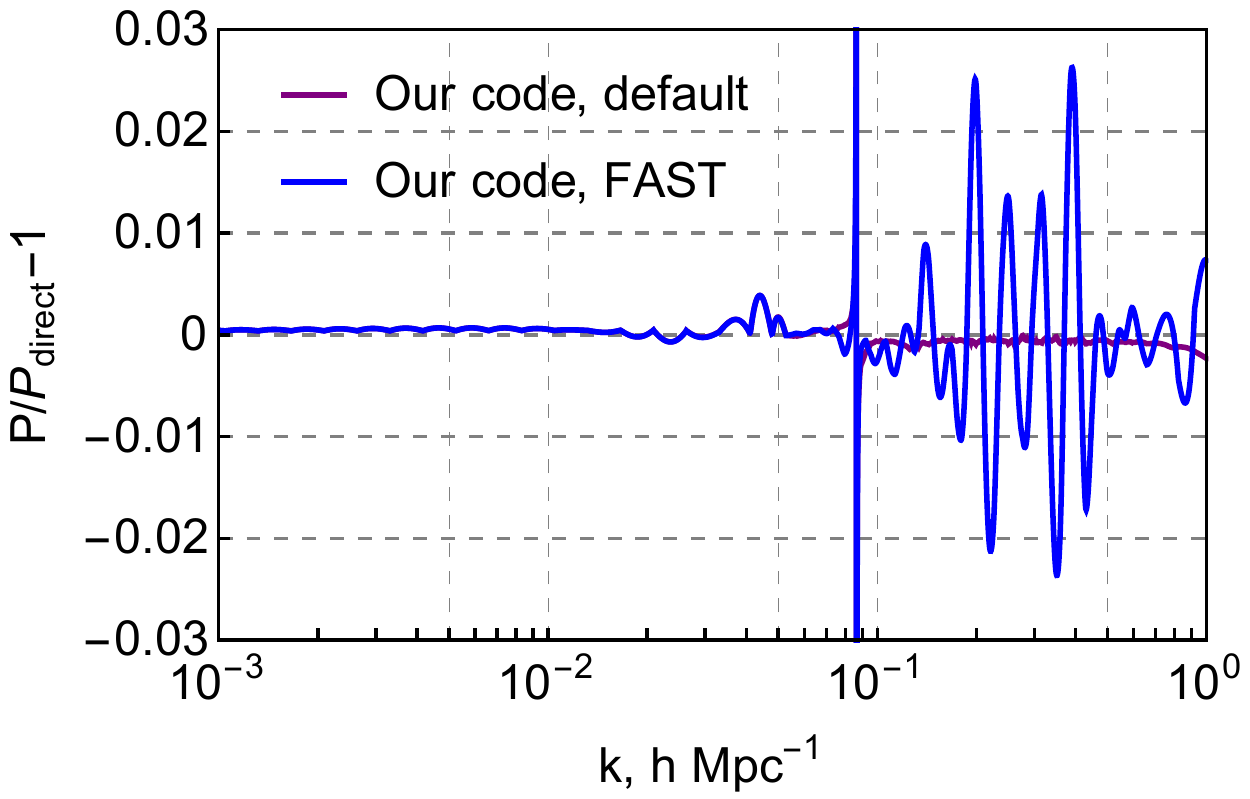}
\end{center}
\caption{\label{fig:errfftlog} 
Residuals between our calculation of the one-loop matter power spectrum contribution and the direct numerical evaluation 
for the default settings (left panel) and in the ``\texttt{FAST}'' mode (right panel).
}
\end{figure}

Let us now discuss the accuracy of our code, using 
the 
one-loop real-space calculations as an example.
The purpose of this comparison is to show that our FFTLog routine 
has comparable precision 
to that of direct numerical integration.

The residuals between the FFTLog-based calculation and the direct numerical evaluation of 
the one-loop matter power spectrum are shown in Fig.~\ref{fig:errfftlog}.
The singularity at $k\approx 0.1~h/$Mpc reflects the fact that the one-loop spectrum crosses zero in this region.
We show the results both for the default precision with $N_{\rm FFTLog}=256$
and for the fast mode with $N_{\rm FFTLog}=128$.
The 
default choice of $N_{\rm FFTLog}=256$ is seen to provide a relative 
accuracy $\sim 0.1\%$ over the range of wavenumbers
$k\lesssim 0.2~h$/Mpc, as
relevant for future galaxy surveys. 
Note that the relative accuracy quoted here does not depend on time; 
the absolute accuracy actually increases 
at high redshifts.

However, the $0.1\%$ accuracy of the one-loop correction 
can be somewhat excessive 
in many cases.
For this reason, for all practical applications the user is encouraged to run the code in the fast mode.
Whilst this 
provides us with a somewhat lower accuracy $\sim 1\%$, there is 
a significant speed gain.
On the one hand, this numerical error is still smaller than the two-loop contribution omitted in our model. 
On the other hand, the one-loop contribution itself must be a small 
correction to the linear power spectrum in order for perturbation theory
to make sense. 
Thus, the $\mathcal{O}(1\%)$ accuracy on the one-loop correction translates into the $\mathcal{O}(0.1\%)$ accuracy 
on the total power spectrum.
Therefore, the fast mode seems to be sufficient for the bulk of practical applications in which the 
one-loop power spectrum is used as a model. 
To explicitly verify this, 
we have rerun the MCMC analysis of the BOSS data from Ref.~\cite{Ivanov:2019pdj} 
and the analysis of the large N-body simulation data from Ref.~\cite{Nishimichi:2020tvu}.
In both cases, 
fast and default modes yielded indistinguishable results.

\subsection{Wiggly-non-Wiggly Splitting}

\begin{figure}[h!]
\begin{center}
\includegraphics[width=0.48\textwidth]{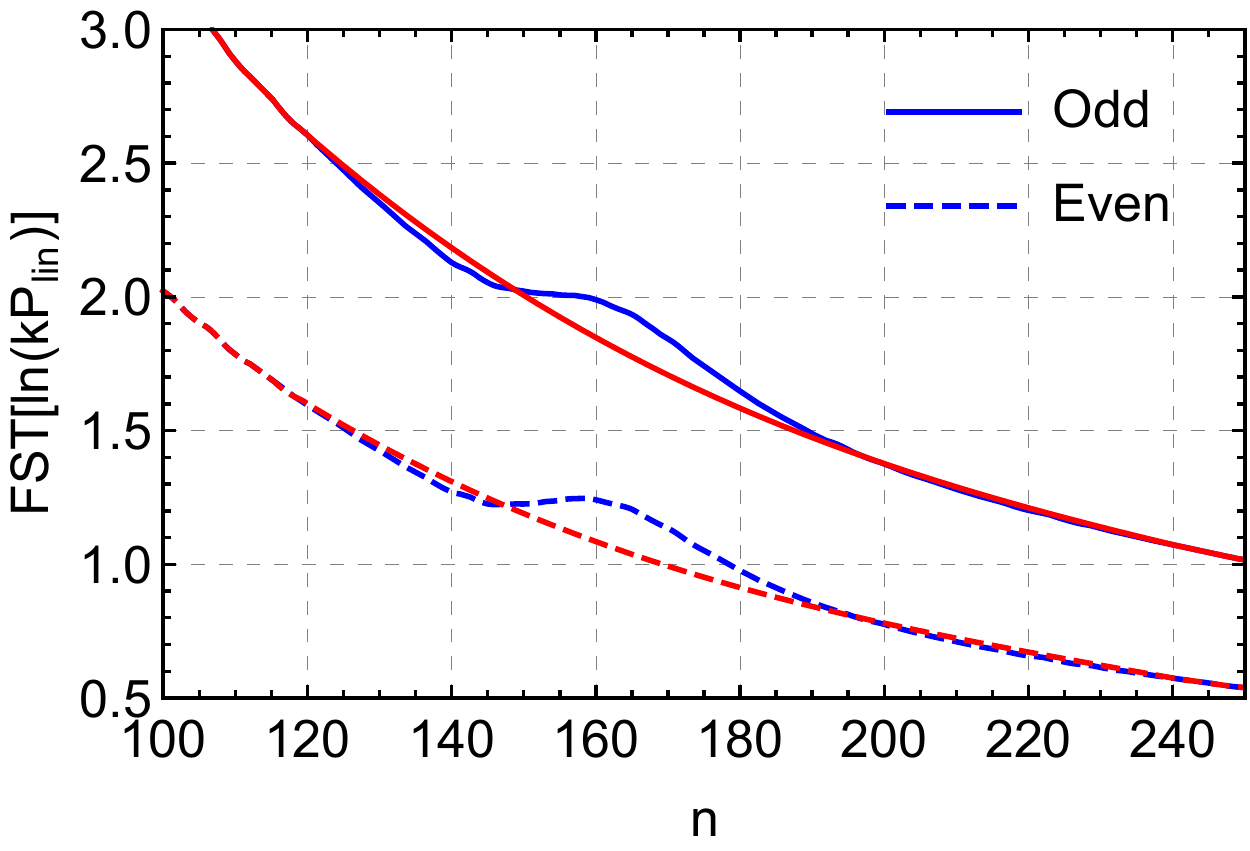}
\includegraphics[width=0.51\textwidth]{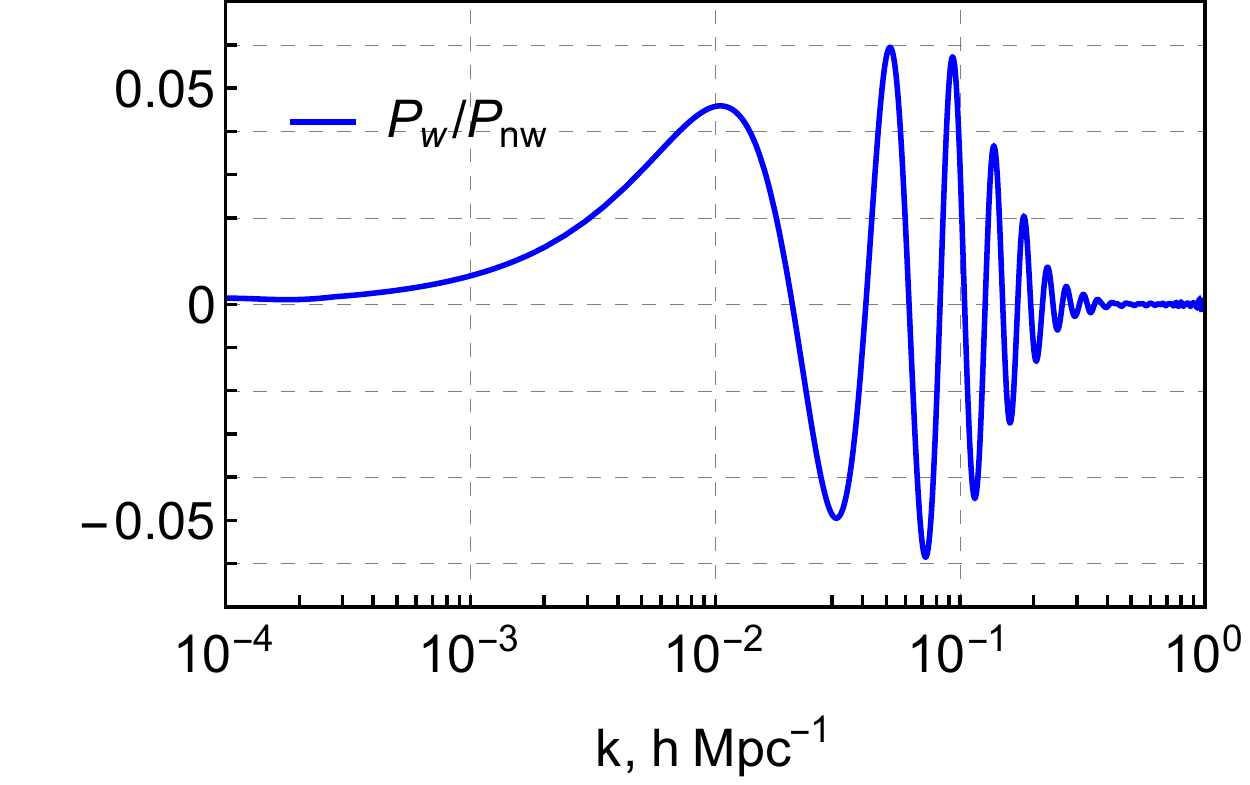}
\end{center}
\caption{\label{fig:wig}
Left: Visualization of our wiggly-smooth splitting algorithm. 
We show even and odd discrete Fast Sine Transform
coefficients of $\ln(kP_\lin(k))$ for the range of indices relevant for the BAO 
before (in blue) and after (in red) the splitting.
Right: The resulting wiggly power spectrum normalized to the smooth one.
}
\end{figure}

The algorithm for wiggly-non-wiggly splitting 
implemented in the code 
is based on the discrete spectral analysis method proposed in Ref.~\cite{Hamann:2010pw}.
The main idea is to Fourier transform the power spectrum to position space,
localize the BAO peak, remove it, and smoothly interpolate the correlation function 
in the previous location of the peak. For computational efficiency this is done by means 
of a discrete Fourier transform. In practice, we do the following:
\begin{enumerate} 
	\item Sample an array of $\ln (kP_\lin(z,k))$ in $2^{16}$ points over the range $[7\cdot 10^{-5},7]\,$Mpc$^{-1}$;
	\item Fast sine transform (FST) this array;
	\item Interpolate the odd and even harmonics using splines;
	\item Remove the harmonics spanning the range of indices $[120,240]$, see Fig.~\ref{fig:wig}. These harmonics correspond to the BAO peak
	for the comoving sound horizon at decoupling: $r_d\sim 150$ Mpc;
	We have found that this choice of the boundaries works well for the variations of $r_d$ in the range $(130,170)$ Mpc;
	\item Interpolate the FST harmonics in the BAO range; 
	\item FST 
	back the new coefficients to recover $\ln (kP_{\text{nw}}(k))$.
\end{enumerate} 

The resulting wiggly power spectrum $P_\text{w}\equiv P_\lin -P_{\text{nw}}$ is shown in the right panel of
Fig.~\ref{fig:wig}. 
It is important to stress that we work in units of Mpc, such that 
the splitting is 
insensitive to $h$. The only cosmology-sensitive part of our procedure is the location of the BAO peak,
which corresponds to the comoving sound horizon~$r_d$. However, $r_d$ is a very weak function of cosmology.
For instance, in $\L$CDM $r_d\propto \omega_m^{-0.25}\omega_b^{-0.12}$ \cite{Ivanov:2019pdj}. 
Given this reason, we use the same frequency cuts 
in the wiggly-non-wiggly procedure
during MCMC scans over different cosmologies. 
Alternatively, we have tried an algorithm which rescales the frequency cuts ``on-the-fly''
according to the value of $r_d$ which is being sampled by the code.
The difference between the two procedures is negligibly small 
and does not affect parameter inference even from the large-volume PT challenge simulation data~\cite{Nishimichi:2020tvu}.

\begin{figure}[h!]
\begin{center}
\includegraphics[width=0.49\textwidth]{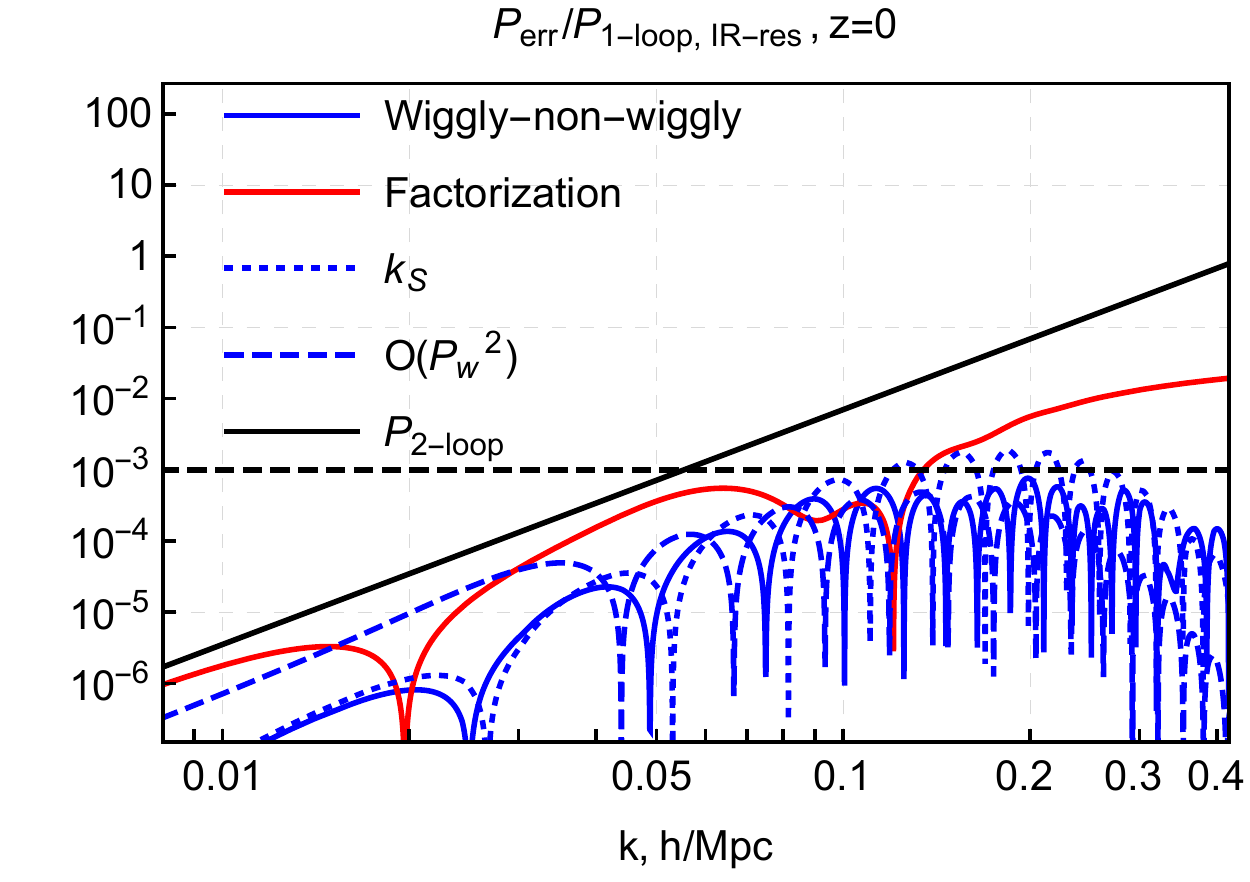}
\includegraphics[width=0.49\textwidth]{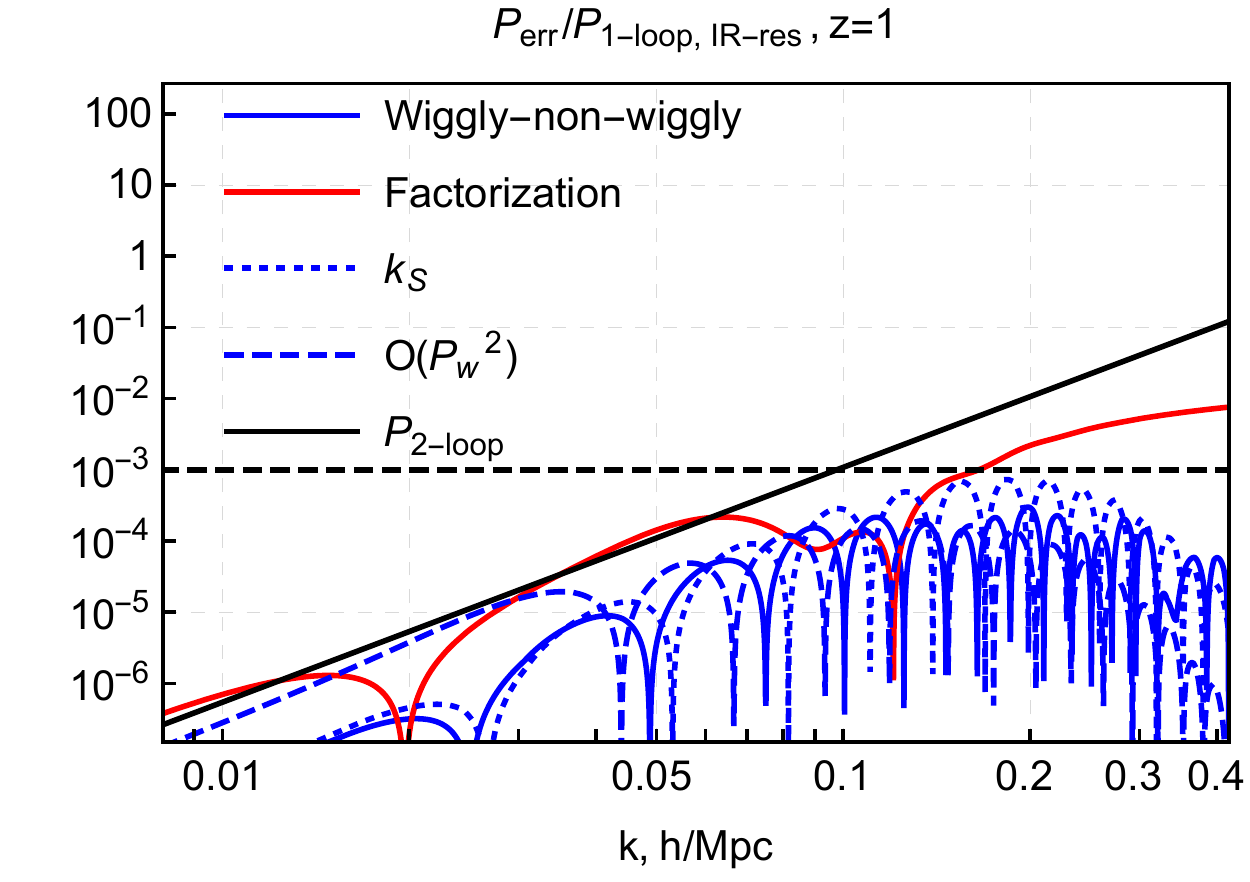}
\end{center}
\caption{\label{fig:err} Errors introduced by various approximations in IR resummation 
relative to the two-loop contribution at $z=0$ (left panel) and $z=1$ (right panel).
See the main text for detail.
}
\end{figure}

\subsection{Error Budget of IR Resummation}

In Section \ref{sec:irresth} we listed various sources of error in our implementation of IR resummation. 
It was argued that these errors are under control, i.e.~their contributions can be 
minimized to arbitrary small values. 
This is a theoretical statement, which may not hold in reality due 
to 
choices made in practical implementation. In this subsection, we will 
explicitly show that the residual error of the one-loop power spectrum including 
IR resummation
is smaller than the two-loop contributions. Let us discuss each problematic ingredient separately.

\textit{Wiggly-non-Wiggly decomposition.}
The error introduced by our splitting procedure is always smaller than the two-loop corrections.
One can argue that this is a generic statement that 
can be generalized to higher orders. 
Indeed, imagine that the BAO were described by an analytic 
harmonic function such 
that one could find an exact analytic expression for $P_{\rm nw,\,true}$.
Imagine now that instead of using this analytic expression we perform a numerical wiggly-non-wiggly decomposition 
that introduces some intrinsic error $\Delta P_{\rm w-nw}\ll P_{\rm lin}$,
\be
P_{\rm nw} =  P_{\rm nw,\,true}+\Delta P_{\rm w-nw}\,,\quad P_{\rm w} =   P_{\rm w,\,true}-\Delta P_{\rm w-nw}\,.
\ee 
Now let us perform a leading order tree-level calculation,
\be
\begin{split}
P_{\rm LO}=P_{\rm nw}+\e^{-\Sigma^2 k^2}P_{\rm w}=P_{\rm nw,\,true}+\e^{-\Sigma^2 k^2}P_{\rm w,\,true}
+\Delta P_{\rm w-nw}(1-\e^{-\Sigma^2 k^2})\,.
\end{split} 
\ee
We see that at small wavenumbers, the wiggly-non-wiggly error cancels when we sum up the wiggly and smooth parts.
The residual error term on the r.h.s.~can be Taylor expanded and compared to the one-loop 
contribution at low $k$'s,
\be 
\Delta P_{\rm w-nw}(1-\e^{-\Sigma^2 k^2})\approx \Sigma^2 k^2 \Delta P_{\rm w-nw}\ll k^2 \sigma_v^2 P_{\rm lin}\,,
\ee
where $\sigma^2_v\approx 36D^2(z)$~[Mpc$/h]^2$ is the variance of the linear displacement field, which controls the
size of the one loop correction at low $k$.
A similar calculation can be repeated at 
one-loop order.
Given this observation, one can argue that as long as the wiggly-non-wiggly splitting error 
$\Delta P_{\rm w-nw}\ll P_{\rm lin}$
is much smaller than the power spectrum itself, the residual error of a $n$-loop 
calculation will be smaller than the $n+1$ loop correction. In Fig.~\ref{fig:err} we show the 
residuals between the two one-loop spectra produced by 
changing the cuts of the Fourier harmonics.
The difference generated by the wiggly-smooth procedure is clearly much smaller than the two-loop 
contribution and can be safely neglected.

\textit{Dependence on $k_S$.} This ambiguity is intrinsic to the IR resummation procedure.
However, it was shown in Refs.~\cite{Blas:2016sfa,Ivanov:2018gjr} that the effect 
reduces at higher loop orders. 
Thus, the error due to the separation scale choice is always under rigorous perturbative control. 
To estimate it, we compute the residuals between the two spectra evaluated with $k_S=0.1~h$/Mpc
and $k_S=0.2~h$/Mpc, and display the result in Fig.~\ref{fig:err}.

\textit{Factorization.}
Another source of error can be the approximate 
treatment of 
IR resummation in redshift space. 
Recall that, in principle, one ought to compute 
anisotropic 3-dimensional integrals, 
where the FFTLog
algorithm cannot be 
directly applied. 
However, it is possible to approximately factorize the BAO damping by neglecting terms which are 
formally either higher order or exponentially small. 
For that, one should, essentially, repeat the same arguments as for the wiggly-non-wiggly decomposition,
see Ref.~\cite{Blas:2016sfa,Ivanov:2018gjr} for more detail.

In practice, we have checked that these terms are indeed smaller than the two-loop
corrections at redshifts relevant for future surveys. 
In order to estimate this error we computed the difference between the full formula for 
the one-loop matter power spectrum \eqref{eq:irres1loop} and its ``factorized'' version \eqref{eq:irres1loopfact}.
Crucially, the residual generated by the factorization is a smooth function without a pronounced BAO
feature. 
The reason behind this is that the factorization mostly affects the $P_{22}$-like integrals, in
which the oscillating residuals are integrated over and hence 
washed out.
The absence of 
features suggests that even if we neglected the theoretical error associated 
with two loops completely, this residual could be absorbed by the counterterms
without biasing cosmological parameters in a real data analysis.

One may wonder what happens if we approximate the IR-resummation of one-loop redshift-space integrals with 
a direction-independent damping exponent (as it is the case in real space). 
Na\"ively, this prescription would guarantee the absence of smooth residuals in the loop integrals.
However, we have found that this approximation leads to non-negligible oscillation residuals in 
the density and velocity spectra, motivating the 
factorization prescription.  
These residuals are mostly produced by $P_{13}$-like integrals, for which the factorization
procedure is exact and fast, thus 
there is no need to use the isotropic damping template.
The most time-consuming process is the IR resummation of the $P_{22}$-like integrals,
for which the difference between the direction-independent and full anisotropic templates was
found to be quite small. 
Approximating the BAO damping of the $P_{22}$ integrands 
with the direction-independent template notably
reduces the computational cost of IR resummation. 
This introduces $\lesssim 0.1\%$ error on the 
full power spectra, which is smaller than the neglected two-loop contributions
on mildly-nonlinear scales.
Even though this approximation seems promising, it has not yet been fully
included in the current version of the code.
Currently, we have implemented 
this method only for the $P_{22}$-like integrals of biased tracers; we  
plan to 
run more thorough tests before implementing it for the matter loop integrals as well.

We stress that we have implemented the full factorization
formula with the anisotropic damping factor 
for the $P_{13}$-like bias integrals produced 
by the operator $\F_{\G_2}$. 
This formula is exact for these types of integrals.

\textit{Corrections of order $\mathcal{O}(P^2_{\rm w})$.}
Finally, we have checked that the terms $\sim P^2_{\text{w}}$ omitted in the IR resummation procedure of Ref.~\cite{Blas:2016sfa} are indeed negligible. In principle, these corrections can be taken into account at zeroth order at no additional cost, but their contribution is so small
that they are irrelevant for all practical applications. 

All the sources of error related to IR resummation are shown in Fig.~\ref{fig:err}. 
We see that the biggest error is introduced by the factorization procedure, but its contribution is 
quite smooth and its slope matches the shape of the two-loop contribution.

\subsection{Evaluation of Redshift Space Multipoles}
When neither IR resummation nor the AP effect are present, the code can be greatly expedited by performing 
the $\mu$-integrals analytically. 
In this case we use the explicit FFTLog matrices directly for the power spectrum multipoles.
This calculation is initiated if the flag 
\mbox{``\texttt{IR resummation = No}''}
is passed to the code. Note the AP effect is not implemented in this case.
If the flag ``\texttt{IR resummation = Yes}'' is passed instead, 
a different routine is performed; 
we separately compute all 
Fourier integrals that multiply different powers of $\mu^2$ (see Eq.~\eqref{eq:P1lzs}) and 
use the following algorithm: 
\begin{enumerate}
\item Compute each loop integral separately for wiggly and non-wiggly components;
First, we evaluate it for the non-wiggly input power spectra only. 
Second, we compute the one-loop integrals with one entry of $P_{\rm w}$ and one entry 
of $P_{\rm nw}$;
\item Suppress the wiggly one-loop spectra with the anisotropic damping factor;
\item Combine these terms 
and add the tree-level IR-resummed part. 
This allows us to arrive at the 
final expression for
$P_{\rm gg}(z,k,\mu)$ given in Eq.~\eqref{eq:irres1loopfact};
\item Map the arguments $(k,\,\mu)\to (k_{\rm obs},\,\mu_{\rm obs})$
as dictated by the AP conversion for an input cosmological model; 
\item Perform the angular integrals
over $\mu_{\rm obs}$ from Eq.~\eqref{eq:APPl} 
using the pre-computed Gaussian quadrature with 40 weights.
\end{enumerate}

Note that there is only one numerical routine
that performs 
the Legendre integrals over $\mu$ 
both for IR-resummation and the AP distortions. 
If one needs to take into account 
the AP-distortions but not IR resummation, one can set
the BAO damping factor, $\Sigma_{\rm tot}$, to zero.
Moreover, the AP effect can be computed only if all three flags ``\texttt{RSD = Yes}'', 
``\texttt{IR resummation = Yes}'', and ``\texttt{AP = Yes}'' are passed to the code.
If necessary, 
one could compute the multipoles induced by the AP effect in real space
by setting the growth factor 
$f=0$ before the RSD module is executed.

\subsection{Neutrino Masses}

Massive neutrinos require some special treatment.
Strictly speaking, our method is not applicable if the growth rate is scale-dependent.
In this case, one cannot use the usual EdS perturbation theory kernels, and the full 
calculation of time-and-scale
dependent Green's functions is needed~\cite{Blas:2014hya,Senatore:2017hyk}. 
However, these references showed that for 
dark matter in real space, the difference
between the full calculation and the EdS approximation is very small for realistic neutrino masses.
This suggests that it is safe to use our EdS-based FFTLog calculations in this case.

To approximately incorporate massive neutrinos in the calculation of biased tracers, 
we use the linear power spectrum for the ``cold dark matter+baryons'' (``cb'') fluid as an input in all loop calculations. 
This prescription has been advocated on the basis of N-body simulations in~\cite{Villaescusa-Navarro:2013pva,Castorina:2013wga,Costanzi:2013bha,Castorina:2013wga,Castorina:2015bma,Villaescusa-Navarro:2017mfx}; 
besides, Refs.~\cite{Raccanelli:2017kht,Vagnozzi:2018pwo} claimed its importance for the neutrino mass measurement.
The ``cb'' power spectrum is a default input of our non-linear module.
If necessary, 
one can use the total matter density via 
the flag 
 ``\texttt{cb=No}.'' 

The situation is more complicated in redshift space. 
Just like in the biased tracer case, 
N-body simulations (e.g.~\cite{Villaescusa-Navarro:2017mfx}) 
suggest that one has to use the linear logarithmic growth
factor $f_{\rm cb}$ of the ``cb'' fluid. 
Then the halo power spectra of N-body simulations approach the Kaiser 
prediction \cite{Kaiser:1987qv} evaluated with the ``cb'' quantities. 
Crucially, 
for observationally allowed 
neutrino masses, the scale-dependence of $f_{\rm cb}$ is around $0.1\%$ on large scales where the
definition of $f_{\rm cb}$ is meaningful.
Strictly speaking, the presence of this scale-dependence invalidates our whole
redshift-space one-loop calculation including IR resummation and calls for a computation 
of the appropriate Green's functions. However, given that this effect is very small,
we will neglect it and use the EdS approximation with a scale-independent approximation for $f_{\rm cb}$.
In principle, one can include the effect of appropriate Green's functions
by perturbatively expanding around the EdS kernels. We leave this for future work.

Overall, in the presence of massive neutrinos, we use the same FFTLog-EdS formulas as before, but apply them to the 
linear ``cb'' power spectrum, including the suppression at short-scales by massive neutrinos' free-streaming.
At any required redshift, the code takes the power spectrum at this exact time, 
such that the linear time-dependence of the neutrino suppression is taken into account.
This approach is justified by N-body simulations of Ref.~\cite{Villaescusa-Navarro:2017mfx}, which showed that the 
leading effect of massive neutrinos is always a suppression of the linear power spectrum,
and any residual scale-dependence of this suppression is insignificant even for volumes as large 
as $100$~(Gpc/$h)^3$. This observation was also confirmed in various forecasts, e.g.~\cite{LoVerde:2016ahu,Chudaykin:2019ock}.
Given these reasons, 
we expect that using the usual FFTLog formulas 
in the presence of massive neutrinos will be a good 
approximation even for future surveys like DESI or Euclid.

\subsection{Non-Standard Extensions of $\Lambda$CDM}

The code in its current form can be used without any limitations for all non-minimal cosmological 
models that do not require modification of the perturbation theory kernels. One such example is the 
Early Dark Energy (EDE) model, in which the standard $\Lambda$CDM early universe physics is significantly
modified in an attempt to resolve the Hubble tension~\cite{Poulin:2018cxd}. \texttt{CLASS-PT} has been already successfully used 
to put the strongest constraints to date on the EDE model from the combination of the CMB and LSS data~\cite{Ivanov:2020ril} 
(see also~\cite{DAmico:2020ods}). 

In principle, \texttt{CLASS-PT} can be extended even to those cases which require
modifications in the mode-coupling kernels, e.g.~modified gravity. 
If these models do not violate the Equivalence Principle,
one has to simply recompute the perturbation theory matrices incorporating the 
new kernels 
from these extended models. In this case, the body of the code does not need to be modified.
If the Equivalence Principle is violated, 
IR resummation must be altered accordingly, see Refs.~\cite{Crisostomi:2019vhj,Lewandowski:2019txi}.

\subsection{Modified CMB Lensing Routine}

In certain situations, it may be useful to have some alternative estimate for non-linear corrections
that can be used instead of HALOFIT for the CMB lensing calculations. 
This is clearly the case for exploration of the non-standard cosmological models  
for which the HALOFIT fitting formula was not calibrated. Since the non-linear corrections relevant for CMB lensing 
are relatively small for angular multipoles $\ell < \text{few}\times 10^3$ \cite{Lewis:2006fu},
one may expect that perturbation theory gives reasonably accurate results for current lensing data 
such as, e.g.~the Planck measurements.
One technical difficulty in applying perturbation theory to CMB lensing 
is the 
significant width of the lensing kernel.
This requires non-linear corrections from many different redshifts,
whose full calculation is very time-consuming. 
However, given that the perturbations of the lensing potential are only very mildly-nonlinear,
and given the statistical errors of current lensing data, 
the accuracy of non-linear corrections around $\sim 1\%$ is tolerable. 
In this case one can adopt the following simple approximation scheme:
\begin{enumerate}
\item Compute the full matter power spectrum $P^{\rm ref}_{\rm 1-loop}$ at some fixed reference redshift $z_{\rm ref}$;
\item Obtain the spectra at different redshifts $z_i$ by rescaling $P^{\rm ref}_{\rm 1-loop}$ with scale-independent 
linear growth factors,
\[
P_{\rm 1-loop}(z_i,k) = \left(\frac{D(z_{i})}{D(z_{\rm ref})}\right)^4 P^{\rm ref}_{\rm 1-loop}(z_{\rm ref},k)\,.
\]
\end{enumerate}
This procedure is exact in EdS if we neglect the time-dependence of IR-resummation and counterterms.
The effect of both is around $1\%$ on mildly non-linear scales and hence can be 
neglected for our purposes.

Our modified lensing module was tested on the Planck 2018 data in Ref.~\cite{Ivanov:2019hqk},
where it was found to give the same result as HALOFIT for $\nu \L$CDM and $\nu \L$CDM+$N_{\rm eff}$ models. However,
we would like to stress that its accuracy has not been extensively
tested for the precision required for future experiments.

\section{Results and Performance}
\label{sec:plots}

In this Section we show some results and discuss the performance of our 
code.
All plots 
are generated with the \texttt{Jupyter} notebook that 
can be downloaded from the GitHub page of the code.
Our timing results were obtained on a MacBook Pro Retina Early 2015 laptop, 
with a 2.7 GHz Intel Core i5 processor and using OS X version 10.11.6. 
In all cases,  
\texttt{CLASS} was run with the \texttt{C} compiler \texttt{gcc-6.1.0}.
Our \texttt{classy} is based on \texttt{Python 2.7.10}, \texttt{numpy 1.14.5}, and \texttt{scipy 0.19.0}.
The results of this Section
will be presented for the non-linear power
spectrum using the 
We use the following nuisance parameters:
\be
\begin{split}
& c_s^2 = 1~[\text{Mpc}/h]^2\,,\quad R_*^2 = c_0 = 5~[\text{Mpc}/h]^2\,,\quad  P_{\rm shot} = 5\times 10^3~[\text{Mpc}/h]^3\,,\\
& b_1 = 2\,,\quad  b_2 = -1\,,\quad b_{{\cal G}_2}=0.1\,,\quad b_{\Gamma_3} = -0.1\,,\\
& c_2 = 15~[\text{Mpc}/h]^2\,,\quad c_4 = -5~[\text{Mpc}/h]^2\,,\quad \tilde{c}_{\nabla^4_\z\delta} = 100~[\text{Mpc}/h]^4 \,.
\end{split} 
\ee
These 
are consistent with the values extracted from high-resolution BOSS mock galaxy catalogs 
and the actual BOSS survey data \cite{Ivanov:2019pdj}. 
We stress that these nuisance parameters should be fitted from the data in any realistic analysis.

\subsection{Examples of Nonlinear Spectra} 

\begin{figure}[h!]
\begin{center}
\includegraphics[width=0.9\textwidth]{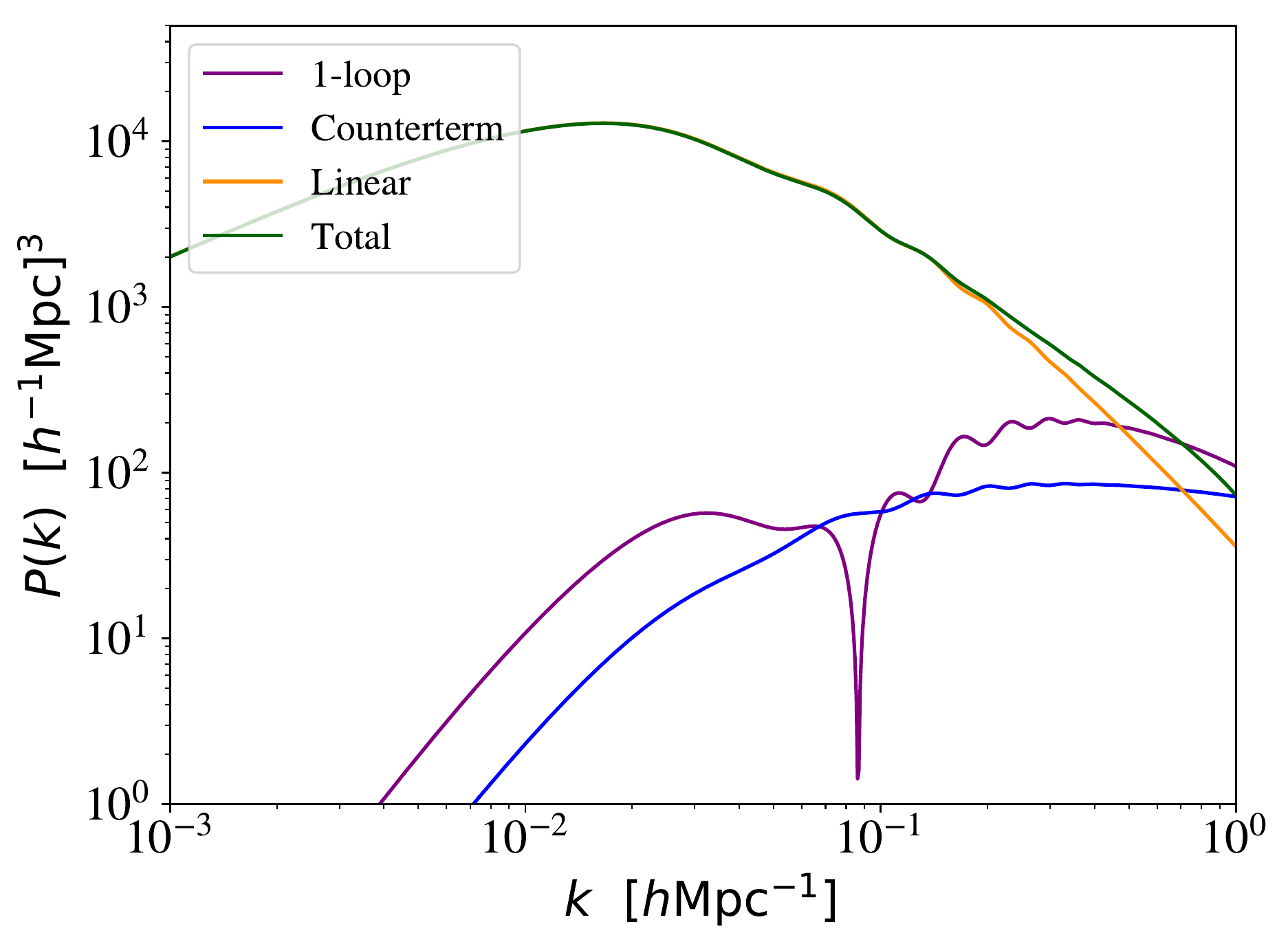}
\end{center}
\caption{\label{fig:pr} 
Breakdown of different contributions to the 
one-loop matter power spectrum of dark matter in real space.
}
\end{figure}

\begin{figure}[h!]
\begin{center}
\includegraphics[width=0.49\textwidth]{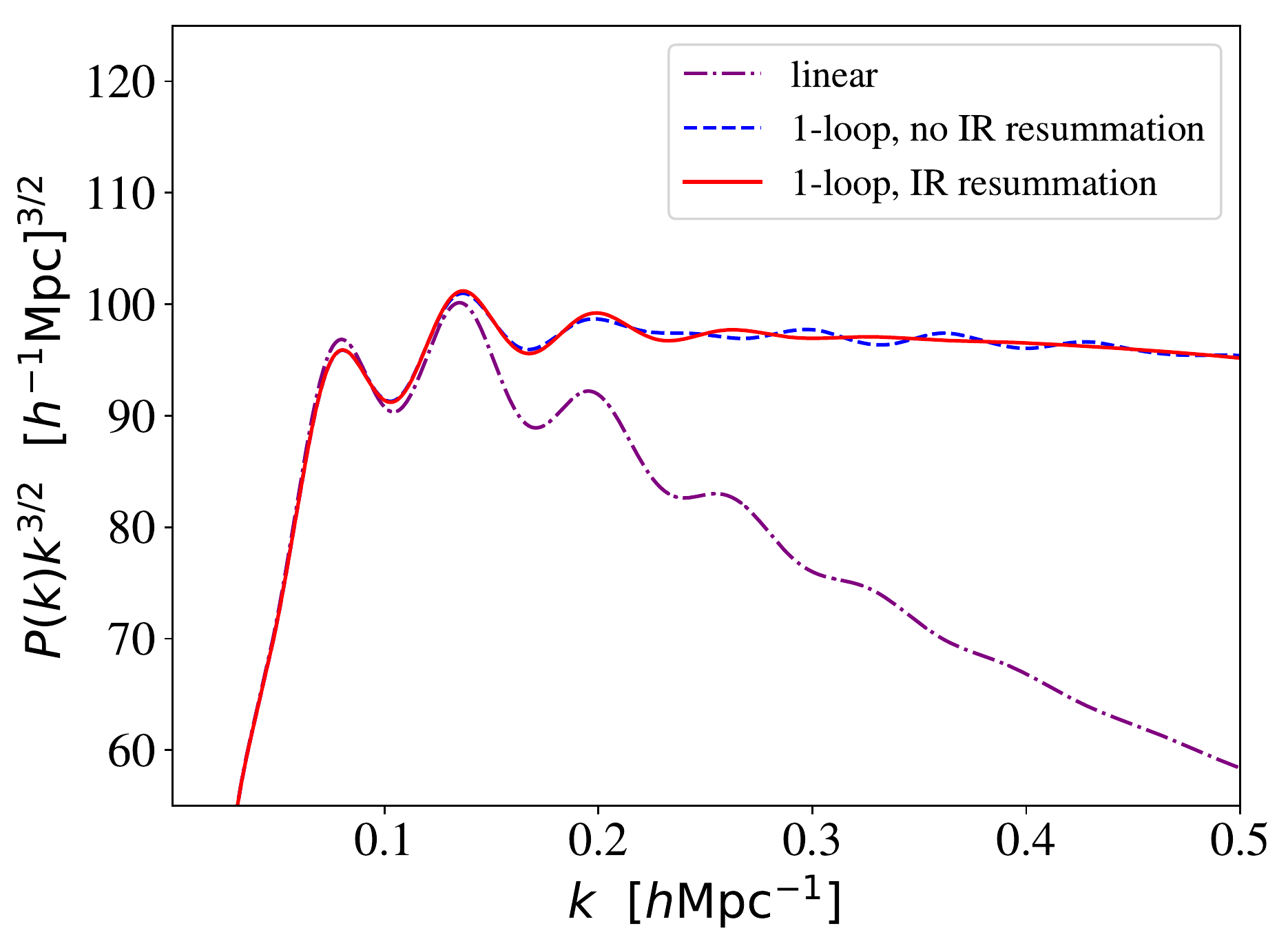}
\includegraphics[width=0.49\textwidth]{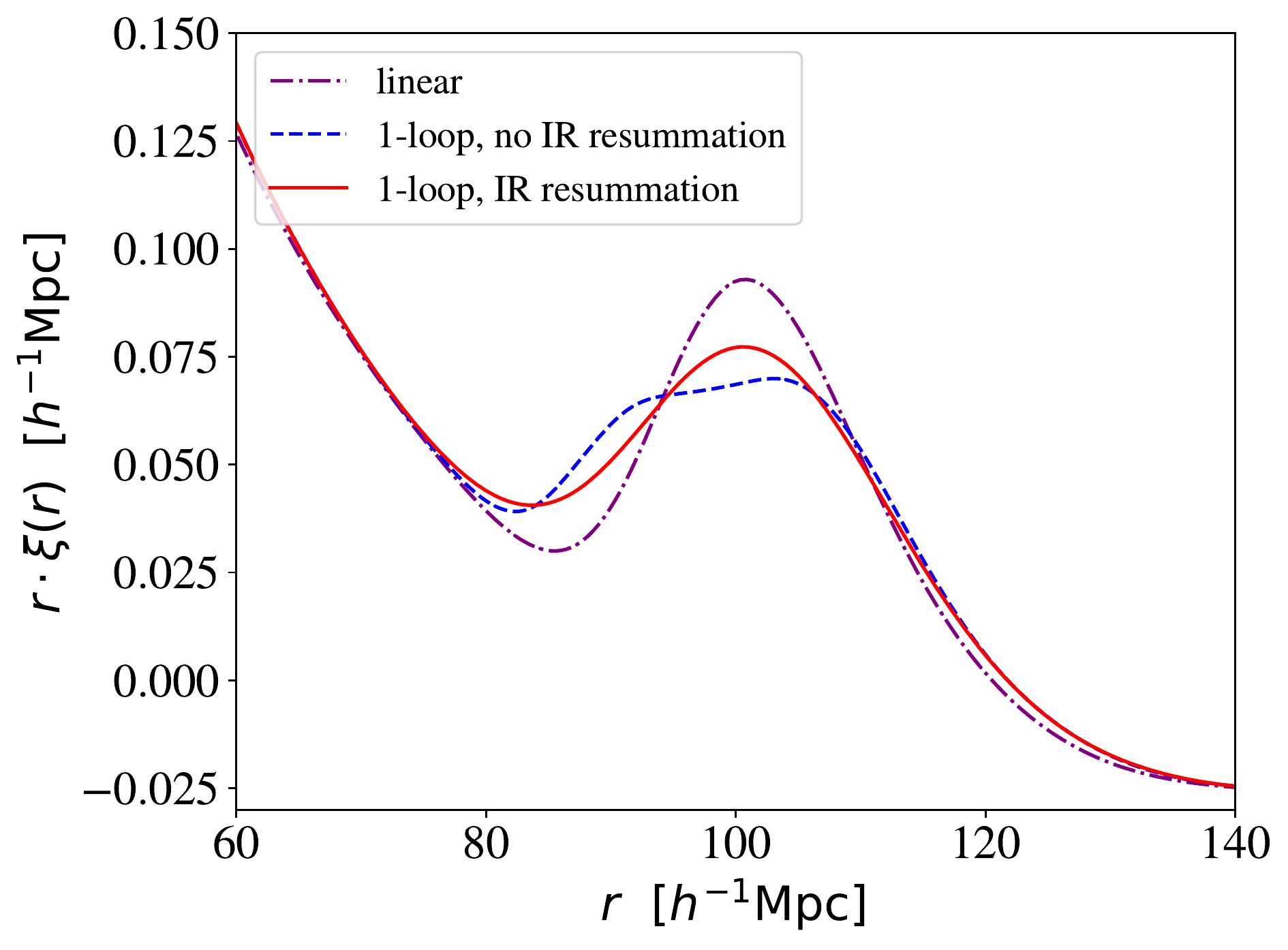}
\end{center}
\caption{\label{fig:res} 
\textit{Left panel:} The total power spectrum with and without IR resummation, along with the 
linear theory prediction. All spectra are multiplied by $k^{3/2}$ for better visualization.
\textit{Right panel:} The position space correlation functions extracted from the same calculations.
}
\end{figure}

Fig.~\ref{fig:pr} shows the breakdown of different contributions to the matter power spectrum in redshift
space without IR resummation. 
In Fig.~\ref{fig:res} we show the effect of IR resummation.
Without this procedure, the one-loop correction fails to capture the shape of 
the BAO wiggles and even their frequency. 
This result is well known in the literature \cite{Baldauf:2015xfa,Blas:2016sfa} 
and it explicitly shows that IR resummation is a necessary ingredient of any realistic non-linear calculation.
For comparison, we also display the linear theory power spectrum.

\begin{figure}[h!]
\begin{center}
\includegraphics[width=0.49\textwidth]{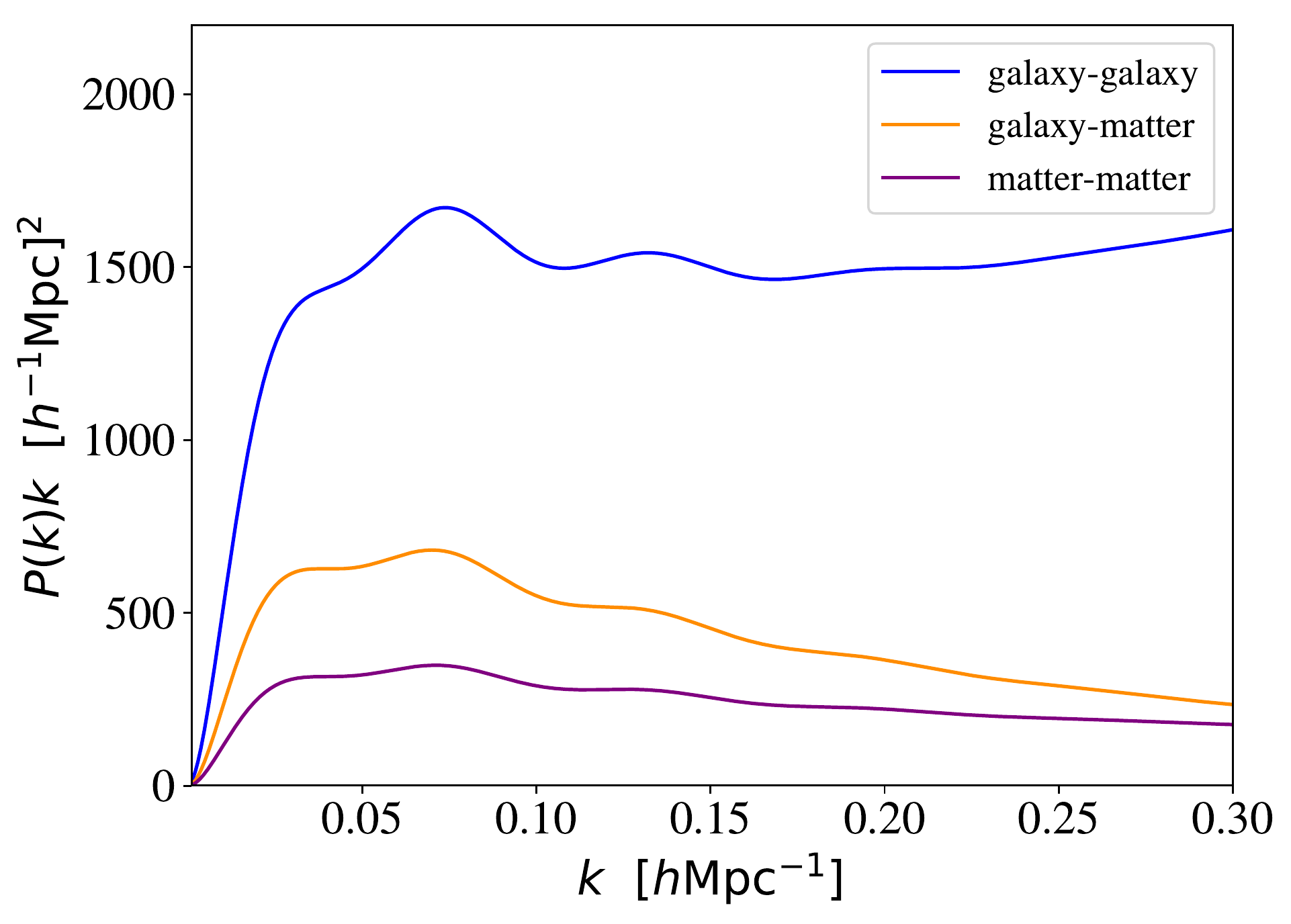}
\includegraphics[width=0.49\textwidth]{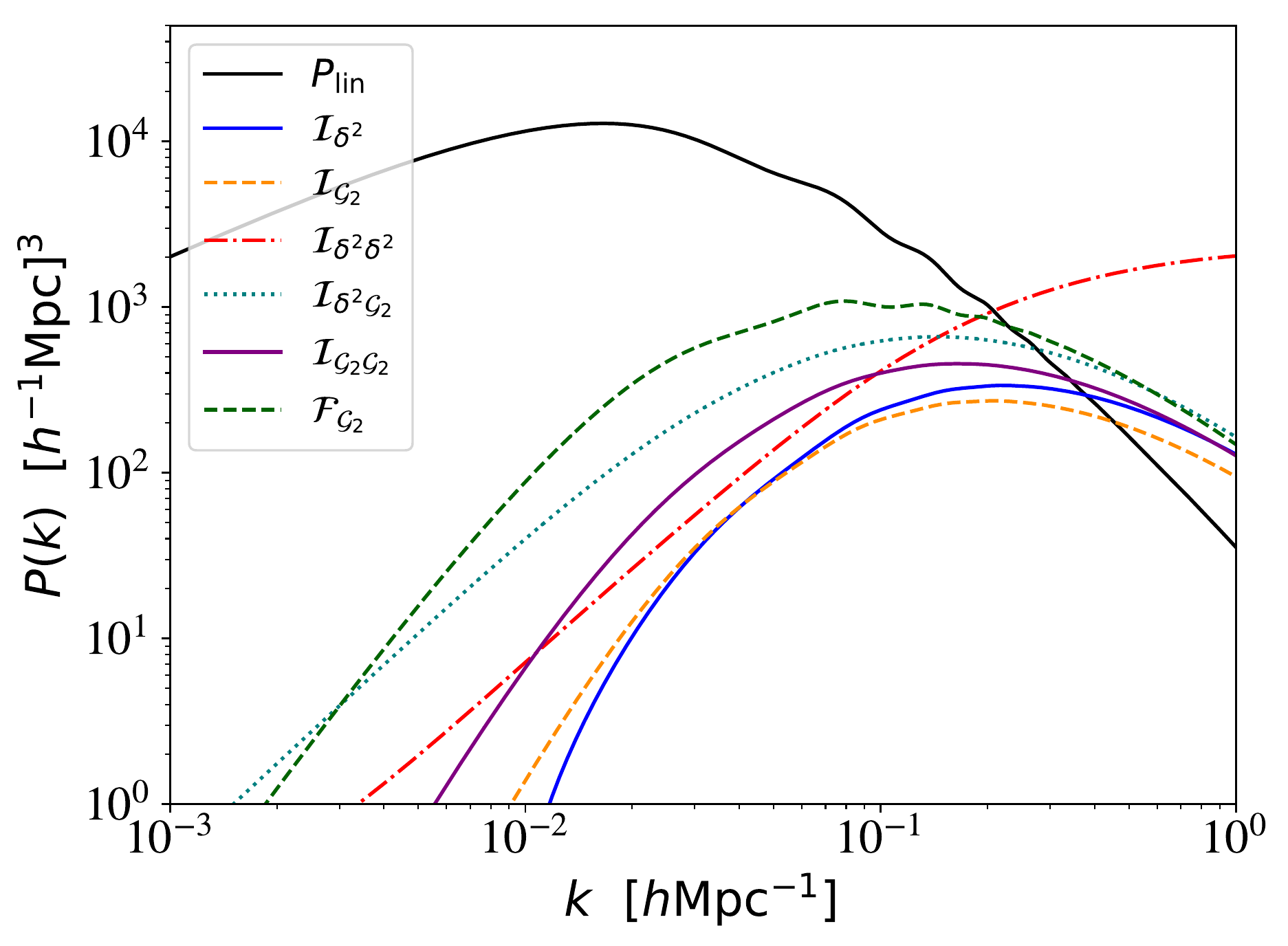}
\end{center}
\caption{\label{fig:pgg} \textit{Left panel:} One loop predictions for the 
matter-matter, matter-galaxy and galaxy-galaxy power 
spectra of the BOSS-like galaxy sample.
\textit{Right panel:} breakdown of different bias contributions to the 
one-loop galaxy power spectrum.
}
\end{figure}
\begin{figure}[h!]
\begin{center}
\includegraphics[width=0.49\textwidth]{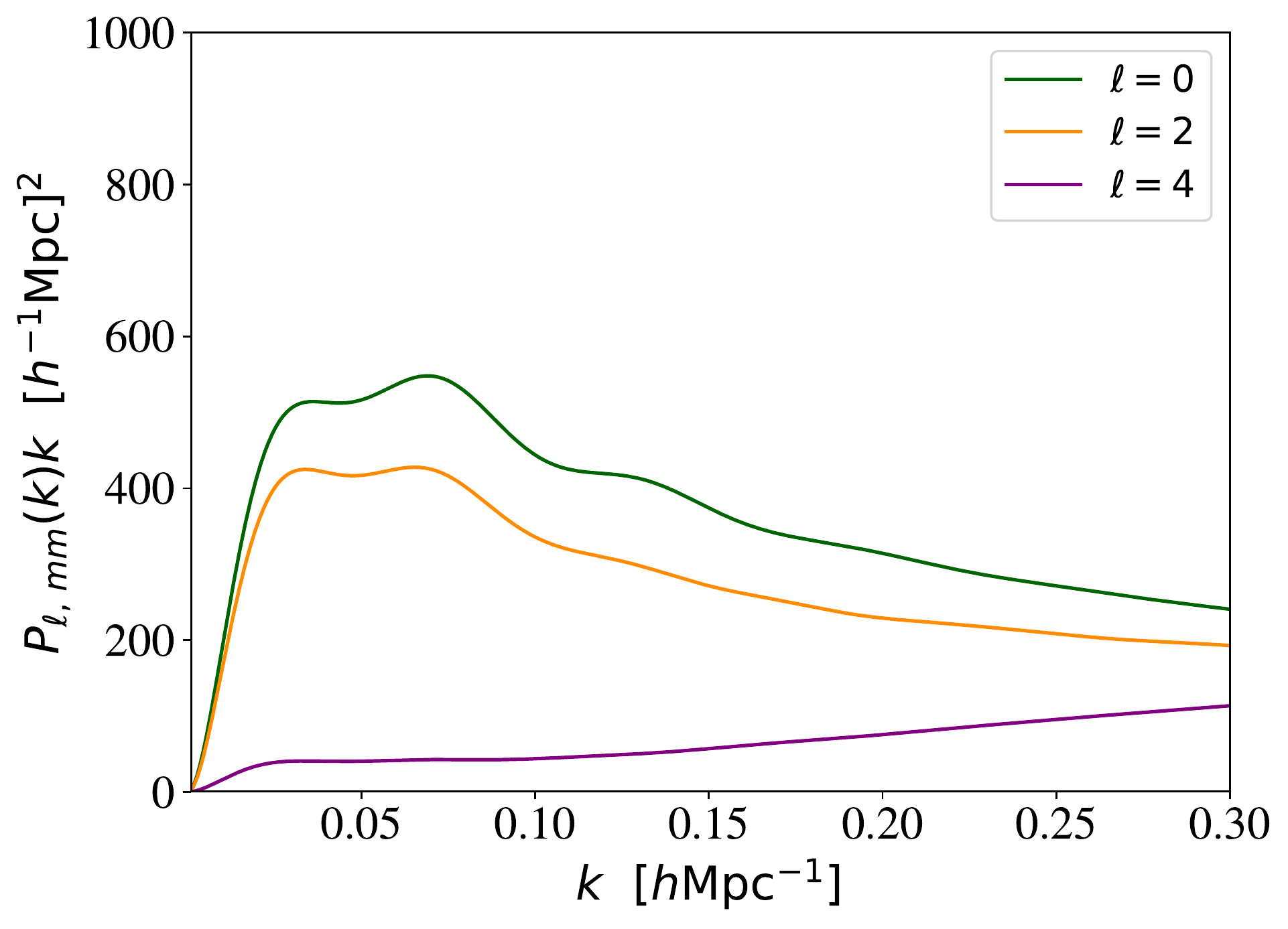}
\includegraphics[width=0.49\textwidth]{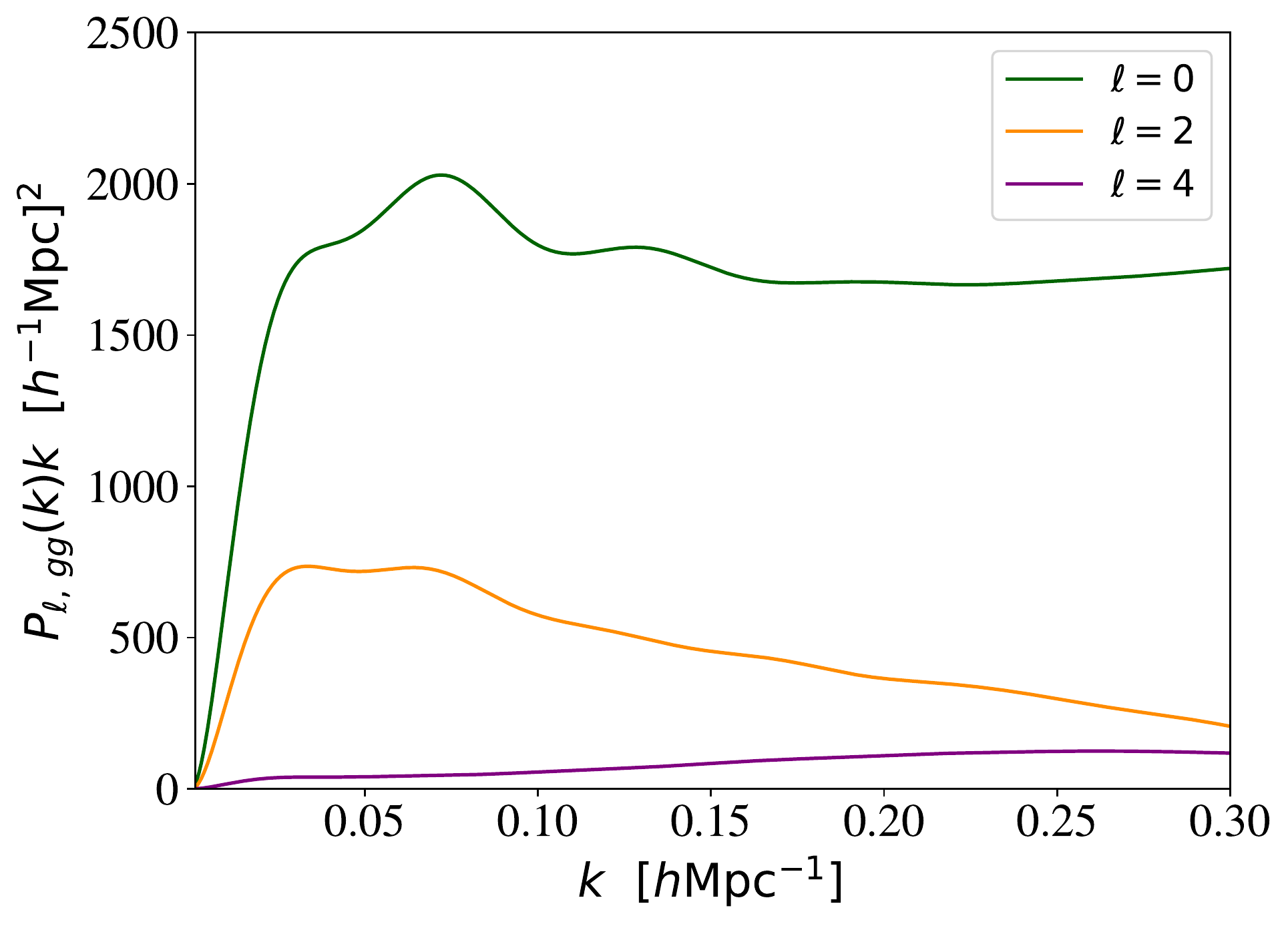}
\end{center}
\caption{\label{fig:prsd} \textit{Left panel:} 
Redshift-space multipoles of the matter power spectrum.
\textit{Right panel:} the same for the galaxy power spectrum of the BOSS-like sample.
}
\end{figure}

In Fig.~\ref{fig:pgg} we show the galaxy-galaxy, galaxy-matter and matter-matter spectra 
in real space (left panel) and the breakdown of different bias loop corrections (right panel).
Fig.~\ref{fig:prsd} displays redshift-space multipoles of dark matter (left panel)
and BOSS galaxies (right panel). Note that in the latter case we have 
included the $\tilde{c}_{\nabla^4_\z\delta}$-counterterm.\footnote{This counterterm 
was obtained by simply multiplying the hexadecapole $P_4$ counterterm 
by $k^2$ 
at no additional computational cost. 
We have checked that a small residual difference between this procedure and the full treatment, 
which appears due to subleading effects in the AP effect and IR resummation, is negligible.
}

\begin{figure}[h!]
\begin{center}
\includegraphics[width=0.49\textwidth]{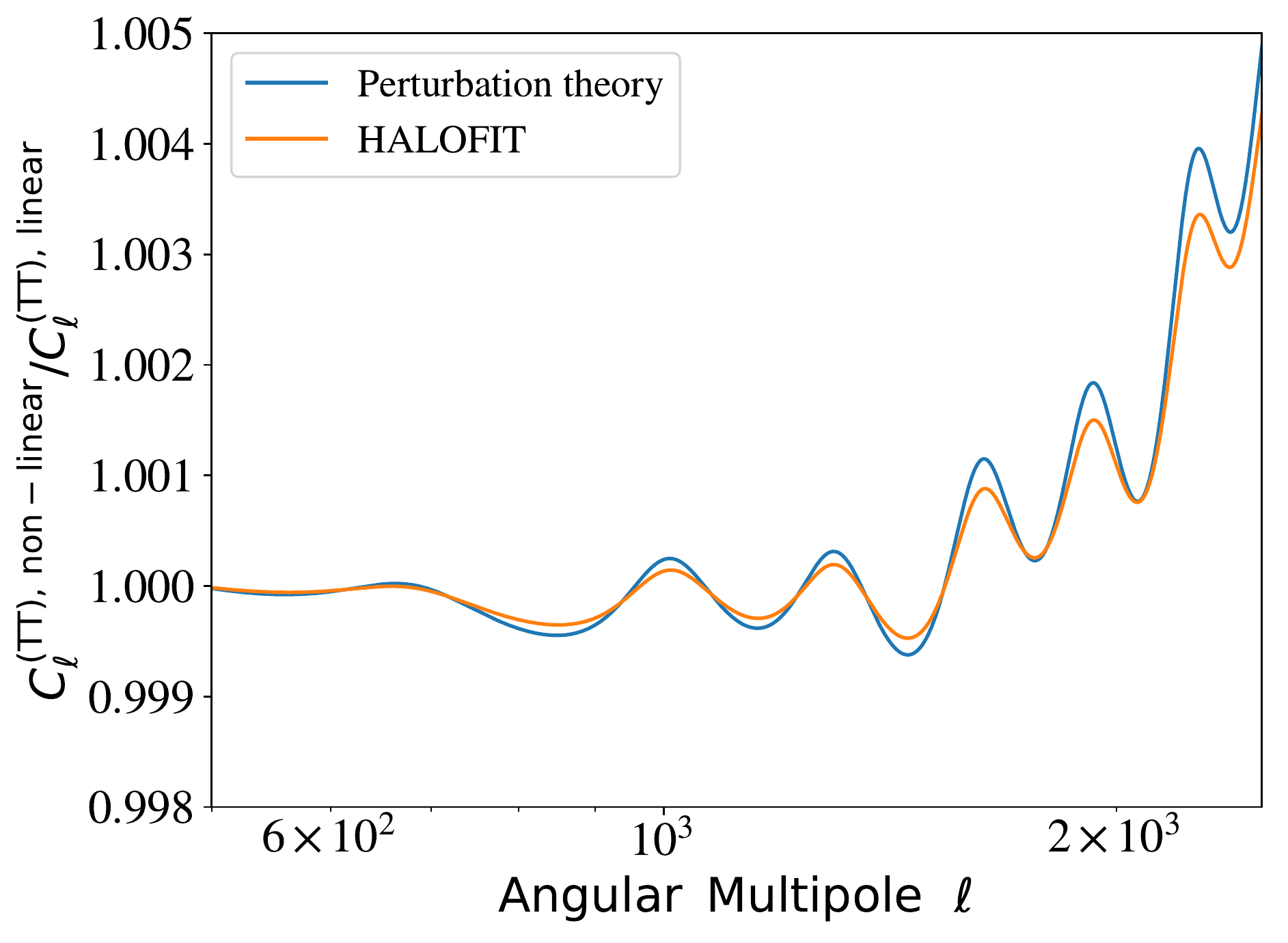}
\includegraphics[width=0.49\textwidth]{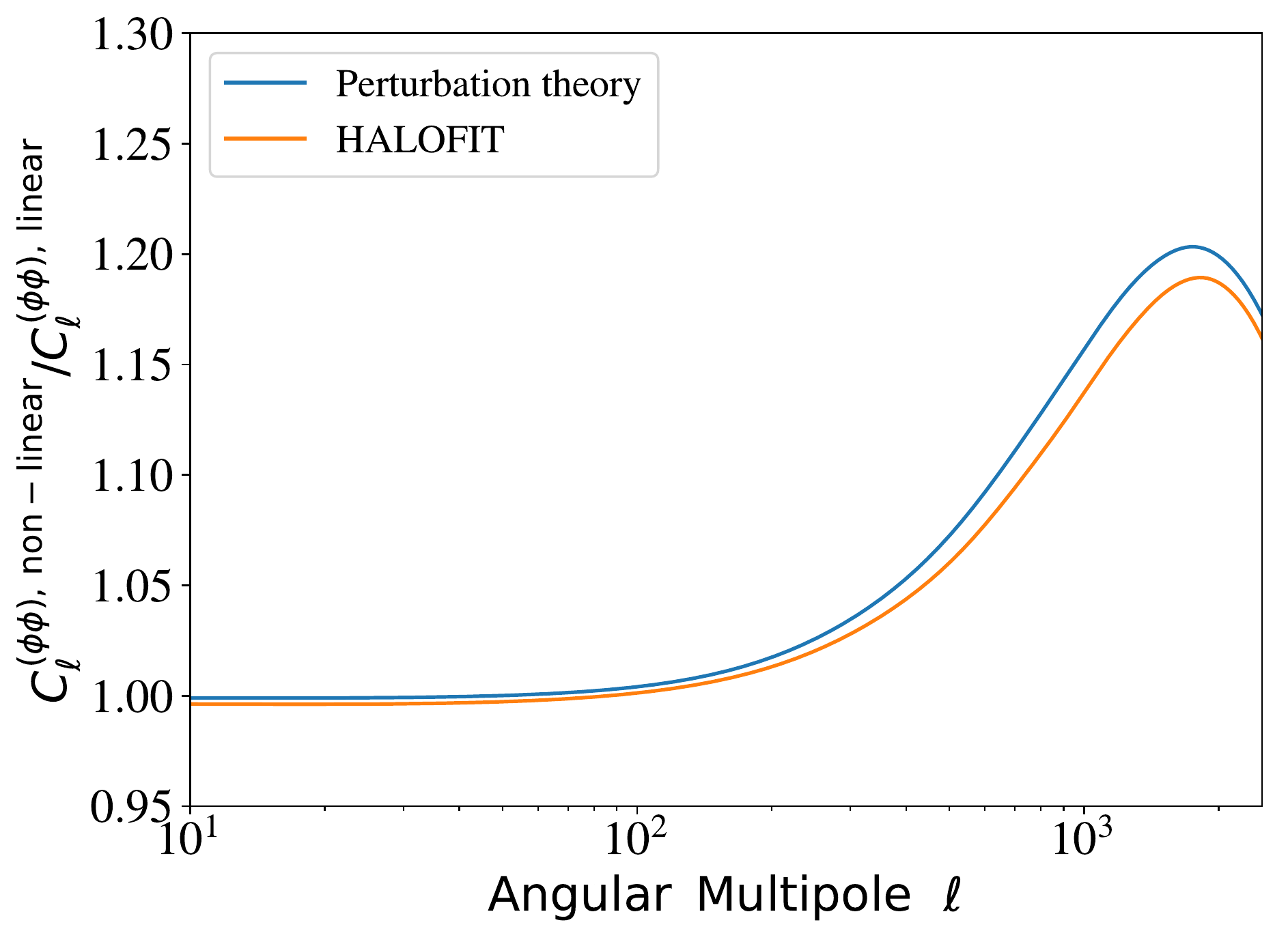}
\includegraphics[width=0.49\textwidth]{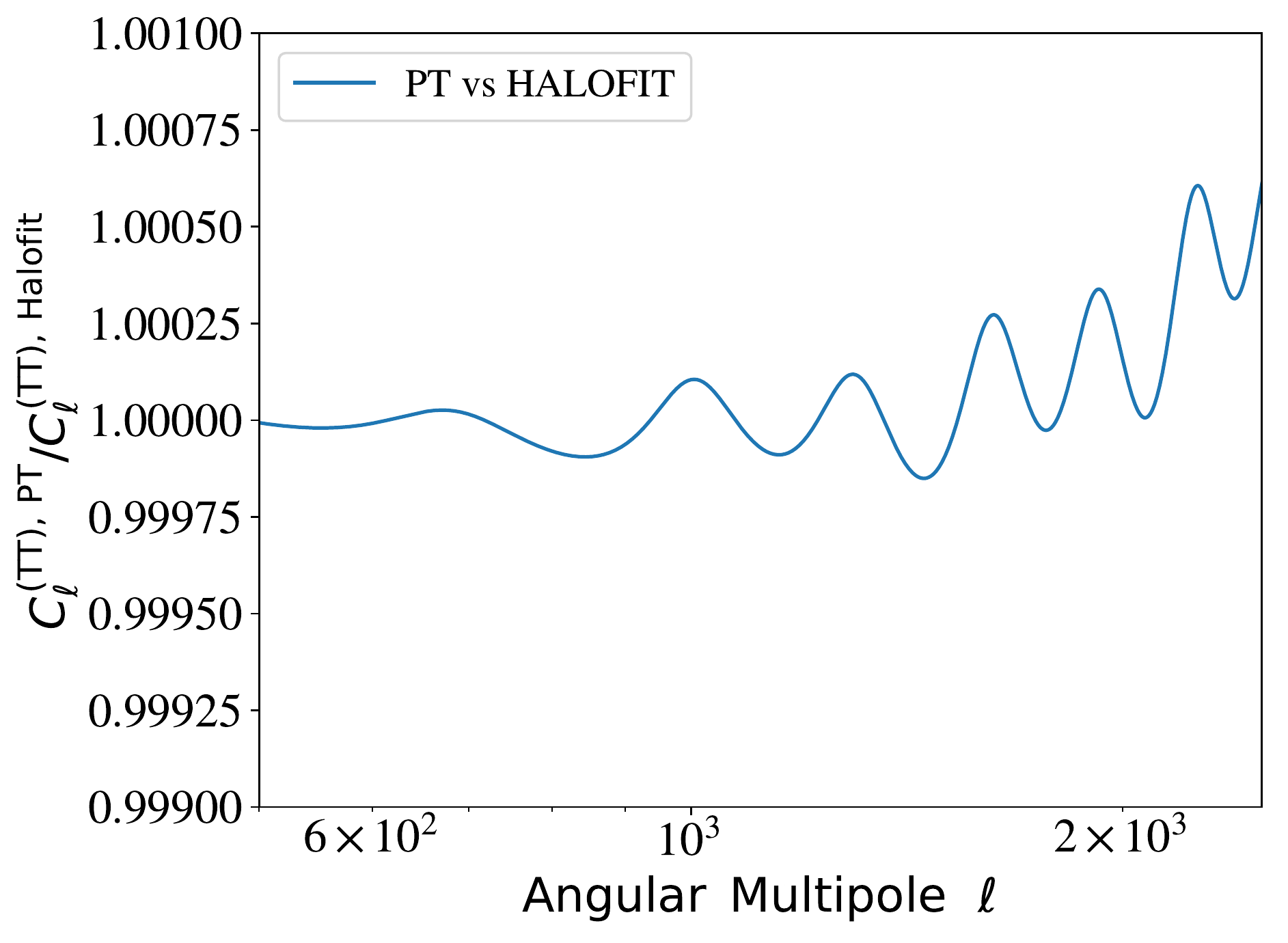}
\includegraphics[width=0.49\textwidth]{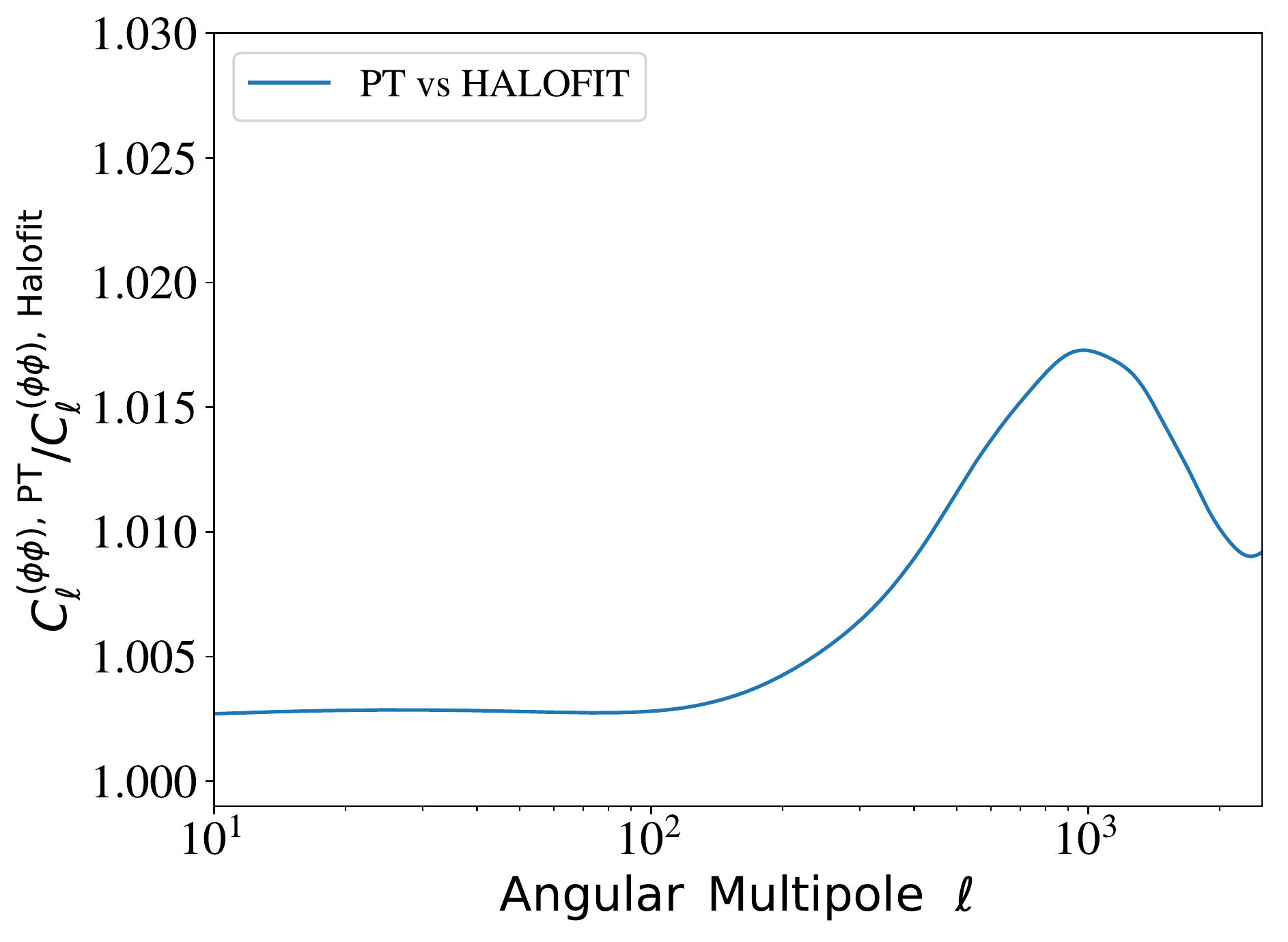}
\end{center}
\caption{\label{fig:cl}  \textit{Upper left panel:} The lensed TT CMB power spectrum
computed with perturbation theory (PT) and HALOFT, normalized to the linear theory prediction.
\textit{Upper right panel:} A similar fraction for the CMB lensing potential power spectrum. \textit{Lower panels:}Ratio of
the non-linear models 
for the lensing potential power spectrum and the lensed CMB TT power spectrum.
}
\end{figure}

Fig.~\ref{fig:cl} shows the lensed temperature (TT) and the CMB lensing potential power spectra
computed in perturbation theory and with HALOFIT (divided by the linear theory prediction), as well as the relative difference between the two. 
One sees that the difference between the PT and HALOFIT predictions is less than $0.1\%$ 
for the lensed TT spectrum and around $\sim 2\%$ at the small-scale part of the $C_\ell^{(\phi\phi)}$
spectrum. These differences can be taken as an estimate for the theoretical error associated 
with the modeling of non-linear corrections. 
We believe that the residual between the PT and HALOFIT spectra can be 
reduced by an appropriate tuning of the dark matter effective sound speed $c_s^2$. 
Moreover, a 
better description can be wrought 
by combining the two methods;
perturbation theory is very accurate on mildly non-linear scale, whereas the N-body 
based fitting formulas capture the leading behavior in the fully non-linear regime.
The exploration of the matter power spectrum on these short scales 
can be done with relatively cheap small-box simulations.
A thorough study of this possibility is left for future work.

\begin{table*}[t!]
  \begin{tabular}{|c||c|c|c|c|c|} \hline
    Run  &  Real space & IR  resum. &  RSD  &  
    IR+RSD & IR+RSD+AP
      \\ [0.2cm]
      \hline 
\multicolumn{6}{|c|}{Default mode}   \\
 \hline 

 \hline 
  Matter &  0.036 (0.036) & 
 0.175 (0.036)
  & 0.375 (0.375) 
  & 0.75 (0.62) & 0.76 (0.63) \\ \hline
  Tracers  & 0.21 (0.21) 
  & 0.35 (0.21)
  & 0.89 (0.89)  
  & 1.27 (1.12) &  1.30 (1.14)\\ 
\hline 
\multicolumn{6}{|c|}{FAST mode}   \\
 \hline 
  Matter &  $6.3~(6.1)\times  10^{-3}$ & 
 0.14 (0.0061)
  & 0.063 (0.061) 
  & 0.22 (0.09)  & 0.22 (0.09)  \\ \hline
  Tracers  & 0.033 (0.034) 
  & 0.17 (0.034)
  & 0.14 (0.14)  
  & 0.31 (0.18) &  0.31 (0.18)\\ 
\hline 
\hline 
\end{tabular}
\caption{Performance of the code for baseline precision runs. 
We show the execution time 
in [sec.] as follows: $t_{\rm full}(t_{\rm FFTLog})$, where 
$t_{\rm full}$ is the full end-to-end time taken by the non-linear module,
and $t_{\rm FFTLog}$ is the time elapsed during the matrix multiplication with FFTLog.
}
\label{table0}
\end{table*}

\subsection{Performance} 

Let us discuss now the performance of our numerical routine.
Table~\ref{table0} displays the run time for various spectra computed by \texttt{CLASS-PT}
in the default (high-precision) and fast modes. 
These values are the typical ones obtained on authors' laptops; some variation is expected 
for different machines. 
Importantly, the execution time reduces roughly by a 
factor of 4 in the fast mode, which uses a grid twice smaller than the default one.
This is a consequence of the fact that the most time-consuming process is matrix multiplication, 
which very roughly scales as $N_{\rm FFTLog}^2$.

We see that basic runs for the real-space power spectrum without IR resummation are quite fast.
Their speed is comparable to that 
of other methods, e.g.~\texttt{FAST-PT}~\cite{McEwen:2016fjn}.
IR resummation and bias tracers 
increase the execution time by a factor of $\sim 5$ separately.
Since these two procedures are independent, this results in an overall speed loss by a factor of $10$ compared to the basic run.
Redshift space distortions affect the calculation in two ways.
First, there are additional convolution integrals that appear in multipole moments.
Second, redshift space requires a more sophisticated IR resummation procedure,
which again increases the number of convolution integrals even further.
When the two effects are combined, the execution time reaches the level 
of 1.3 second for high precision settings and 0.3 seconds in the fast mode. 
The inclusion of the AP effect does not notably affect the speed.

\subsection{Cautionary Remarks}
\label{sec:remark}

There are several caveats to be borne in mind when using our code. 

First, at face value, the code can be used for any beyond-$\Lambda$CDM cosmology provided that the structure 
of the perturbation theory kernels is not modified.
This is the case for the bulk of extended models explored by Planck \cite{Aghanim:2018eyx}. 
However, the numerical implementation choices made in
the code have not been 
extensively tested for cosmological models that are \textit{extremely} different from 
the Planck best-fitting cosmology. 
Some of our choices, i.e. the frequency cuts in the wiggly-non-wiggly decomposition,
would have to be reconsidered if someone wants to 
explore, say a model with a large numbers of neutrino species, e.g.~$N_{\rm eff} = 42$.

\begin{figure}[h!]
\begin{center}
\includegraphics[width=0.49\textwidth]{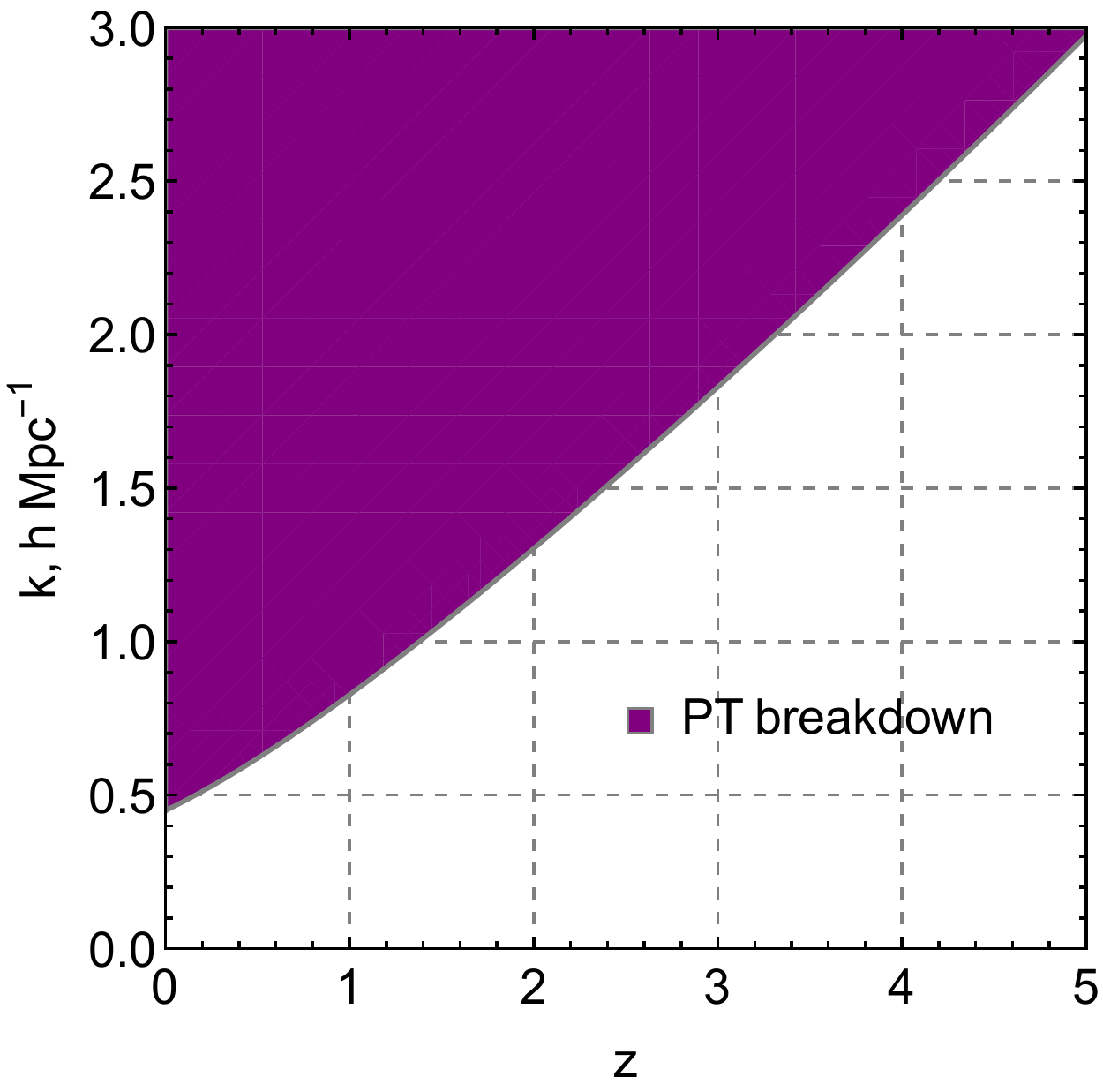}
\end{center}
\caption{\label{fig:kmax} Regime of applicability of the one-loop corrections coputed by our code.
}
\end{figure}

Second, many implementation choices in our code were made to maximize precision on large scales $k\leq 1~h$/Mpc.
Our baseline realization of the non-linear calculation \textit{must not be used} for $k\gtrsim 3~h$/Mpc.
Therefore, our code is not suitable for small-scale galaxy clustering or some lensing calculations 
where a significant amount of signal comes from the highly 
nonlinear scales. One also must carefully choose 
$k_{\rm max}$ as a function of redshift, recalling that 
the loop corrections reduce at high-$z$, allowing smaller scales to be probed. 
The maximal wavenumber to which 
our code can be used to extract information
from the matter clustering corresponds to the 
scale where 
the loop expansion blows up, i.e. whence the two-loop 
correction becomes comparable to the tree-level prediction. Using the fit to the two-loop power spectrum from Refs.~\cite{Baldauf:2016sjb,Chudaykin:2019ock}, this scale can be estimated as
\be
k_{\rm NL}(z)=0.45 [D(z)]^{-\frac{4}{3.3}}~h\Mpc^{-1}\,. 
\ee
The use of non-linear corrections computed with our code is, strictly speaking, justified 
only for $k<k_{\rm NL}$. The corresponding validity domain is shown 
in Fig.~\ref{fig:kmax} as a function of redshift.\footnote{
Note that even if the two-loop corrections are not directly included in the model, one can still use the one-loop perturbation theory prediction evaluated by our code at high $k_{\rm max}$ provided that the two-loop corrections are included in the theoretical error covariance~\cite{Baldauf:2016sjb,Chudaykin:2019ock}.
}

One obvious caveat is that our code does not include 
relativistic corrections and wide-angle effects, and therefore it should be used with care 
on very large scales. Furthermore, it does not have corrections to the linear bias 
due to 
local primordial non-Gaussianities. All of these corrections do not require 
nonlinear calculations and can be easily added if necessary.


\section{Application to the BOSS Data}
\label{sec:boss}

In this Section, we illustrate an application of our code to the analysis of 
the final BOSS data release \cite{Alam:2016hwk}. 
To this
end, we interface \texttt{CLASS-PT} with the MCMC sampler \texttt{montepython v3.0} \cite{Audren:2012wb,Brinckmann:2018cvx}.
We will analyze a full-shape likelihood built out of the publicly available 
BOSS data and products taken from\footnote{\href{https://fbeutler.github.io/hub/hub.html}{
\textcolor{blue}{https://fbeutler.github.io/hub/hub.html}} \\
\href{https://github.com/fbeutler/fbeutler.github.io/tree/master/hub}{
\textcolor{blue}{https://github.com/fbeutler/fbeutler.github.io/tree/master/hub}}\,.
}, see Refs.~\cite{Beutler:2016ixs,Beutler:2016arn} for more detail.
This likelihood has already been used in Refs.~\cite{Ivanov:2019pdj,Ivanov:2019hqk,Philcox:2020vvt},
where one can find all technical details. We repeat it here just as an illustration. 

As an aside, we would like to mention that the shape of the galaxy power spectrum
has been used for cosmological parameter measurements since the dawn of galaxy surveys, see e.g. \cite{Tegmark:1997rp,Tegmark:1997yq,Tegmark:2003uf,Cole:2005sx,Blake:2006kv,Percival:2006gt}.
This practice, however, has been abandoned in the recent full-shape analyses that 
are based on the methodology borrowed from the BAO measurements~\cite{Beutler:2013yhm,Alam:2016hwk,Beutler:2016arn}.
These analyses infer distance information by studying how the AP effect 
distorts some fixed-shape power spectrum template. 
The fixed template method is also adopted in the measurement of rms velocity fluctuation $f\sigma_8$.
This method has a number of limitations which can compromise the cosmological analysis
of future high-precision data~\cite{Chudaykin:2019ock,Ivanov:2019pdj}.\footnote{First, it can lead to biased results. 
In $\Lambda$CDM the power spectrum shape is fixed by $\omega_b,\omega_{cdm}$ and $n_s$, 
which are measured very precisely from the CMB data.
However, future surveys will probe the shape parameters with 
precision comparable to that of the CMB \cite{Font-Ribera:2013rwa,Chudaykin:2019ock}.  
Fixing these parameters instead of marginalizing over them can result in bias and underestimation of errors.
Second, the power spectrum shape is dictated by physics at recombination, 
and hence the shape priors imply very strong priors on the early universe. 
Third, the distances measured with the fixed template method cannot be easily related 
to parameters of particular models. 
Fourth, this method works only for the cosmological models where the 
shape of the matter power spectrum remains unaltered after recombination.
Strictly speaking, even the standard $\L$CDM model with massive neutrinos violates this assumption
because the linear growth factor is scale-dependent~\cite{Lesgourgues:2006nd}.}

An alternative to the fixed shape approach is to return to the methodology 
of measuring the cosmological parameters of a given model from the full power spectrum.
This is the standard method adopted in the analyses of the CMB data \cite{Aghanim:2018eyx}. 
An important advantage of this method is its universality: it can be applied to any model including beyond $\Lambda$CDM cosmologies.
In this Section, for illustration purposes, we present the constraints obtained in this way for the base $\Lambda$CDM model. 
It is straightforward to repeat this analysis for more complicated beyond-$\L$CDM models,
see e.g.~\cite{DAmico:2020kxu} for the analysis within $w$CDM and \cite{Ivanov:2019hqk,Philcox:2020vvt} for 
$\nu\Lambda$CDM and $\nu\Lambda$CDM+$N_{\rm eff}$. 
We stress that the key novelty of our analysis is the most advanced theoretical model 
for the nonlinear power spectrum. In other aspects our method closely follows the ones proposed and used decades ago.

Our likelihood covers
the pre-reconstructed redshift-space power spectra of BOSS 
galaxies across two non-overlapping redshift bins, $0.2<z<0.5$
and $0.5<z<0.75$ from two patches of the sky (North Galactic Cap and South Galactic Cap, NGC and SGC).
We use the momentum range $[0.01,0.25]~h/$Mpc, 
which is stable w.r.t. instrumental systematics and two-loop corrections, that are omitted in our theory model.
We fit the BOSS galaxy power spectra assuming the base flat $\L$CDM model, fixing
the tilt of the primordial power spectrum of scalar fluctuations $n_s$
and the physical baryon density $\omega_b$
to the Planck 2018 best-fit values \cite{Aghanim:2018eyx};
\be
\label{eq:nspr}
n_s = 0.9649\,,\quad \omega_b = 0.02237\,.
\ee
In principle, we can also scan over these parameters in our chains, see e.g.~\cite{Ivanov:2019pdj,Philcox:2020vvt},
but do not do it here for simplicity.
The role of the priors \eqref{eq:nspr} and their impact on parameter inference have been thoroughly investigated in Ref.~\cite{Ivanov:2019pdj}.
Following Ref.~\cite{Aghanim:2018eyx}, we approximate the neutrino sector with only one massive eigenstate 
and fix its mass to the lowest value allowed by the oscillation experiments, $m_\nu = 0.06$ eV.
This choice is made purely
for demonstration purposes.
We believe that it is more appropriate to scan over this unknown parameter,
as 
is done in Refs.~\cite{Ivanov:2019pdj,Ivanov:2019hqk,Philcox:2020vvt}.

Our MCMC chains sample the remaining cosmological parameters of the minimal $\L$CDM model: the physical 
density of dark matter $\omega_{cdm}$; the Hubble constant $H_0$; 
the amplitude of primordial 
scalar fluctuations $A_{\rm s}$. We do not assume any priors on these parameters.
To be more precise, we scan over $A^{1/2}_s$ normalized to the Planck best-fit value,
\be
A^{1/2} =\text{norm}\equiv \left(\frac{A_{\rm s}}{A_{\rm s,\,Planck}} \right)^{1/2}\,.
\ee
This choice allows us to treat $A_s$ 
as a nuisance parameters and quickly scan over it, which leads to better convergence.\footnote{Modulo IR resummation, the non-linear power spectra depend on $A_s$ through a simple rescaling,
\be
P_{\rm tree}(A_s) =  \frac{A_s}{A_{s,\,{\rm ref}}}P_{\rm tree}(A_{s,\,{\rm ref}})\,,\quad  
P_{\rm 1-loop}(A_s) =  \left(\frac{A_s}{A_{s,\,{\rm ref}}}\right)^2 P_{\rm 1-loop}(A_{s,\,{\rm ref}})
\ee
Once IR resummation is taken into account, this rescaling is, strictly speaking, inexact
because $A_s$ also controls the 
amplitude of the BAO damping scale. However, variations of the damping scale are
analogous to changes of the separation 
scale $k_S$, which is a higher order effect. 
Thus, using the rescaling of $A_s$ is accurate up to two-loop contributions, which are omitted in our model anyway. 
} We have run our analysis both in the fast and default modes and obtained identical results.

\begin{figure}[h!]
\includegraphics[width=1.0\textwidth]{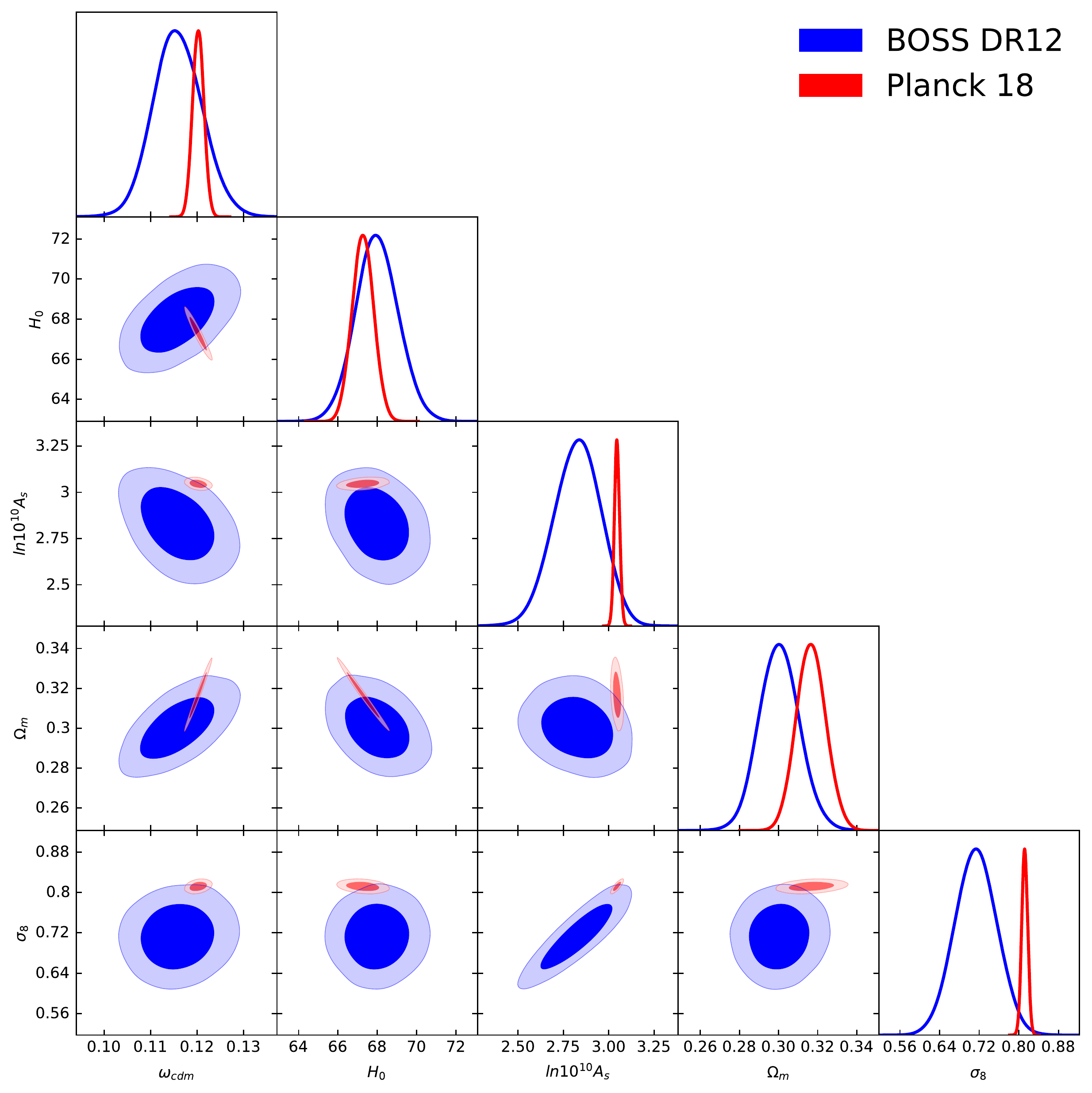}
\caption{ The posterior distribution for $\omega_{cdm},~H_0,~\ln(10^{10}A_s)$
and derived parameters $\Omega_m,~\sigma_8$ inferred from the joint BOSS DR12 full-shape likelihood. 
For comparison we also show the Planck 2018 posterior (red contours)
for the same model (base flat $\L$CDM).
$H_0$ is quoted in units [km/s/Mpc].
}    
\label{fig:final}
\end{figure}
\begin{table}[ht!]
\begin{center}
\begin{tabular}{|c|c|c|} 
 \hline
BOSS DR12& best-fit & mean  $\pm 1\sigma$  \\ [0.5ex] 
 \hline\hline
 $\omega_{cdm}$  & $0.1169$ & $0.1159_{-0.0054}^{+0.0050}$  \\ \hline
 $H_0$  & $67.91$ & $67.98_{-1.1}^{+1.1}$  \\ \hline 
  $\ln(10^{10}A_s)$  & $2.83$ & $2.82^{+0.13}_{-0.13}$  \\ \hline \hline
    $\Omega_{m}$ & $ 0.3034$ & $0.3006_{-0.010}^{+0.010}$ \\ \hline
 $\sigma_8$ & $ 0.716$ & $ 0.710^{+ 0.043}_{- 0.043}$  \\ \hline
\end{tabular}
\begin{tabular}{|c|c|c|} 
 \hline
Planck 2018& best-fit & mean  $\pm 1\sigma$  \\ [0.5ex] 
 \hline\hline
 $\omega_{cdm}$  & $0.1204$ & $0.1202_{-0.0012}^{+0.0012}$  \\ \hline
 $H_0$  & $67.29$ & $67.28_{-0.55}^{+0.53}$  \\ \hline
  $\ln(10^{10}A_s)$ & $3.04$ & $3.045_{-0.015}^{+0.014}$  \\ \hline \hline
   $\Omega_{m}$ & $ 0.3168$ & $0.3166^{+0.0075}_{-0.0075}$  \\ \hline
  $\sigma_8$ & $0.8099$ & $0.8117_{-0.006}^{+0.0057}$  \\ \hline
\end{tabular}
\caption{
The results of our MCMC analysis for the joint BOSS DR12 full-shape likelihood (left panel).
For comparison we also show the results from the final Planck data release \cite{Aghanim:2018eyx}
(right table)
for the same cosmological model 
used in our analysis (base $\L$CDM with 
a fixed neutrino mass).
$H_0$ is quoted in [km/s/Mpc] units.
}
 \label{tab:final}
\end{center}
\end{table}

As far as the nuisance parameters are concerned, we fit them for each galaxy sample separately, due to different calibration and selection functions.
We have seven nuisance parameters in total (sampling all in the ``fast'' mode from Ref.~\cite{Lewis:2013hha} alongside $A^{1/2}$~): 
linear bias $b_1$, 
quadratic bias $b_2$, 
tidal bias $b_{{\cal G}_2}$, shot noise $P_{\rm shot}$
and three counterterms $c_0,c_2,\tilde{c}$. 
The cubic bias $b_{\Gamma_3}$ is set to zero.
The detailed description of these parameters can be found in Ref.~\cite{Ivanov:2019pdj}.
We chose the following priors for the bias parameters:\footnote{We use the notation $\mathcal{N}(\text{mean},\text{variance})$
for the Gaussian prior.
}
 \begin{equation}
   \begin{split}
   &  b_1A^{1/2}\in (1,4)\,,\quad    b_2A^{1/2}\sim \mathcal{N}(0,1)\,,\\
 &    b_{\mathcal{G}_2}A^{1/2}\sim  \mathcal{N}(0,1)\,,\quad   P_{\rm shot}\times [h^{-1} \Mpc ]^{-3}\sim  \mathcal{N}(5000,5000^2) \,,\\
 \end{split}   
 \end{equation}
 and the counterterms:
 \begin{equation}
   \begin{split}
   &  c_0,c_2\times [h^{-1}\Mpc ]^{-2}\sim  \mathcal{N}(0,30^2) \,,\quad \tilde{c}  \times [h^{-1} \Mpc]^{-4}\sim  \mathcal{N}(500,500^2)\,,\\
 \end{split}   
 \end{equation}
 which are selected such that the corresponding shapes do not exceed the linear theory spectra
 on the scales used for the fit.
 Furthermore, 
 the priors for the bias parameters are motivated by the coevolution model and results of N-body simulations.
 Alternatively, one could fix the priors on the nuisance parameters using the method proposed in Ref.~\cite{Nishimichi:2020tvu}.
 We discuss the 
 treatment of nuisance parameters, including their priors
 and measurements, in Appendix~\ref{app:bias}.
 All in all, the priors for nuisance parameters do not significantly 
 affect the constraints on cosmological parameters. 

The results of our analysis\footnote{
The plot and marginalized limits are produced with the \texttt{getdist} package (available at \href{https://getdist.readthedocs.io/en/latest/}{
\textcolor{blue}{https://getdist.readthedocs.io/en/latest/}})~\cite{Lewis:2019xzd},
which is part of the \texttt{CosmoMC} code \cite{Lewis:2002ah,Lewis:2013hha}. 
} are shown in Fig.~\ref{fig:final} and in Table~\ref{tab:final}, 
where we separated the 
directly sampled 
parameters $(\omega_{cdm},H_0,A_s)$ from 
the derived ones $(\Omega_m,\sigma_8)$.
These constraints agree well with those 
reported in Ref.~\cite{Ivanov:2019pdj}, although 
the priors used in our present analysis are slightly different. 
Note that presented BOSS constraints should always be
taken in conjunction with the priors on $\omega_b,n_s$ and $m_\nu$ made in our analysis.
For comparison, we also show the results of our analysis of the baseline Planck 2018 likelihood \cite{Aghanim:2019ame} for the same cosmological model.\footnote{We stress that in our Planck analysis we also varied $n_s$, $\omega_b$ (and 
the reionization depth $\tau_{\rm reio}$), which should be contrasted with our BOSS analysis, where 
$n_s$, $\omega_b$ were fixed.}

We publicly release our BOSS \texttt{Montepython} likelihoods in a separate repository
\href{https://github.com/Michalychforever/lss_montepython}{
\textcolor{blue}{https://github.com/Michalychforever/lss\_montepython}}. 
The likelihoods are available is 
the standard and optimized versions. The latter includes
the Fourier-space window function 
treatment
and analytic marginalization 
over the nuisance parameters $\{c_0,c_2,\tilde{c},P_{\rm shot}\}$
along the lines of Ref.~\cite{DAmico:2019fhj}.


\section{Conclusions}
\label{sec:concl}

In this paper we have presented a new open-source extension of the Boltzmann 
solver \texttt{CLASS} that incorporates one-loop perturbation theory calculations.
This module, called \texttt{CLASS-PT}, computes Fourier-space power spectra of matter and biased tracers 
in real and redshift space. 
It contains 
all ingredients required for the application to data: 
IR resummation to describe 
the non-linear evolution of the BAO wiggles; non-linear bias prescriptions; UV counterterms that 
capture the effects of poorly known short-scale physics on large scales, such as 
fingers-of-God and 
baryonic feedback.
We stress that the main advantage of perturbation theory over other 
approaches is that it guarantees high precision on wavenumbers smaller than
the nonlinear scale~ $k_{\rm NL}\sim 0.5~h$/Mpc at $z = 0$.
Many complicated phenomena that operate on short scales 
drastically
simplify in
the long-wavelength limit, 
where they can be consistently and accurately taken into account.

The current execution time of 
\texttt{CLASS-PT} is fast enough to make the Markov Chain Monte Carlo analysis
of redshift-space clustering data feasible.
The code was already used for these purposes in References~\cite{Chudaykin:2019ock,Ivanov:2019pdj,Ivanov:2019hqk,Philcox:2020vvt,Nishimichi:2020tvu}. 

The realization of nonlinear perturbation theory as a module directly 
inside the Boltzmann code \texttt{CLASS} has many advantages.
It is clearly structured, easy to modify, and designed to avoid hard-coding.
Moreover, it can be readily interfaced with other software, 
e.g.~conventional MCMC 
samplers such as \texttt{Montepython} 
\cite{Audren:2012wb,Brinckmann:2018cvx} and \texttt{cobaya}\footnote{\href{https://github.com/CobayaSampler/cobaya}{
\textcolor{blue}{https://github.com/CobayaSampler/cobaya}}
}.
The \texttt{CLASS} code is one of the standard tools
established in cosmology. 
By writing our module directly as 
part of \texttt{CLASS}, it was our aim
to make the nonlinear cosmological perturbation theory calculations more available to 
the broad community: now, 
all users familiar with \texttt{CLASS} can easily perform
these calculations.

Like many things, \texttt{CLASS-PT} is ever-evolving.
The first avenue of improvement 
is devoted to the improvement of efficiency 
and accuracy of our calculation. 
We believe that some implementation choices used in the current version of
\texttt{CLASS-PT} may not be not optimal and will certainly 
be revisited in the future.

The second line of research is aimed at incorporating new non-linear effects.
In particular, the FFTLog algorithm is convenient for the implementation 
of two-loop power spectrum and one-loop bispectrum 
calculations~\cite{Simonovic:2017mhp}. 
Moreover, we plan to implement the observer-dependent convolution integrals 
describing selection effects such as intrinsic alignment of galaxies (see Ref.~\cite{Desjacques:2018pfv} and references therein).
Additionally, it is important to accurately take into account 
corrections due to the scale-dependent growth introduced, e.g.~by massive
neutrinos. We leave these research directions for future work.

\section*{Acknowledgments}

We are indebted to
Florian Beutler for providing public access to the BOSS DR12 measurements
and various data products.
We thank Nils Sch\"oneberg for sharing with us his implementation 
of FFT before the official release of \texttt{CLASS-matter}. 

We are grateful to Konstanin Dolgikh, Evan McDonough and Colin Hill 
for beta-testing the code and their useful feedback.

We would like to thank Sergey Sibiryakov, Zvonimir Vlah and Matias Zaldarriaga for useful discussions. 
A.~C. and M.~I. are partly supported by the RFBR grant 20-02-00982.

\appendix 

\section{Redshift-Space FFTLog Master Integrals}
\label{app:rsdfft}

In this Appendix we present explicit expressions for the functions 
appearing 
in the integrals \eqref{eq:intAs}.
The coefficients of the integrals with one and two insertions
of loop momenta read
\be 
\begin{split}
& A_1(\nu_1,\nu_2)=\frac{1}{2} ({\sf I}(\nu_1-1,\nu_2)-{\sf I}(\nu_1,\nu_2-1)+{\sf I}(\nu_1,\nu_2))\,,\\
& A_2(\nu_1,\nu_2)=-\frac{1}{8}\Big({\sf I}(\nu_1,\nu_2)+{\sf I}(\nu_1,\nu_2-2)+{\sf I}(\nu_1-2,\nu_2)\\
&~~~~~~~~~~-2{\sf I}(\nu_1,\nu_2-1)-2{\sf I}(\nu_1-1,\nu_2)-2{\sf I}(\nu_1-1,\nu_2-1)\Big)\,,\\
& B_2(\nu_1,\nu_2)=3\Big({\sf I}(\nu_1,\nu_2)+{\sf I}(\nu_1,\nu_2-2)+{\sf I}(\nu_1-2,\nu_2)
+\frac{2}{3}{\sf I}(\nu_1,\nu_2-1) \\
&~~~~~~~~~~-2{\sf I}(\nu_1-1,\nu_2)-2{\sf I}(\nu_1-1,\nu_2-1)\Big)\,.\\
\end{split}
\ee
For the integrals with three insertions one finds
\be 
\begin{split}
& A_3(\nu_1,\nu_2) =-\frac{3}{16}\Big(
{\sf I}(\nu_1,\nu_2)
+{\sf I}(\nu_1-3,\nu_2)
-3 {\sf I}(\nu_1-2,\nu_2-1)
-{\sf I}(\nu_1-2,\nu_2)
+3 {\sf I}(\nu_1-1,\nu_2-2)\\
&-2 {\sf I}(\nu_1-1,\nu_2-1)
-{\sf I}(\nu_1-1,\nu_2)
-{\sf I}(\nu_1,\nu_2-3)
+3 {\sf I}(\nu_1,\nu_2-2)
-3 {\sf I}(\nu_1,\nu_2-1)
\Big)\,,
\\
& B_3(\nu_1,\nu_2) =
\frac{1}{16} \Big(5 
{\sf I}(\nu_1-3,\nu_2)
-15 {\sf I}(\nu_1-2,\nu_2-1)
+3 {\sf I}(\nu_1-2,\nu_2)
+15 {\sf I}(\nu_1-1,\nu_2-2)\\
& -18 {\sf I}(\nu_1-1,\nu_2-1)
+3 {\sf I}(\nu_1-1,\nu_2)
-5 {\sf I}(\nu_1,\nu_2-3)
+15 {\sf I}(\nu_1,\nu_2-2)\\
&
-15 {\sf I}(\nu_1,\nu_2-1)
+5 {\sf I}(\nu_1,\nu_2)\Big)\,.
\end{split}
\ee
Finally, the integrals with four insertions of the loop momentum yield
\be
\begin{split}
& A_4(\nu_1,\nu_2)=
\frac{3}{128} \Big(
{\sf I}(\nu_1-4,\nu_2)
-4 {\sf I}(\nu_1-3,\nu_2-1)
-4 {\sf I}(\nu_1-3,\nu_2)
+6 {\sf I}(\nu_1-2,\nu_2-2)\\
& +4 {\sf I}(\nu_1-2,\nu_2-1)
+6 {\sf I}(\nu_1-2,\nu_2)
-4 {\sf I}(\nu_1-1,\nu_2-3)
+4 {\sf I}(\nu_1-1,\nu_2-2)\\
& +4 {\sf I}(\nu_1-1,\nu_2-1)
-4 {\sf I}(\nu_1-1,\nu_2)
+{\sf I}(\nu_1,\nu_2-4)
-4 {\sf I}(\nu_1,\nu_2-3) \\
& +6 {\sf I}(\nu_1,\nu_2-2)
-4 {\sf I}(\nu_1,\nu_2-1)
+{\sf I}(\nu_1,\nu_2)\Big)\,,
\end{split}
\ee
\be
\begin{split}
&B_4(\nu_1,\nu_2)=-\frac{3}{64} \Big(
5 {\sf I}(\nu_1-4,\nu_2)
-20 {\sf I}(\nu_1-3,\nu_2-1)
-4{\sf I}(\nu_1-3,\nu_2)\\&
+30 {\sf I}(\nu_1-2,\nu_2-2)
-12 {\sf I}(\nu_1-2,\nu_2-1)
-2 {\sf I}(\nu_1-2,\nu_2)
-20 {\sf I}(\nu_1-1,\nu_2-3)\\&
+36 {\sf I}(\nu_1-1,\nu_2-2)
-12 {\sf I}(\nu_1-1,\nu_2-1)
-4 {\sf I}(\nu_1-1,\nu_2)
+5 {\sf I}(\nu_1,\nu_2-4)
-20 {\sf I}(\nu_1,\nu_2-3)\\&
+30 {\sf I}(\nu_1,\nu_2-2)
-20 {\sf I}(\nu_1,\nu_2-1)
+5 {\sf I}(\nu_1,\nu_2)\Big)\,,
\end{split}
\ee
\be
\begin{split}
&C_4(\nu_1,\nu_2)=\frac{1}{128} \Big(
35 {\sf I}(\nu_1-4,\nu_2)
-140 {\sf I}(\nu_1-3,\nu_2-1)
+20 {\sf I}(\nu_1-3,\nu_2)
+210 {\sf I}(\nu_1-2,\nu_2-2)\\&
-180 {\sf I}(\nu_1-2,\nu_2-1)
+18 {\sf I}(\nu_1-2,\nu_2)
-140 {\sf I}(\nu_1-1,\nu_2-3)
+300 {\sf I}(\nu_1-1,\nu_2-2)\\&
-180 {\sf I}(\nu_1-1,\nu_2-1)
+20 {\sf I}(\nu_1-1,\nu_2)
+35 {\sf I}(\nu_1,\nu_2-4)
-140 {\sf I}(\nu_1,\nu_2-3)\\&
+210 {\sf I}(\nu_1,\nu_2-2)
-140 {\sf I}(\nu_1,\nu_2-1)
+35{\sf I}(\nu_1,\nu_2)
\Big)\,.
\end{split}
\ee

\section{Brief Installation Manual}
\label{app:manual}

\texttt{CLASS-PT} is compatible with both \texttt{python v2} and \texttt{v3}.
It is installed and configured in 8 easy steps:

\begin{enumerate}
	\item Download the \texttt{OpenBLAS} library from \texttt{http://www.openblas.net/}
	\item Extract the library in a folder and configure the package by executing 
\[
\$~\texttt{gmake CC=gcc FC=gfortran} 
\]
in that folder.
\item Install the package via
\[
\$~\texttt{make install PREFIX=path/to/OpenBLAS}
\]
\item Download and unpack \texttt{CLASS-PT}. 
\item Change the path to \texttt{OpenBLAS} in 
\texttt{CLASS-PT/Makefile} 
to your actual path to the compiled library \texttt{path/to/OpenBLAS/lib/libopenblas.a}
\item Update the paths to \texttt{path/to/OpenBLAS/lib/libopenblas.a}
in the \texttt{extra\_link\_args} of \texttt{CLASS-PT/python/setup.py}
\item Compile \texttt{CLASS-PT} as usual by typing 
\be 
\begin{split}
&\$~\texttt{make clean}\\
\notag
&\$~\texttt{make}
\end{split}
\ee
\item Reap the rewards of lightning-fast perturbation theory with 
\texttt{CLASS-PT} and \texttt{classy}!
\end{enumerate}

\section{Treatment of nuisance parameters}
\label{app:bias}

Let us first discuss our choice of priors for the nuisance parameters. 
We assume a flat non-informative prior on $b_1A^{1/2}\in (1,4)$. 
Since the satellite fraction 
of the BOSS galaxy sample is quite small, most of the galaxies are centrals 
and hence 
should trace the properties of the host
dark matter halos. The measurements of $b_2$ and $b_{{\cal G}_2}$ from N-body simulations \cite{Lazeyras:2017hxw} yield
\be
\label{eq:b2bg2}
b_2\approx -0.6\,,\quad  b_{{\cal G}_2}\approx -0.3 \quad \text{for} \quad
b_1\approx 2\,.
\ee
Note that the values for $b_{{\cal G}_2}$ 
are also consistent with the predictions of the coevolution model~\cite{Desjacques:2016bnm}.
On general grounds, the bias parameters are expected to be $\mathcal{O}(1)$
in the EFT, which motivates the priors \mbox{$b_2 A^{1/2},b_{{\cal G}_2}A^{1/2}\sim \mathcal{N}(0,1)$}. Note that we have inserted $A^{1/2}\approx 1$ in the definition of our sample parameters
because this choice leads to somewhat better convergence of the MCMC chains.

As far as the higher-derivative counterterms $c_0$ and $c_2$ are concerned,
they are, in general, also expected to be 
\be 
c_0,c_2=\mathcal{O}(1)\times k^{-2}_{\rm NL}
\ee 
The non-linear scale in redshift space depends on the
velocity dispersion of the BOSS galaxies, which can be quite large. Indeed, previous BOSS full-shape analyses report $\sigma_v \sim 5$ Mpc/$h$ \cite{Beutler:2016arn}, which is several times larger than the real-space estimate \mbox{$k^{-1}_{\rm NL}\sim 2$ Mpc/$h$}. It is important to stress that
the quoted measurement of $\sigma_v$ from Ref.~\cite{Beutler:2016arn}
results from an application of a simplified fitting function, and the actual velocity
dispersion can be different if one is using the full EFT model.  
Nevertheless, we adopt the following priors 
for the counterterms that are wide enough to accommodate a large velocity dispersion,
\be
\label{eq:c0c2}
c_0,c_2\sim \mathcal{N}(0,30^2)~[\text{Mpc}/h]^2\,. 
\ee

The prior for the next-to-leading order RSD counterterm $\tilde{c}$ 
is more subtle. Na\"ively, this contribution has the order of the two-loop correction and hence has not been originally included in the one-loop EFT theory model~\cite{Senatore:2014vja,Lewandowski:2015ziq}. However, due to the strong fingers-of-God found in the BOSS galaxy sample, the coefficient $\tilde{c}$ turned out to be enhanced compared to the naive EFT estimates. Dedicated analyses of the BOSS mock catalogs and the real data \cite{Ivanov:2019pdj} gave 
\be 
\label{eq:ct}
\tilde{c}\sim \sigma_v^4\sim [5~\Mpc/h]^4 \sim 500~[\Mpc/h]^4 \,.
\ee
This motivates 
the prior $\tilde{c}\sim \mathcal{N}(500,500^2)~[\Mpc/h]^4$

Finally, as far as the constant shot noise contribution $P_{\rm shot}$
is concerned, its true value is expected to deviate from the Poissonian 
prediction $\bar n^{-1}$ due to exclusion effects \cite{Baldauf:2013hka}
and fiber collisions \cite{Hahn:2016kiy}. The latter are not possible to predict from first principles 
and 
are difficult 
to model even in mock catalogs~\cite{Kitaura:2015uqa}. Thus, we adopted a more practical data-driven approach; 
first finding the best-fit values for $P_{\rm shot}$ from the data itself and then imposing a large conservative prior centered at this best-fit value. The Poissonian part $\bar n^{-1}$ has already been subtracted from the power spectrum estimator of our data. We found the residual shot noise contribution best-fits to be roughly
\be 
P_{\rm shot}\sim 5\cdot 10^3~[\Mpc/h]^3\,,
\ee
for all the data chunks studied in this paper. This motivated 
the prior $P_{\rm shot}\sim \mathcal{N} (5,5^2)\cdot 10^3~[\Mpc/h]^3$, which is
wide enough to accommodate absolute deviations from $\bar n^{-1}$ 
across all data samples, in particular, in the high-z SGC sample, whose Poissonian shot noise is quite large. Notice that the residual $P_{\rm shot}$ can, in principle, be negative. This fact is reflected in our prior.

\begin{figure}[ht!]
\includegraphics[width=1\textwidth]{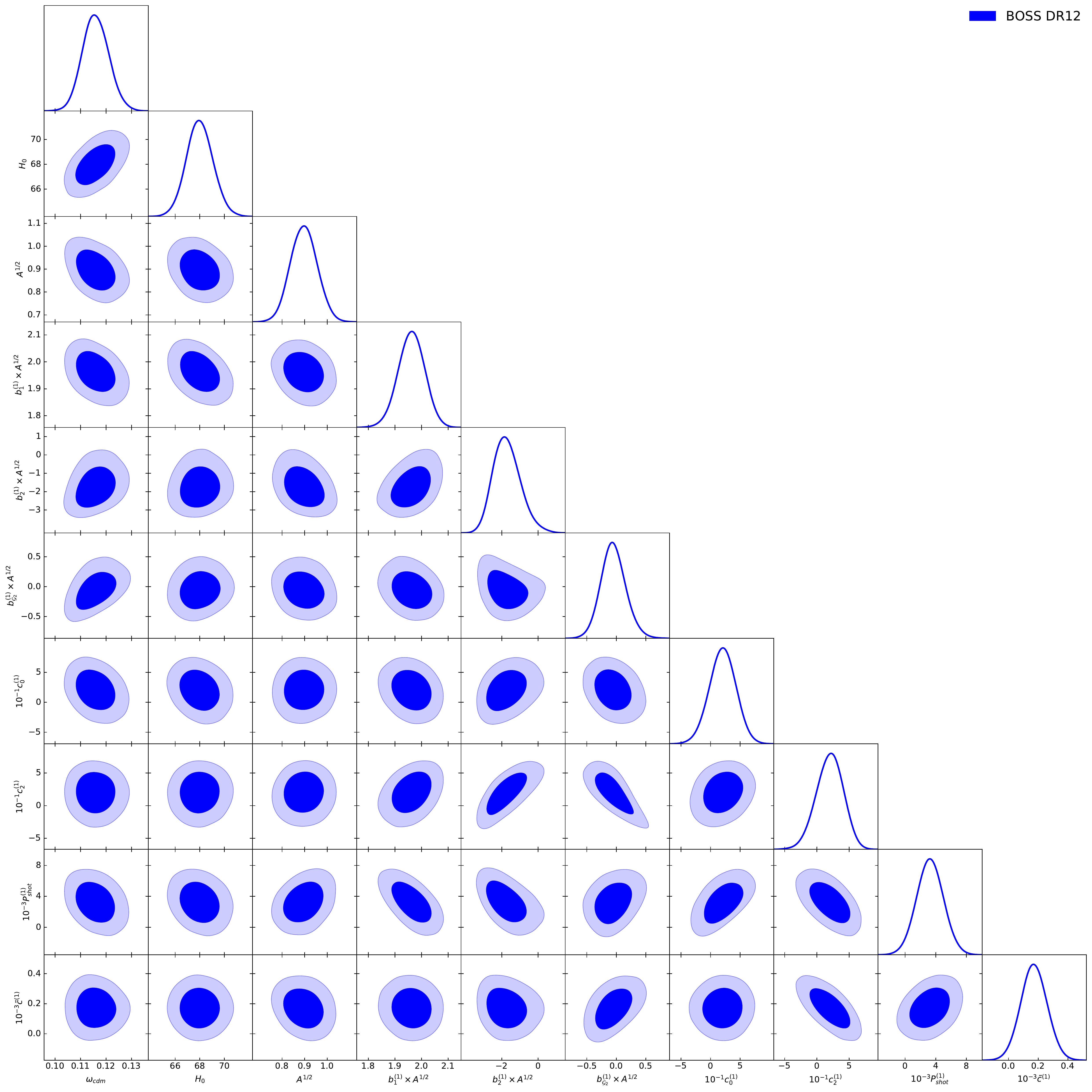}
\caption{ 
The posterior distribution for $\omega_{cdm},~H_0,~A^{1/2}\equiv(A_s/A_{s,\,{\rm Planck}})^{1/2}$ and the high-z NGC nuisance parameters inferred from the joint BOSS DR12 full-shape likelihood. 
}    
\label{fig:ngcz3}
\end{figure}

\begin{figure}[ht!]
\includegraphics[width=1\textwidth]{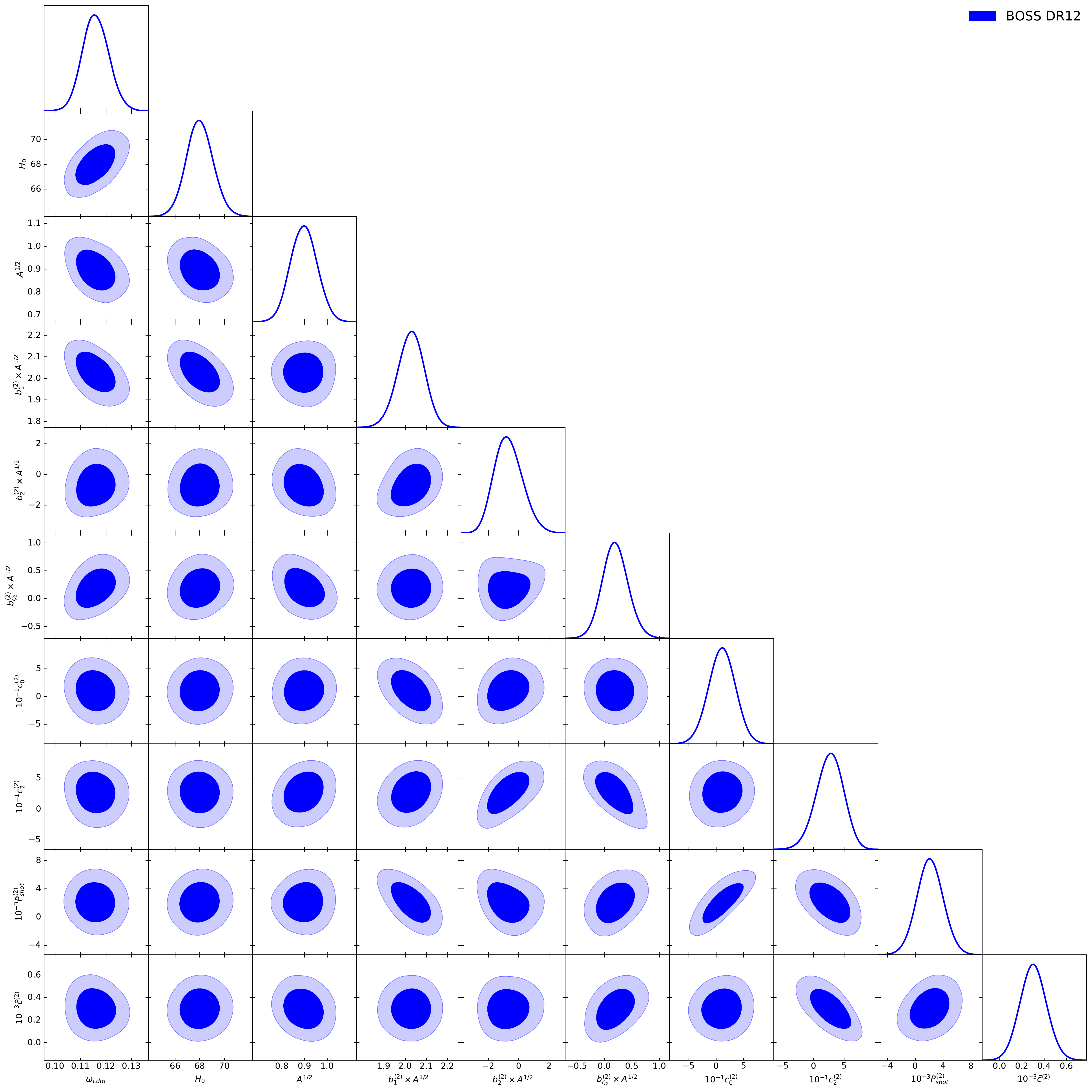}
\caption{ 
As Fig.~\ref{fig:ngcz3} but for the high-z SGC region. 
}    
\label{fig:sgcz3}
\end{figure}

\begin{figure}[ht!]
\includegraphics[width=1\textwidth]{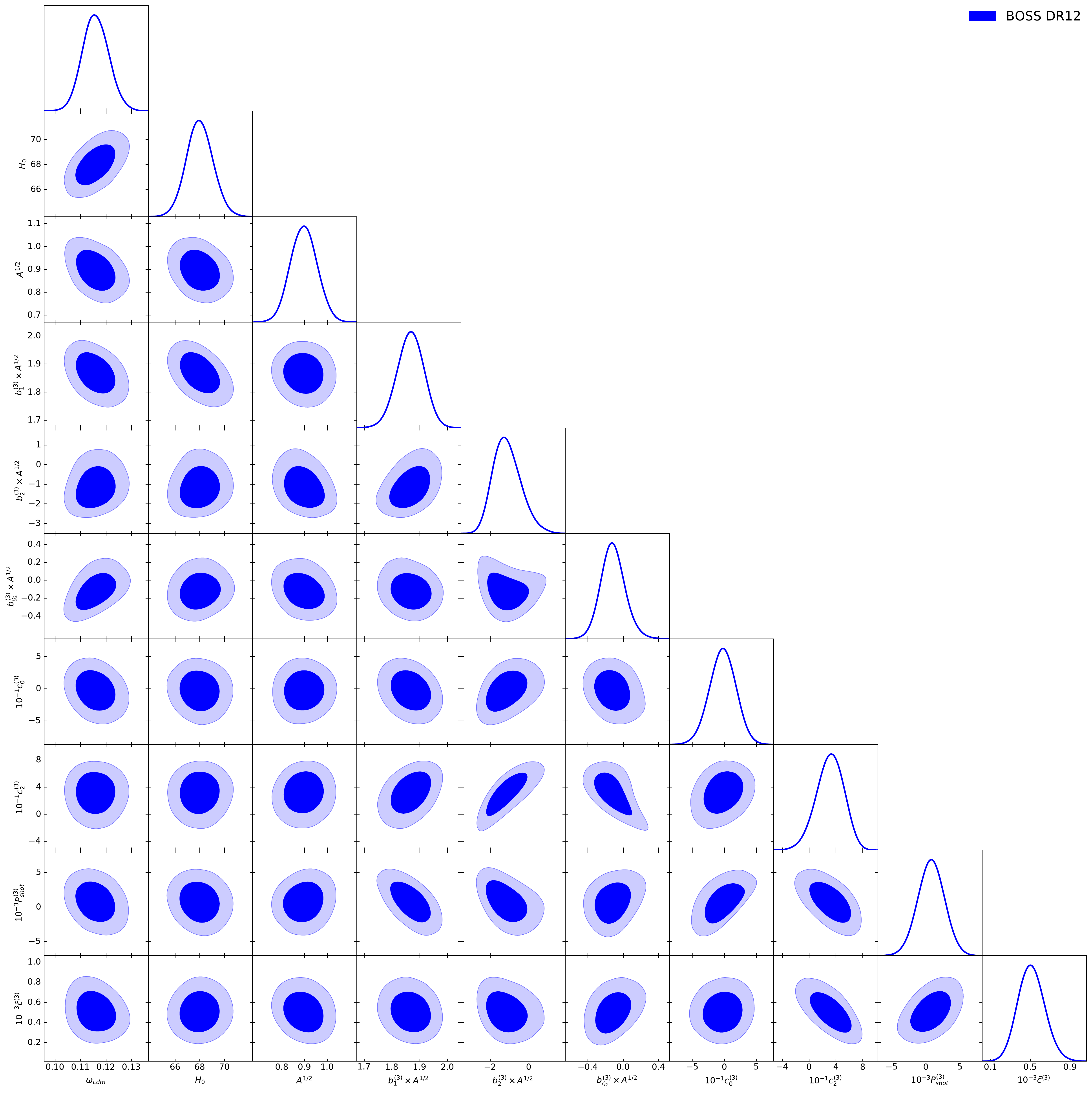}
\caption{ 
As Fig.~\ref{fig:ngcz3} but for the low-z NGC region.
}    
\label{fig:ngcz1}
\end{figure}

\begin{figure}[ht!]
\includegraphics[width=1\textwidth]{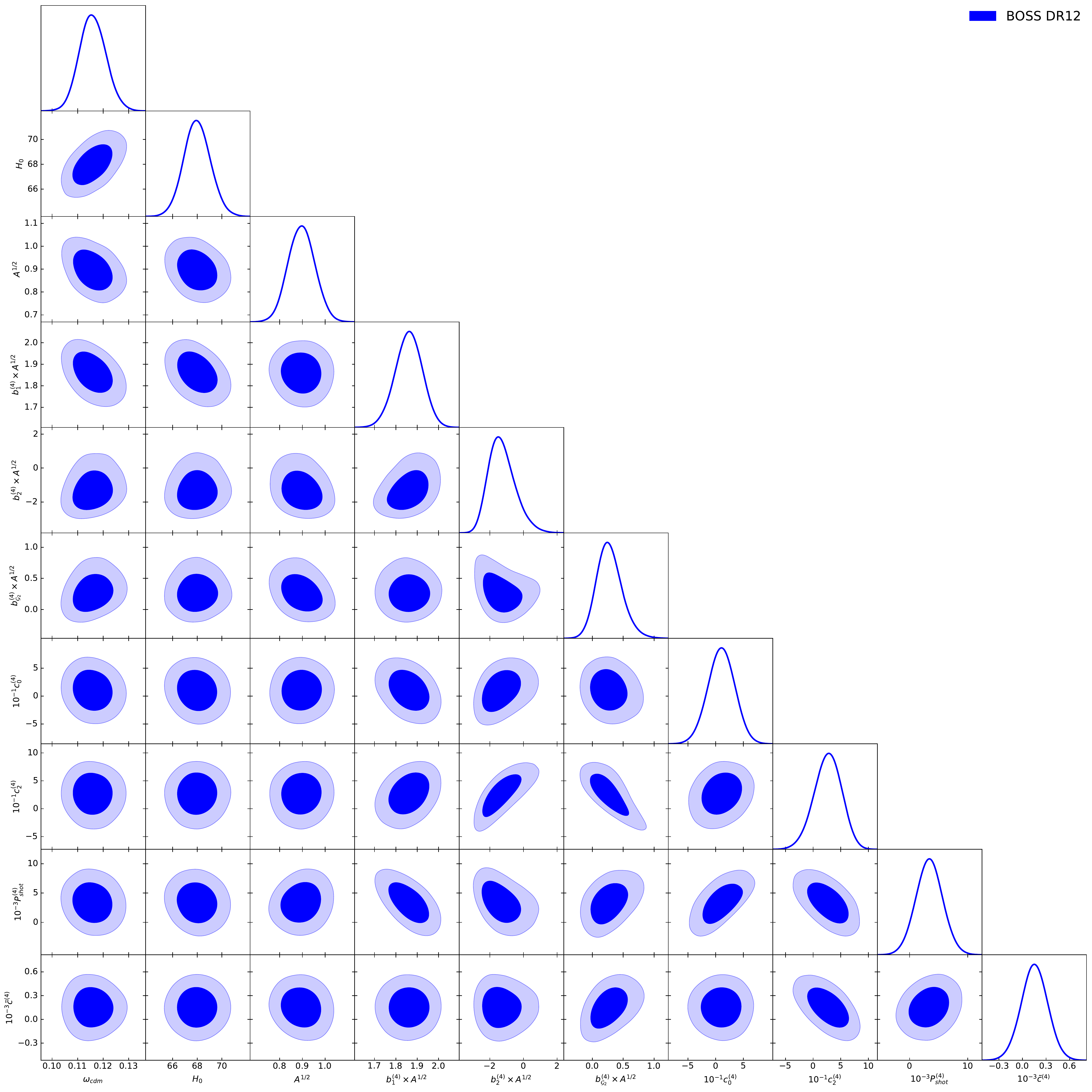}
\caption{ 
As Fig.~\ref{fig:ngcz3} but for the low-z SGC region.
}    
\label{fig:sgcz1}
\end{figure}

We present the optimal values of the nuisance parameters found in our MCMC analysis in Tab.~\ref{tab:full} and in Figs~\ref{fig:ngcz3} (high-z NGC),~\ref{fig:sgcz3} (high-z SGC),~\ref{fig:ngcz1} (low-z NGC),~\ref{fig:sgcz1} (low-z SGC). 
We differentiate between nuisance parameters for different BOSS data samples
with the following superscripts:
\be
\text{(1) = high-z NGC, (2) = high-z SGC, (3) = low-z NGC, (4) = low-z SGC}
\ee
We stress that these parameters are obtained from a joint fit, i.e.~the cosmological parameters are assumed to be the same across all samples.
One can see that the best-fit values are in good agreement
with the ones expected from the BOSS galaxy sample. 
The measured values of $c_0,c_2$ and $\tilde{c}$ are indeed consistent
with the estimate for the velocity dispersion effects \eqref{eq:c0c2}, \eqref{eq:ct}. 
Moreover, the best-fitting values of $b_2$ and $b_{\mathcal{G}_2}$
agree with the values found in the N-body simulations for the host halos similar to those of the BOSS sample \eqref{eq:b2bg2}. 
It would be interesting to further investigate if these values are also compatible with biases inferred from the bispectrum~\cite{Gil-Marin:2014sta,Gil-Marin:2014baa}.

\begin{table}[ht!]
\begin{center}
\begin{tabular}{|l|c|c|c|c|}
 \hline
Param & best-fit & mean$\pm\sigma$ & 95\% lower & 95\% upper \\ \hline
$\omega_{cdm }$ &$0.1169$ & $0.1159_{-0.0054}^{+0.0050}$ & $0.1061$ & $0.1258$ \\
$H_0$ &$67.91$ & $67.98_{-1.1}^{+1.1}$ & $65.81$ & $70.18$ \\
$A^{1/2}$ &$0.9499$ & $0.9069_{-0.054}^{+0.058}$ & $0.7931$ & $1.016$ \\
$\Omega_{m }$ &$0.3034$ & $0.3006_{-0.010}^{+0.010}$ & $0.2812$ & $0.3207$ \\
$\sigma_8$ &$0.716$ & $0.710_{-0.043}^{+0.043}$ & $0.627$ & $0.794$ \\ \hline
$b^{(1)}_{1 }A^{1/2}$ &$1.949$ & $1.958_{-0.048}^{+0.05}$ & $1.86$ & $2.056$ \\
$b^{(1)}_{2 }A^{1/2}$ &$-1.722$ & $-1.766_{-0.79}^{+0.62}$ & $-3.145$ & $-0.2932$ \\
$b^{(1)}_{{\mathcal{G}_2} }A^{1/2}$ &$-0.02759$ & $-0.04753_{-0.22}^{+0.19}$ & $-0.466$ & $0.3902$ \\
$10^{-1}c^{(1)}_{{0} }$ &$1.794$ & $1.991_{-2.2}^{+2.3}$ & $-2.438$ & $6.494$ \\
$10^{-1}c^{(1)}_{{2} }$ &$2.192$ & $1.949_{-1.9}^{+2.2}$ & $-2.216$ & $5.992$ \\
$10^{-3}P^{(1)}_{{\rm shot} }$ &$3.308$ & $3.344_{-1.8}^{+1.7}$ & $-0.1072$ & $6.83$ \\
$10^{-3}\tilde{c}^{(1)}$ &$0.1775$ & $0.1684_{-0.09}^{+0.086}$ & $-0.008266$ & $0.348$ \\
$b^{(2)}_{1 }A^{1/2}$ &$2.015$ & $2.024_{-0.06}^{+0.065}$ & $1.897$ & $2.148$ \\
$b^{(2)}_{2 }A^{1/2}$ &$-0.5063$ & $-0.7053_{-1}^{+0.83}$ & $-2.475$ & $1.179$ \\
$b^{(2)}_{{\mathcal{G}_2} }A^{1/2}$ &$0.05212$ & $0.1852_{-0.24}^{+0.22}$ & $-0.2893$ & $0.6625$ \\
$10^{-1}c^{(2)}_{{0} }$ &$1.284$ & $1.081_{-2.4}^{+2.5}$ & $-3.826$ & $5.964$ \\
$10^{-1}c^{(2)}_{{2} }$ &$4.206$ & $2.711_{-2.1}^{+2.4}$ & $-1.846$ & $7.03$ \\
$10^{-3}P^{(2)}_{{\rm shot} }$ &$1.352$ & $2.141_{-2}^{+1.9}$ & $-1.707$ & $6.119$ \\
$10^{-3}\tilde{c}^{(2)}$ &$0.2429$ & $0.2977_{-0.12}^{+0.12}$ & $0.05633$ & $0.5421$ \\
$b^{(3)}_{1 }$ &$1.813$ & $1.864_{-0.048}^{+0.049}$ & $1.768$ & $1.961$ \\
$b^{(3)}_{2 }$ &$-1.985$ & $-1.164_{-0.8}^{+0.61}$ & $-2.488$ & $0.2811$ \\
$b^{(3)}_{{\mathcal{G}_2} }$ &$-0.05043$ & $-0.1177_{-0.14}^{+0.12}$ & $-0.3899$ & $0.1657$ \\
$10^{-1}c^{(3)}_{{0} }$ &$-1.577$ & $-0.3099_{-2.1}^{+2.1}$ & $-4.457$ & $3.89$ \\
$10^{-1}c^{(3)}_{{2} }$ &$1.14$ & $3.118_{-1.9}^{+2.2}$ & $-1.019$ & $7.129$ \\
$10^{-3}P^{(3)}_{{\rm shot} }$ &$1.927$ & $0.7849_{-1.9}^{+2}$ & $-3.205$ & $4.718$ \\
$10^{-3}\tilde{c}^{(3)}$ &$0.5257$ & $0.5022_{-0.14}^{+0.13}$ & $0.235$ & $0.7746$ \\
$b^{(4)}_{1 }A^{1/2}$ &$1.848$ & $1.857_{-0.062}^{+0.065}$ & $1.729$ & $1.985$ \\
$b^{(4)}_{2 }A^{1/2}$ &$-1.078$ & $-1.301_{-0.9}^{+0.63}$ & $-2.761$ & $0.3134$ \\
$b^{(4)}_{{\mathcal{G}_2} }A^{1/2}$ &$0.232$ & $0.2693_{-0.23}^{+0.18}$ & $-0.1374$ & $0.7091$ \\
$10^{-1}c^{(4)}_{{0} }$ &$2.036$ & $1.032_{-2.4}^{+2.5}$ & $-3.908$ & $5.88$ \\
$10^{-1}c^{(4)}_{{2} }$ &$3.319$ & $2.637_{-2.4}^{+2.6}$ & $-2.412$ & $7.552$ \\
$10^{-3}P^{(4)}_{{\rm shot} }$ &$3.849$ & $3.467_{-2.4}^{+2.4}$ & $-1.243$ & $8.23$ \\
$10^{-3}\tilde{c}^{(4)}$ &$0.1109$ & $0.1476_{-0.17}^{+0.17}$ & $-0.1991$ & $0.4959$ \\
\hline
 \end{tabular} 
 \caption{
The results of our MCMC analysis for the joint BOSS DR12 full-shape likelihood including all nuisance parameters.
We use the following units:
[km/s/Mpc] for $H_0$, [Mpc/$h$]$^2$ for $c_0,c_2$, [Mpc/$h$]$^4$ for $\tilde{c}$,
[Mpc/$h$]$^3$ for $P_{\rm shot}$. The upper group displays the cosmological parameters that are considered to be the same for all data chunks.
}
 \label{tab:full}
 \end{center}
 \end{table} 


\bibliographystyle{JHEP}
\bibliography{short}

\providecommand{\href}[2]{#2}\begingroup\raggedright\begin{thebibliography}{100}

\bibitem{Aghanim:2018eyx}
{\scshape Planck} collaboration, N.~Aghanim et~al., \emph{{Planck 2018 results.
  VI. Cosmological parameters}},
  \href{https://arxiv.org/abs/1807.06209}{{\ttfamily 1807.06209}}.

\bibitem{Seljak:1996is}
U.~Seljak and M.~Zaldarriaga, \emph{{A Line of sight integration approach to
  cosmic microwave background anisotropies}},
  \href{https://doi.org/10.1086/177793}{\emph{Astrophys. J.} {\bfseries 469}
  (1996) 437} [\href{https://arxiv.org/abs/astro-ph/9603033}{{\ttfamily
  astro-ph/9603033}}].

\bibitem{Lewis:2002ah}
A.~Lewis and S.~Bridle, \emph{{Cosmological parameters from CMB and other data:
  A Monte Carlo approach}},
  \href{https://doi.org/10.1103/PhysRevD.66.103511}{\emph{Phys. Rev.}
  {\bfseries D66} (2002) 103511}
  [\href{https://arxiv.org/abs/astro-ph/0205436}{{\ttfamily
  astro-ph/0205436}}].

\bibitem{Blas:2011rf}
D.~Blas, J.~Lesgourgues and T.~Tram, \emph{{The Cosmic Linear Anisotropy
  Solving System (CLASS) II: Approximation schemes}},
  \href{https://doi.org/10.1088/1475-7516/2011/07/034}{\emph{JCAP} {\bfseries
  1107} (2011) 034} [\href{https://arxiv.org/abs/1104.2933}{{\ttfamily
  1104.2933}}].

\bibitem{Bellini:2017avd}
E.~Bellini et~al., \emph{{Comparison of Einstein-Boltzmann solvers for testing
  general relativity}},
  \href{https://doi.org/10.1103/PhysRevD.97.023520}{\emph{Phys. Rev.}
  {\bfseries D97} (2018) 023520}
  [\href{https://arxiv.org/abs/1709.09135}{{\ttfamily 1709.09135}}].

\bibitem{Brinckmann:2018cvx}
T.~Brinckmann and J.~Lesgourgues, \emph{{MontePython 3: boosted MCMC sampler
  and other features}},
  \href{https://doi.org/10.1016/j.dark.2018.100260}{\emph{Phys. Dark Univ.}
  {\bfseries 24} (2019) 100260}
  [\href{https://arxiv.org/abs/1804.07261}{{\ttfamily 1804.07261}}].

\bibitem{Alam:2016hwk}
{\scshape BOSS} collaboration, S.~Alam et~al., \emph{{The clustering of
  galaxies in the completed SDSS-III Baryon Oscillation Spectroscopic Survey:
  cosmological analysis of the DR12 galaxy sample}},
  \href{https://doi.org/10.1093/mnras/stx721}{\emph{Mon. Not. Roy. Astron.
  Soc.} {\bfseries 470} (2017) 2617}
  [\href{https://arxiv.org/abs/1607.03155}{{\ttfamily 1607.03155}}].

\bibitem{Laureijs:2011gra}
{\scshape EUCLID} collaboration, R.~Laureijs et~al., \emph{{Euclid Definition
  Study Report}},  \href{https://arxiv.org/abs/1110.3193}{{\ttfamily
  1110.3193}}.

\bibitem{Amendola:2016saw}
L.~Amendola et~al., \emph{{Cosmology and fundamental physics with the Euclid
  satellite}}, \href{https://doi.org/10.1007/s41114-017-0010-3}{\emph{Living
  Rev. Rel.} {\bfseries 21} (2018) 2}
  [\href{https://arxiv.org/abs/1606.00180}{{\ttfamily 1606.00180}}].

\bibitem{Aghamousa:2016zmz}
{\scshape DESI} collaboration, A.~Aghamousa et~al., \emph{{The DESI Experiment
  Part I: Science,Targeting, and Survey Design}},
  \href{https://arxiv.org/abs/1611.00036}{{\ttfamily 1611.00036}}.

\bibitem{Takahashi:2012em}
R.~Takahashi, M.~Sato, T.~Nishimichi, A.~Taruya and M.~Oguri, \emph{{Revising
  the Halofit Model for the Nonlinear Matter Power Spectrum}},
  \href{https://doi.org/10.1088/0004-637X/761/2/152}{\emph{Astrophys. J.}
  {\bfseries 761} (2012) 152}
  [\href{https://arxiv.org/abs/1208.2701}{{\ttfamily 1208.2701}}].

\bibitem{Audren:2012wb}
B.~Audren, J.~Lesgourgues, K.~Benabed and S.~Prunet, \emph{{Conservative
  Constraints on Early Cosmology: an illustration of the Monte Python
  cosmological parameter inference code}},
  \href{https://doi.org/10.1088/1475-7516/2013/02/001}{\emph{JCAP} {\bfseries
  1302} (2013) 001} [\href{https://arxiv.org/abs/1210.7183}{{\ttfamily
  1210.7183}}].

\bibitem{Ivanov:2019pdj}
M.~M. Ivanov, M.~Simonovic and M.~Zaldarriaga, \emph{{Cosmological Parameters
  from the BOSS Galaxy Power Spectrum}},
  \href{https://arxiv.org/abs/1909.05277}{{\ttfamily 1909.05277}}.

\bibitem{Ivanov:2019hqk}
M.~M. Ivanov, M.~Simonović and M.~Zaldarriaga, \emph{{Cosmological Parameters
  and Neutrino Masses from the Final Planck and Full-Shape BOSS Data}},
  \href{https://arxiv.org/abs/1912.08208}{{\ttfamily 1912.08208}}.

\bibitem{Philcox:2020vvt}
O.~H.~E. Philcox, M.~M. Ivanov, M.~Simonović and M.~Zaldarriaga,
  \emph{{Combining Full-Shape and BAO Analyses of Galaxy Power Spectra: A 1.6\%
  CMB-independent constraint on H0}},
  \href{https://arxiv.org/abs/2002.04035}{{\ttfamily 2002.04035}}.

\bibitem{Nishimichi:2020tvu}
T.~Nishimichi, G.~D'Amico, M.~M. Ivanov, L.~Senatore, M.~Simonovic, M.~Takada
  et~al., \emph{{Blinded challenge for precision cosmology with large-scale
  structure: results from effective field theory for the redshift-space galaxy
  power spectrum}},  \href{https://arxiv.org/abs/2003.08277}{{\ttfamily
  2003.08277}}.

\bibitem{Chudaykin:2019ock}
A.~Chudaykin and M.~M. Ivanov, \emph{{Measuring neutrino masses with
  large-scale structure: Euclid forecast with controlled theoretical error}},
  \href{https://doi.org/10.1088/1475-7516/2019/11/034}{\emph{JCAP} {\bfseries
  1911} (2019) 034} [\href{https://arxiv.org/abs/1907.06666}{{\ttfamily
  1907.06666}}].

\bibitem{Carlson:2009it}
J.~Carlson, M.~White and N.~Padmanabhan, \emph{{A critical look at cosmological
  perturbation theory techniques}},
  \href{https://doi.org/10.1103/PhysRevD.80.043531}{\emph{Phys. Rev.}
  {\bfseries D80} (2009) 043531}
  [\href{https://arxiv.org/abs/0905.0479}{{\ttfamily 0905.0479}}].

\bibitem{Tassev:2013zua}
S.~Tassev, \emph{{N-point Statistics of Large-Scale Structure in the Zel'dovich
  Approximation}},
  \href{https://doi.org/10.1088/1475-7516/2014/06/012}{\emph{JCAP} {\bfseries
  1406} (2014) 012} [\href{https://arxiv.org/abs/1311.6316}{{\ttfamily
  1311.6316}}].

\bibitem{Bertolini:2015fya}
D.~Bertolini, K.~Schutz, M.~P. Solon, J.~R. Walsh and K.~M. Zurek,
  \emph{{Non-Gaussian Covariance of the Matter Power Spectrum in the Effective
  Field Theory of Large Scale Structure}},
  \href{https://doi.org/10.1103/PhysRevD.93.123505}{\emph{Phys. Rev.}
  {\bfseries D93} (2016) 123505}
  [\href{https://arxiv.org/abs/1512.07630}{{\ttfamily 1512.07630}}].

\bibitem{Schmittfull:2016jsw}
M.~Schmittfull, Z.~Vlah and P.~McDonald, \emph{{Fast large scale structure
  perturbation theory using one-dimensional fast Fourier transforms}},
  \href{https://doi.org/10.1103/PhysRevD.93.103528}{\emph{Phys.\ Rev.\ D}
  {\bfseries 93} (2016) 103528}
  [\href{https://arxiv.org/abs/1603.04405}{{\ttfamily 1603.04405}}].

\bibitem{McEwen:2016fjn}
J.~E. McEwen, X.~Fang, C.~M. Hirata and J.~A. Blazek, \emph{{FAST-PT: a novel
  algorithm to calculate convolution integrals in cosmological perturbation
  theory}}, \href{https://doi.org/10.1088/1475-7516/2016/09/015}{\emph{JCAP}
  {\bfseries 1609} (2016) 015}
  [\href{https://arxiv.org/abs/1603.04826}{{\ttfamily 1603.04826}}].

\bibitem{Fang:2016wcf}
X.~Fang, J.~A. Blazek, J.~E. McEwen and C.~M. Hirata, \emph{{FAST-PT II: an
  algorithm to calculate convolution integrals of general tensor quantities in
  cosmological perturbation theory}},
  \href{https://doi.org/10.1088/1475-7516/2017/02/030}{\emph{JCAP} {\bfseries
  1702} (2017) 030} [\href{https://arxiv.org/abs/1609.05978}{{\ttfamily
  1609.05978}}].

\bibitem{Simonovic:2017mhp}
M.~Simonović, T.~Baldauf, M.~Zaldarriaga, J.~J. Carrasco and J.~A. Kollmeier,
  \emph{{Cosmological perturbation theory using the FFTLog: formalism and
  connection to QFT loop integrals}},
  \href{https://doi.org/10.1088/1475-7516/2018/04/030}{\emph{JCAP} {\bfseries
  1804} (2018) 030} [\href{https://arxiv.org/abs/1708.08130}{{\ttfamily
  1708.08130}}].

\bibitem{DAmico:2020kxu}
G.~D'Amico, L.~Senatore and P.~Zhang, \emph{{Limits on $w$CDM from the EFTofLSS
  with the PyBird code}},  \href{https://arxiv.org/abs/2003.07956}{{\ttfamily
  2003.07956}}.

\bibitem{1970A&A.....5...84Z}
Y.~B. {Zel'Dovich}, \emph{{Reprint of 1970A\&amp;A.....5...84Z. Gravitational
  instability: an approximate theory for large density perturbations.}},
  {\emph{Astronomy and Astrophysics} {\bfseries 500} (1970) 13}.

\bibitem{Scoccimarro:1995if}
R.~Scoccimarro and J.~Frieman, \emph{{Loop corrections in nonlinear
  cosmological perturbation theory}},
  \href{https://doi.org/10.1086/192306}{\emph{Astrophys.\ J.\ Suppl.}
  {\bfseries 105} (1996) 37}
  [\href{https://arxiv.org/abs/astro-ph/9509047}{{\ttfamily
  astro-ph/9509047}}].

\bibitem{Scoccimarro:1996se}
R.~Scoccimarro and J.~Frieman, \emph{{Loop corrections in nonlinear
  cosmological perturbation theory 2. Two point statistics and
  selfsimilarity}}, \href{https://doi.org/10.1086/178177}{\emph{Astrophys.\ J.}
  {\bfseries 473} (1996) 620}
  [\href{https://arxiv.org/abs/astro-ph/9602070}{{\ttfamily
  astro-ph/9602070}}].

\bibitem{Scoccimarro:1997st}
R.~Scoccimarro, S.~Colombi, J.~N. Fry, J.~A. Frieman, E.~Hivon and A.~Melott,
  \emph{{Nonlinear evolution of the bispectrum of cosmological perturbations}},
  \href{https://doi.org/10.1086/305399}{\emph{Astrophys. J.} {\bfseries 496}
  (1998) 586} [\href{https://arxiv.org/abs/astro-ph/9704075}{{\ttfamily
  astro-ph/9704075}}].

\bibitem{Bernardeau:2001qr}
F.~Bernardeau, S.~Colombi, E.~Gaztanaga and R.~Scoccimarro, \emph{{Large scale
  structure of the universe and cosmological perturbation theory}},
  \href{https://doi.org/10.1016/S0370-1573(02)00135-7}{\emph{Phys. Rept.}
  {\bfseries 367} (2002) 1}
  [\href{https://arxiv.org/abs/astro-ph/0112551}{{\ttfamily
  astro-ph/0112551}}].

\bibitem{Crocce:2005xy}
M.~Crocce and R.~Scoccimarro, \emph{{Renormalized cosmological perturbation
  theory}}, \href{https://doi.org/10.1103/PhysRevD.73.063519}{\emph{Phys.\
  Rev.\ D} {\bfseries 73} (2006) 063519}
  [\href{https://arxiv.org/abs/astro-ph/0509418}{{\ttfamily
  astro-ph/0509418}}].

\bibitem{Blas:2013aba}
D.~Blas, M.~Garny and T.~Konstandin, \emph{{Cosmological perturbation theory at
  three-loop order}},
  \href{https://doi.org/10.1088/1475-7516/2014/01/010}{\emph{JCAP} {\bfseries
  1401} (2014) 010} [\href{https://arxiv.org/abs/1309.3308}{{\ttfamily
  1309.3308}}].

\bibitem{Crocce:2005xz}
M.~Crocce and R.~Scoccimarro, \emph{{Memory of initial conditions in
  gravitational clustering}},
  \href{https://doi.org/10.1103/PhysRevD.73.063520}{\emph{Phys.\ Rev.\ D}
  {\bfseries 73} (2006) 063520}
  [\href{https://arxiv.org/abs/astro-ph/0509419}{{\ttfamily
  astro-ph/0509419}}].

\bibitem{McQuinn:2015tva}
M.~McQuinn and M.~White, \emph{{Cosmological perturbation theory in 1+1
  dimensions}},
  \href{https://doi.org/10.1088/1475-7516/2016/01/043}{\emph{JCAP} {\bfseries
  01} (2016) 043} [\href{https://arxiv.org/abs/1502.07389}{{\ttfamily
  1502.07389}}].

\bibitem{Baumann:2010tm}
D.~Baumann, A.~Nicolis, L.~Senatore and M.~Zaldarriaga, \emph{{Cosmological
  Non-Linearities as an Effective Fluid}},
  \href{https://doi.org/10.1088/1475-7516/2012/07/051}{\emph{JCAP} {\bfseries
  1207} (2012) 051} [\href{https://arxiv.org/abs/1004.2488}{{\ttfamily
  1004.2488}}].

\bibitem{Carrasco:2012cv}
J.~J.~M. Carrasco, M.~P. Hertzberg and L.~Senatore, \emph{{The Effective Field
  Theory of Cosmological Large Scale Structures}},
  \href{https://doi.org/10.1007/JHEP09(2012)082}{\emph{JHEP} {\bfseries 09}
  (2012) 082} [\href{https://arxiv.org/abs/1206.2926}{{\ttfamily 1206.2926}}].

\bibitem{Carrasco:2013mua}
J.~J.~M. Carrasco, S.~Foreman, D.~Green and L.~Senatore, \emph{{The Effective
  Field Theory of Large Scale Structures at Two Loops}},
  \href{https://doi.org/10.1088/1475-7516/2014/07/057}{\emph{JCAP} {\bfseries
  07} (2014) 057} [\href{https://arxiv.org/abs/1310.0464}{{\ttfamily
  1310.0464}}].

\bibitem{Baldauf:2016sjb}
T.~Baldauf, M.~Mirbabayi, M.~Simonović and M.~Zaldarriaga, \emph{{LSS
  constraints with controlled theoretical uncertainties}},
  \href{https://arxiv.org/abs/1602.00674}{{\ttfamily 1602.00674}}.

\bibitem{Peloso:2013zw}
M.~Peloso and M.~Pietroni, \emph{{Galilean invariance and the consistency
  relation for the nonlinear squeezed bispectrum of large scale structure}},
  \href{https://doi.org/10.1088/1475-7516/2013/05/031}{\emph{JCAP} {\bfseries
  05} (2013) 031} [\href{https://arxiv.org/abs/1302.0223}{{\ttfamily
  1302.0223}}].

\bibitem{Kehagias:2013yd}
A.~Kehagias and A.~Riotto, \emph{{Symmetries and Consistency Relations in the
  Large Scale Structure of the Universe}},
  \href{https://doi.org/10.1016/j.nuclphysb.2013.05.009}{\emph{Nucl.\ Phys.\ B}
  {\bfseries 873} (2013) 514}
  [\href{https://arxiv.org/abs/1302.0130}{{\ttfamily 1302.0130}}].

\bibitem{Creminelli:2013mca}
P.~Creminelli, J.~Noreña, M.~Simonović and F.~Vernizzi, \emph{{Single-Field
  Consistency Relations of Large Scale Structure}},
  \href{https://doi.org/10.1088/1475-7516/2013/12/025}{\emph{JCAP} {\bfseries
  12} (2013) 025} [\href{https://arxiv.org/abs/1309.3557}{{\ttfamily
  1309.3557}}].

\bibitem{Eisenstein:2006nj}
D.~J. Eisenstein, H.-j. Seo and M.~J. White, \emph{{On the Robustness of the
  Acoustic Scale in the Low-Redshift Clustering of Matter}},
  \href{https://doi.org/10.1086/518755}{\emph{Astrophys. J.} {\bfseries 664}
  (2007) 660} [\href{https://arxiv.org/abs/astro-ph/0604361}{{\ttfamily
  astro-ph/0604361}}].

\bibitem{Crocce:2007dt}
M.~Crocce and R.~Scoccimarro, \emph{{Nonlinear Evolution of Baryon Acoustic
  Oscillations}}, \href{https://doi.org/10.1103/PhysRevD.77.023533}{\emph{Phys.
  Rev.} {\bfseries D77} (2008) 023533}
  [\href{https://arxiv.org/abs/0704.2783}{{\ttfamily 0704.2783}}].

\bibitem{Sugiyama:2013gza}
N.~S. Sugiyama and D.~N. Spergel, \emph{{How does non-linear dynamics affect
  the baryon acoustic oscillation?}},
  \href{https://doi.org/10.1088/1475-7516/2014/02/042}{\emph{JCAP} {\bfseries
  1402} (2014) 042} [\href{https://arxiv.org/abs/1306.6660}{{\ttfamily
  1306.6660}}].

\bibitem{Senatore:2014via}
L.~Senatore and M.~Zaldarriaga, \emph{{The IR-resummed Effective Field Theory
  of Large Scale Structures}},
  \href{https://doi.org/10.1088/1475-7516/2015/02/013}{\emph{JCAP} {\bfseries
  1502} (2015) 013} [\href{https://arxiv.org/abs/1404.5954}{{\ttfamily
  1404.5954}}].

\bibitem{Baldauf:2015xfa}
T.~Baldauf, M.~Mirbabayi, M.~Simonović and M.~Zaldarriaga, \emph{{Equivalence
  Principle and the Baryon Acoustic Peak}},
  \href{https://doi.org/10.1103/PhysRevD.92.043514}{\emph{Phys. Rev.}
  {\bfseries D92} (2015) 043514}
  [\href{https://arxiv.org/abs/1504.04366}{{\ttfamily 1504.04366}}].

\bibitem{Vlah:2015zda}
Z.~Vlah, U.~Seljak, M.~Y. Chu and Y.~Feng, \emph{{Perturbation theory,
  effective field theory, and oscillations in the power spectrum}},
  \href{https://doi.org/10.1088/1475-7516/2016/03/057}{\emph{JCAP} {\bfseries
  1603} (2016) 057} [\href{https://arxiv.org/abs/1509.02120}{{\ttfamily
  1509.02120}}].

\bibitem{Blas:2015qsi}
D.~Blas, M.~Garny, M.~M. Ivanov and S.~Sibiryakov, \emph{{Time-Sliced
  Perturbation Theory for Large Scale Structure I: General Formalism}},
  \href{https://doi.org/10.1088/1475-7516/2016/07/052}{\emph{JCAP} {\bfseries
  1607} (2016) 052} [\href{https://arxiv.org/abs/1512.05807}{{\ttfamily
  1512.05807}}].

\bibitem{Blas:2016sfa}
D.~Blas, M.~Garny, M.~M. Ivanov and S.~Sibiryakov, \emph{{Time-Sliced
  Perturbation Theory II: Baryon Acoustic Oscillations and Infrared
  Resummation}},
  \href{https://doi.org/10.1088/1475-7516/2016/07/028}{\emph{JCAP} {\bfseries
  1607} (2016) 028} [\href{https://arxiv.org/abs/1605.02149}{{\ttfamily
  1605.02149}}].

\bibitem{Senatore:2017pbn}
L.~Senatore and G.~Trevisan, \emph{{On the IR-Resummation in the EFTofLSS}},
  \href{https://doi.org/10.1088/1475-7516/2018/05/019}{\emph{JCAP} {\bfseries
  1805} (2018) 019} [\href{https://arxiv.org/abs/1710.02178}{{\ttfamily
  1710.02178}}].

\bibitem{Ivanov:2018gjr}
M.~M. Ivanov and S.~Sibiryakov, \emph{{Infrared Resummation for Biased Tracers
  in Redshift Space}},
  \href{https://doi.org/10.1088/1475-7516/2018/07/053}{\emph{JCAP} {\bfseries
  1807} (2018) 053} [\href{https://arxiv.org/abs/1804.05080}{{\ttfamily
  1804.05080}}].

\bibitem{Lewandowski:2018ywf}
M.~Lewandowski and L.~Senatore, \emph{{An analytic implementation of the
  IR-resummation for the BAO peak}},
  \href{https://doi.org/10.1088/1475-7516/2020/03/018}{\emph{JCAP} {\bfseries
  03} (2020) 018} [\href{https://arxiv.org/abs/1810.11855}{{\ttfamily
  1810.11855}}].

\bibitem{McDonald:2009dh}
P.~McDonald and A.~Roy, \emph{{Clustering of dark matter tracers: generalizing
  bias for the coming era of precision LSS}},
  \href{https://doi.org/10.1088/1475-7516/2009/08/020}{\emph{JCAP} {\bfseries
  0908} (2009) 020} [\href{https://arxiv.org/abs/0902.0991}{{\ttfamily
  0902.0991}}].

\bibitem{Assassi:2014fva}
V.~Assassi, D.~Baumann, D.~Green and M.~Zaldarriaga, \emph{{Renormalized Halo
  Bias}}, \href{https://doi.org/10.1088/1475-7516/2014/08/056}{\emph{JCAP}
  {\bfseries 1408} (2014) 056}
  [\href{https://arxiv.org/abs/1402.5916}{{\ttfamily 1402.5916}}].

\bibitem{Senatore:2014eva}
L.~Senatore, \emph{{Bias in the Effective Field Theory of Large Scale
  Structures}},
  \href{https://doi.org/10.1088/1475-7516/2015/11/007}{\emph{JCAP} {\bfseries
  1511} (2015) 007} [\href{https://arxiv.org/abs/1406.7843}{{\ttfamily
  1406.7843}}].

\bibitem{Lewandowski:2014rca}
M.~Lewandowski, A.~Perko and L.~Senatore, \emph{{Analytic Prediction of
  Baryonic Effects from the EFT of Large Scale Structures}},
  \href{https://doi.org/10.1088/1475-7516/2015/05/019}{\emph{JCAP} {\bfseries
  1505} (2015) 019} [\href{https://arxiv.org/abs/1412.5049}{{\ttfamily
  1412.5049}}].

\bibitem{Mirbabayi:2014zca}
M.~Mirbabayi, F.~Schmidt and M.~Zaldarriaga, \emph{{Biased Tracers and Time
  Evolution}}, \href{https://doi.org/10.1088/1475-7516/2015/07/030}{\emph{JCAP}
  {\bfseries 1507} (2015) 030}
  [\href{https://arxiv.org/abs/1412.5169}{{\ttfamily 1412.5169}}].

\bibitem{Desjacques:2016bnm}
V.~Desjacques, D.~Jeong and F.~Schmidt, \emph{{Large-Scale Galaxy Bias}},
  \href{https://doi.org/10.1016/j.physrep.2017.12.002}{\emph{Phys. Rept.}
  {\bfseries 733} (2018) 1} [\href{https://arxiv.org/abs/1611.09787}{{\ttfamily
  1611.09787}}].

\bibitem{Senatore:2014vja}
L.~Senatore and M.~Zaldarriaga, \emph{{Redshift Space Distortions in the
  Effective Field Theory of Large Scale Structures}},
  \href{https://arxiv.org/abs/1409.1225}{{\ttfamily 1409.1225}}.

\bibitem{Perko:2016puo}
A.~Perko, L.~Senatore, E.~Jennings and R.~H. Wechsler, \emph{{Biased Tracers in
  Redshift Space in the EFT of Large-Scale Structure}},
  \href{https://arxiv.org/abs/1610.09321}{{\ttfamily 1610.09321}}.

\bibitem{Matsubara:2007wj}
T.~Matsubara, \emph{{Resumming Cosmological Perturbations via the Lagrangian
  Picture: One-loop Results in Real Space and in Redshift Space}},
  \href{https://doi.org/10.1103/PhysRevD.77.063530}{\emph{Phys. Rev.}
  {\bfseries D77} (2008) 063530}
  [\href{https://arxiv.org/abs/0711.2521}{{\ttfamily 0711.2521}}].

\bibitem{Porto:2013qua}
R.~A. Porto, L.~Senatore and M.~Zaldarriaga, \emph{{The Lagrangian-space
  Effective Field Theory of Large Scale Structures}},
  \href{https://doi.org/10.1088/1475-7516/2014/05/022}{\emph{JCAP} {\bfseries
  1405} (2014) 022} [\href{https://arxiv.org/abs/1311.2168}{{\ttfamily
  1311.2168}}].

\bibitem{Carlson:2012bu}
J.~Carlson, B.~Reid and M.~White, \emph{{Convolution Lagrangian perturbation
  theory for biased tracers}},
  \href{https://doi.org/10.1093/mnras/sts457}{\emph{Mon.\ Not.\ Roy.\ Astron.\
  Soc.} {\bfseries 429} (2013) 1674}
  [\href{https://arxiv.org/abs/1209.0780}{{\ttfamily 1209.0780}}].

\bibitem{Vlah:2015sea}
Z.~Vlah, M.~White and A.~Aviles, \emph{{A Lagrangian effective field theory}},
  \href{https://doi.org/10.1088/1475-7516/2015/09/014}{\emph{JCAP} {\bfseries
  09} (2015) 014} [\href{https://arxiv.org/abs/1506.05264}{{\ttfamily
  1506.05264}}].

\bibitem{Vlah:2016bcl}
Z.~Vlah, E.~Castorina and M.~White, \emph{{The Gaussian streaming model and
  convolution Lagrangian effective field theory}},
  \href{https://doi.org/10.1088/1475-7516/2016/12/007}{\emph{JCAP} {\bfseries
  12} (2016) 007} [\href{https://arxiv.org/abs/1609.02908}{{\ttfamily
  1609.02908}}].

\bibitem{Modi:2017wds}
C.~Modi, M.~White and Z.~Vlah, \emph{{Modeling CMB lensing cross correlations
  with CLEFT}},
  \href{https://doi.org/10.1088/1475-7516/2017/08/009}{\emph{JCAP} {\bfseries
  08} (2017) 009} [\href{https://arxiv.org/abs/1706.03173}{{\ttfamily
  1706.03173}}].

\bibitem{Schmittfull:2018yuk}
M.~Schmittfull, M.~Simonović, V.~Assassi and M.~Zaldarriaga, \emph{{Modeling
  Biased Tracers at the Field Level}},
  \href{https://doi.org/10.1103/PhysRevD.100.043514}{\emph{Phys.\ Rev.\ D}
  {\bfseries 100} (2019) 043514}
  [\href{https://arxiv.org/abs/1811.10640}{{\ttfamily 1811.10640}}].

\bibitem{Baldauf:2015aha}
T.~Baldauf, L.~Mercolli and M.~Zaldarriaga, \emph{{Effective field theory of
  large scale structure at two loops: The apparent scale dependence of the
  speed of sound}},
  \href{https://doi.org/10.1103/PhysRevD.92.123007}{\emph{Phys. Rev.}
  {\bfseries D92} (2015) 123007}
  [\href{https://arxiv.org/abs/1507.02256}{{\ttfamily 1507.02256}}].

\bibitem{Pietroni:2008jx}
M.~Pietroni, \emph{{Flowing with Time: a New Approach to Nonlinear Cosmological
  Perturbations}},
  \href{https://doi.org/10.1088/1475-7516/2008/10/036}{\emph{JCAP} {\bfseries
  0810} (2008) 036} [\href{https://arxiv.org/abs/0806.0971}{{\ttfamily
  0806.0971}}].

\bibitem{Fasiello:2016qpn}
M.~Fasiello and Z.~Vlah, \emph{{Nonlinear fields in generalized cosmologies}},
  \href{https://doi.org/10.1103/PhysRevD.94.063516}{\emph{Phys. Rev.}
  {\bfseries D94} (2016) 063516}
  [\href{https://arxiv.org/abs/1604.04612}{{\ttfamily 1604.04612}}].

\bibitem{delaBella:2017qjy}
L.~F. de~la Bella, D.~Regan, D.~Seery and S.~Hotchkiss, \emph{{The matter power
  spectrum in redshift space using effective field theory}},
  \href{https://doi.org/10.1088/1475-7516/2017/11/039}{\emph{JCAP} {\bfseries
  1711} (2017) 039} [\href{https://arxiv.org/abs/1704.05309}{{\ttfamily
  1704.05309}}].

\bibitem{Lewandowski:2015ziq}
M.~Lewandowski, L.~Senatore, F.~Prada, C.~Zhao and C.-H. Chuang, \emph{{EFT of
  large scale structures in redshift space}},
  \href{https://doi.org/10.1103/PhysRevD.97.063526}{\emph{Phys. Rev. D}
  {\bfseries 97} (2018) 063526}
  [\href{https://arxiv.org/abs/1512.06831}{{\ttfamily 1512.06831}}].

\bibitem{Jackson:2008yv}
J.~C. Jackson, \emph{{Fingers of God: A critique of Rees' theory of primoridal
  gravitational radiation}},
  \href{https://doi.org/10.1093/mnras/156.1.1P}{\emph{Mon. Not. Roy. Astron.
  Soc.} {\bfseries 156} (1972) 1P}
  [\href{https://arxiv.org/abs/0810.3908}{{\ttfamily 0810.3908}}].

\bibitem{Kaiser:1987qv}
N.~Kaiser, \emph{{Clustering in real space and in redshift space}}, {\emph{Mon.
  Not. Roy. Astron. Soc.} {\bfseries 227} (1987) 1}.

\bibitem{DAmico:2019fhj}
G.~D'Amico, J.~Gleyzes, N.~Kokron, D.~Markovic, L.~Senatore, P.~Zhang et~al.,
  \emph{{The Cosmological Analysis of the SDSS/BOSS data from the Effective
  Field Theory of Large-Scale Structure}},
  \href{https://arxiv.org/abs/1909.05271}{{\ttfamily 1909.05271}}.

\bibitem{1978ApJ...221....1D}
M.~{Davis}, M.~J. {Geller} and J.~{Huchra}, \emph{{The local mean mass density
  of the universe: new methods for studying galaxy clustering.}},
  \href{https://doi.org/10.1086/156000}{\emph{Astrophys.\ J.} {\bfseries 221}
  (1978) 1}.

\bibitem{Matsubara:1996nf}
T.~Matsubara and Y.~Suto, \emph{{Cosmological redshift distortion of
  correlation functions as a probe of the density parameter and the
  cosmological constant}},
  \href{https://doi.org/10.1086/310290}{\emph{Astrophys.\ J.} {\bfseries 470}
  (1996) L1} [\href{https://arxiv.org/abs/astro-ph/9604142}{{\ttfamily
  astro-ph/9604142}}].

\bibitem{Ballinger:1996cd}
W.~Ballinger, J.~Peacock and A.~Heavens, \emph{{Measuring the cosmological
  constant with redshift surveys}},
  \href{https://doi.org/10.1093/mnras/282.3.877}{\emph{Mon.\ Not.\ Roy.\
  Astron.\ Soc.} {\bfseries 282} (1996) 877}
  [\href{https://arxiv.org/abs/astro-ph/9605017}{{\ttfamily
  astro-ph/9605017}}].

\bibitem{Heinesen:2018hnh}
A.~Heinesen, C.~Blake, Y.-Z. Li and D.~L. Wiltshire, \emph{{Baryon acoustic
  oscillation methods for generic curvature: Application to the SDSS-III Baryon
  Oscillation Spectroscopic Survey}},
  \href{https://doi.org/10.1088/1475-7516/2019/03/003}{\emph{JCAP} {\bfseries
  1903} (2019) 003} [\href{https://arxiv.org/abs/1811.11963}{{\ttfamily
  1811.11963}}].

\bibitem{Heinesen:2019phg}
A.~Heinesen, C.~Blake and D.~L. Wiltshire, \emph{{Quantifying the accuracy of
  the Alcock-Paczy\&nacute;ski scaling of baryon acoustic oscillation
  measurements}},
  \href{https://doi.org/10.1088/1475-7516/2020/01/038}{\emph{JCAP} {\bfseries
  2001} (2020) 038} [\href{https://arxiv.org/abs/1908.11508}{{\ttfamily
  1908.11508}}].

\bibitem{Alcock:1979mp}
C.~Alcock and B.~Paczynski, \emph{{An evolution free test for non-zero
  cosmological constant}},
  \href{https://doi.org/10.1038/281358a0}{\emph{Nature} {\bfseries 281} (1979)
  358}.

\bibitem{Schoneberg:2018fis}
N.~Schöneberg, M.~Simonović, J.~Lesgourgues and M.~Zaldarriaga, \emph{{Beyond
  the traditional Line-of-Sight approach of cosmological angular statistics}},
  \href{https://doi.org/10.1088/1475-7516/2018/10/047}{\emph{JCAP} {\bfseries
  1810} (2018) 047} [\href{https://arxiv.org/abs/1807.09540}{{\ttfamily
  1807.09540}}].

\bibitem{Hamilton:1999uv}
A.~J.~S. Hamilton, \emph{{Uncorrelated modes of the nonlinear power spectrum}},
  \href{https://doi.org/10.1046/j.1365-8711.2000.03071.x}{\emph{Mon. Not. Roy.
  Astron. Soc.} {\bfseries 312} (2000) 257}
  [\href{https://arxiv.org/abs/astro-ph/9905191}{{\ttfamily
  astro-ph/9905191}}].

\bibitem{Schmittfull:2016yqx}
M.~Schmittfull and Z.~Vlah, \emph{{FFT-PT: Reducing the two-loop large-scale
  structure power spectrum to low-dimensional radial integrals}},
  \href{https://doi.org/10.1103/PhysRevD.94.103530}{\emph{Phys. Rev.}
  {\bfseries D94} (2016) 103530}
  [\href{https://arxiv.org/abs/1609.00349}{{\ttfamily 1609.00349}}].

\bibitem{Slepian:2018vds}
Z.~Slepian, \emph{{On Decoupling the Integrals of Cosmological Perturbation
  Theory}},  \href{https://arxiv.org/abs/1812.02728}{{\ttfamily 1812.02728}}.

\bibitem{Hamann:2010pw}
J.~Hamann, S.~Hannestad, J.~Lesgourgues, C.~Rampf and Y.~Y.~Y. Wong,
  \emph{{Cosmological parameters from large scale structure - geometric versus
  shape information}},
  \href{https://doi.org/10.1088/1475-7516/2010/07/022}{\emph{JCAP} {\bfseries
  1007} (2010) 022} [\href{https://arxiv.org/abs/1003.3999}{{\ttfamily
  1003.3999}}].

\bibitem{Blas:2014hya}
D.~Blas, M.~Garny, T.~Konstandin and J.~Lesgourgues, \emph{{Structure formation
  with massive neutrinos: going beyond linear theory}},
  \href{https://doi.org/10.1088/1475-7516/2014/11/039}{\emph{JCAP} {\bfseries
  1411} (2014) 039} [\href{https://arxiv.org/abs/1408.2995}{{\ttfamily
  1408.2995}}].

\bibitem{Senatore:2017hyk}
L.~Senatore and M.~Zaldarriaga, \emph{{The Effective Field Theory of
  Large-Scale Structure in the presence of Massive Neutrinos}},
  \href{https://arxiv.org/abs/1707.04698}{{\ttfamily 1707.04698}}.

\bibitem{Villaescusa-Navarro:2013pva}
F.~Villaescusa-Navarro, F.~Marulli, M.~Viel, E.~Branchini, E.~Castorina,
  E.~Sefusatti et~al., \emph{{Cosmology with massive neutrinos I: towards a
  realistic modeling of the relation between matter, haloes and galaxies}},
  \href{https://doi.org/10.1088/1475-7516/2014/03/011}{\emph{JCAP} {\bfseries
  1403} (2014) 011} [\href{https://arxiv.org/abs/1311.0866}{{\ttfamily
  1311.0866}}].

\bibitem{Castorina:2013wga}
E.~Castorina, E.~Sefusatti, R.~K. Sheth, F.~Villaescusa-Navarro and M.~Viel,
  \emph{{Cosmology with massive neutrinos II: on the universality of the halo
  mass function and bias}},
  \href{https://doi.org/10.1088/1475-7516/2014/02/049}{\emph{JCAP} {\bfseries
  1402} (2014) 049} [\href{https://arxiv.org/abs/1311.1212}{{\ttfamily
  1311.1212}}].

\bibitem{Costanzi:2013bha}
M.~Costanzi, F.~Villaescusa-Navarro, M.~Viel, J.-Q. Xia, S.~Borgani,
  E.~Castorina et~al., \emph{{Cosmology with massive neutrinos III: the halo
  mass function andan application to galaxy clusters}},
  \href{https://doi.org/10.1088/1475-7516/2013/12/012}{\emph{JCAP} {\bfseries
  1312} (2013) 012} [\href{https://arxiv.org/abs/1311.1514}{{\ttfamily
  1311.1514}}].

\bibitem{Castorina:2015bma}
E.~Castorina, C.~Carbone, J.~Bel, E.~Sefusatti and K.~Dolag, \emph{{DEMNUni:
  The clustering of large-scale structures in the presence of massive
  neutrinos}}, \href{https://doi.org/10.1088/1475-7516/2015/07/043}{\emph{JCAP}
  {\bfseries 1507} (2015) 043}
  [\href{https://arxiv.org/abs/1505.07148}{{\ttfamily 1505.07148}}].

\bibitem{Villaescusa-Navarro:2017mfx}
F.~Villaescusa-Navarro, A.~Banerjee, N.~Dalal, E.~Castorina, R.~Scoccimarro,
  R.~Angulo et~al., \emph{{The imprint of neutrinos on clustering in
  redshift-space}},
  \href{https://doi.org/10.3847/1538-4357/aac6bf}{\emph{Astrophys. J.}
  {\bfseries 861} (2018) 53}
  [\href{https://arxiv.org/abs/1708.01154}{{\ttfamily 1708.01154}}].

\bibitem{Raccanelli:2017kht}
A.~Raccanelli, L.~Verde and F.~Villaescusa-Navarro, \emph{{Biases from neutrino
  bias: to worry or not to worry?}},
  \href{https://doi.org/10.1093/mnras/sty2162}{\emph{Mon. Not. Roy. Astron.
  Soc.} {\bfseries 483} (2019) 734}
  [\href{https://arxiv.org/abs/1704.07837}{{\ttfamily 1704.07837}}].

\bibitem{Vagnozzi:2018pwo}
S.~Vagnozzi, T.~Brinckmann, M.~Archidiacono, K.~Freese, M.~Gerbino,
  J.~Lesgourgues et~al., \emph{{Bias due to neutrinos must not uncorrect'd
  go}}, \href{https://doi.org/10.1088/1475-7516/2018/09/001}{\emph{JCAP}
  {\bfseries 1809} (2018) 001}
  [\href{https://arxiv.org/abs/1807.04672}{{\ttfamily 1807.04672}}].

\bibitem{LoVerde:2016ahu}
M.~LoVerde, \emph{{Neutrino mass without cosmic variance}},
  \href{https://doi.org/10.1103/PhysRevD.93.103526}{\emph{Phys. Rev.}
  {\bfseries D93} (2016) 103526}
  [\href{https://arxiv.org/abs/1602.08108}{{\ttfamily 1602.08108}}].

\bibitem{Poulin:2018cxd}
V.~Poulin, T.~L. Smith, T.~Karwal and M.~Kamionkowski, \emph{{Early Dark Energy
  Can Resolve The Hubble Tension}},
  \href{https://doi.org/10.1103/PhysRevLett.122.221301}{\emph{Phys. Rev. Lett.}
  {\bfseries 122} (2019) 221301}
  [\href{https://arxiv.org/abs/1811.04083}{{\ttfamily 1811.04083}}].

\bibitem{Ivanov:2020ril}
M.~M. Ivanov, E.~McDonough, J.~C. Hill, M.~Simonovi\'c, M.~W. Toomey,
  S.~Alexander et~al., \emph{{Constraining Early Dark Energy with Large-Scale
  Structure}},  \href{https://arxiv.org/abs/2006.11235}{{\ttfamily
  2006.11235}}.

\bibitem{DAmico:2020ods}
G.~D'Amico, L.~Senatore, P.~Zhang and H.~Zheng, \emph{{The Hubble Tension in
  Light of the Full-Shape Analysis of Large-Scale Structure Data}},
  \href{https://arxiv.org/abs/2006.12420}{{\ttfamily 2006.12420}}.

\bibitem{Crisostomi:2019vhj}
M.~Crisostomi, M.~Lewandowski and F.~Vernizzi, \emph{{Consistency relations for
  large-scale structure in modified gravity and the matter bispectrum}},
  \href{https://arxiv.org/abs/1909.07366}{{\ttfamily 1909.07366}}.

\bibitem{Lewandowski:2019txi}
M.~Lewandowski, \emph{{Violation of the consistency relations for large-scale
  structure with dark energy}},
  \href{https://arxiv.org/abs/1912.12292}{{\ttfamily 1912.12292}}.

\bibitem{Lewis:2006fu}
A.~Lewis and A.~Challinor, \emph{{Weak gravitational lensing of the CMB}},
  \href{https://doi.org/10.1016/j.physrep.2006.03.002}{\emph{Phys. Rept.}
  {\bfseries 429} (2006) 1}
  [\href{https://arxiv.org/abs/astro-ph/0601594}{{\ttfamily
  astro-ph/0601594}}].

\bibitem{Beutler:2016ixs}
{\scshape BOSS} collaboration, F.~Beutler et~al., \emph{{The clustering of
  galaxies in the completed SDSS-III Baryon Oscillation Spectroscopic Survey:
  baryon acoustic oscillations in the Fourier space}},
  \href{https://doi.org/10.1093/mnras/stw2373}{\emph{Mon. Not. Roy. Astron.
  Soc.} {\bfseries 464} (2017) 3409}
  [\href{https://arxiv.org/abs/1607.03149}{{\ttfamily 1607.03149}}].

\bibitem{Beutler:2016arn}
{\scshape BOSS} collaboration, F.~Beutler et~al., \emph{{The clustering of
  galaxies in the completed SDSS-III Baryon Oscillation Spectroscopic Survey:
  Anisotropic galaxy clustering in Fourier-space}},
  \href{https://doi.org/10.1093/mnras/stw3298}{\emph{Mon. Not. Roy. Astron.
  Soc.} {\bfseries 466} (2017) 2242}
  [\href{https://arxiv.org/abs/1607.03150}{{\ttfamily 1607.03150}}].

\bibitem{Tegmark:1997rp}
M.~Tegmark, \emph{{Measuring cosmological parameters with galaxy surveys}},
  \href{https://doi.org/10.1103/PhysRevLett.79.3806}{\emph{Phys. Rev. Lett.}
  {\bfseries 79} (1997) 3806}
  [\href{https://arxiv.org/abs/astro-ph/9706198}{{\ttfamily
  astro-ph/9706198}}].

\bibitem{Tegmark:1997yq}
M.~Tegmark, A.~J. Hamilton, M.~A. Strauss, M.~S. Vogeley and A.~S. Szalay,
  \emph{{Measuring the galaxy power spectrum with future redshift surveys}},
  \href{https://doi.org/10.1086/305663}{\emph{Astrophys. J.} {\bfseries 499}
  (1998) 555} [\href{https://arxiv.org/abs/astro-ph/9708020}{{\ttfamily
  astro-ph/9708020}}].

\bibitem{Tegmark:2003uf}
{\scshape SDSS} collaboration, M.~Tegmark et~al., \emph{{The 3-D power spectrum
  of galaxies from the SDSS}},
  \href{https://doi.org/10.1086/382125}{\emph{Astrophys. J.} {\bfseries 606}
  (2004) 702} [\href{https://arxiv.org/abs/astro-ph/0310725}{{\ttfamily
  astro-ph/0310725}}].

\bibitem{Cole:2005sx}
{\scshape 2dFGRS} collaboration, S.~Cole et~al., \emph{{The 2dF Galaxy Redshift
  Survey: Power-spectrum analysis of the final dataset and cosmological
  implications}},
  \href{https://doi.org/10.1111/j.1365-2966.2005.09318.x}{\emph{Mon. Not. Roy.
  Astron. Soc.} {\bfseries 362} (2005) 505}
  [\href{https://arxiv.org/abs/astro-ph/0501174}{{\ttfamily
  astro-ph/0501174}}].

\bibitem{Blake:2006kv}
C.~Blake, A.~Collister, S.~Bridle and O.~Lahav, \emph{{Cosmological baryonic
  and matter densities from 600,000 SDSS Luminous Red Galaxies with photometric
  redshifts}},
  \href{https://doi.org/10.1111/j.1365-2966.2006.11263.x}{\emph{Mon. Not. Roy.
  Astron. Soc.} {\bfseries 374} (2007) 1527}
  [\href{https://arxiv.org/abs/astro-ph/0605303}{{\ttfamily
  astro-ph/0605303}}].

\bibitem{Percival:2006gt}
W.~J. Percival et~al., \emph{{The shape of the SDSS DR5 galaxy power
  spectrum}}, \href{https://doi.org/10.1086/510615}{\emph{Astrophys. J.}
  {\bfseries 657} (2007) 645}
  [\href{https://arxiv.org/abs/astro-ph/0608636}{{\ttfamily
  astro-ph/0608636}}].

\bibitem{Beutler:2013yhm}
{\scshape BOSS} collaboration, F.~Beutler et~al., \emph{{The clustering of
  galaxies in the SDSS-III Baryon Oscillation Spectroscopic Survey: Testing
  gravity with redshift-space distortions using the power spectrum
  multipoles}}, \href{https://doi.org/10.1093/mnras/stu1051}{\emph{Mon. Not.
  Roy. Astron. Soc.} {\bfseries 443} (2014) 1065}
  [\href{https://arxiv.org/abs/1312.4611}{{\ttfamily 1312.4611}}].

\bibitem{Font-Ribera:2013rwa}
A.~Font-Ribera, P.~McDonald, N.~Mostek, B.~A. Reid, H.-J. Seo and A.~Slosar,
  \emph{{DESI and other dark energy experiments in the era of neutrino mass
  measurements}},
  \href{https://doi.org/10.1088/1475-7516/2014/05/023}{\emph{JCAP} {\bfseries
  05} (2014) 023} [\href{https://arxiv.org/abs/1308.4164}{{\ttfamily
  1308.4164}}].

\bibitem{Lesgourgues:2006nd}
J.~Lesgourgues and S.~Pastor, \emph{{Massive neutrinos and cosmology}},
  \href{https://doi.org/10.1016/j.physrep.2006.04.001}{\emph{Phys. Rept.}
  {\bfseries 429} (2006) 307}
  [\href{https://arxiv.org/abs/astro-ph/0603494}{{\ttfamily
  astro-ph/0603494}}].

\bibitem{Lewis:2013hha}
A.~Lewis, \emph{{Efficient sampling of fast and slow cosmological parameters}},
  \href{https://doi.org/10.1103/PhysRevD.87.103529}{\emph{Phys. Rev.}
  {\bfseries D87} (2013) 103529}
  [\href{https://arxiv.org/abs/1304.4473}{{\ttfamily 1304.4473}}].

\bibitem{Lewis:2019xzd}
A.~Lewis, \emph{{GetDist: a Python package for analysing Monte Carlo samples}},
   \href{https://arxiv.org/abs/1910.13970}{{\ttfamily 1910.13970}}.

\bibitem{Aghanim:2019ame}
{\scshape Planck} collaboration, N.~Aghanim et~al., \emph{{Planck 2018 results.
  V. CMB power spectra and likelihoods}},
  \href{https://arxiv.org/abs/1907.12875}{{\ttfamily 1907.12875}}.

\bibitem{Desjacques:2018pfv}
V.~Desjacques, D.~Jeong and F.~Schmidt, \emph{{The Galaxy Power Spectrum and
  Bispectrum in Redshift Space}},
  \href{https://doi.org/10.1088/1475-7516/2018/12/035}{\emph{JCAP} {\bfseries
  1812} (2018) 035} [\href{https://arxiv.org/abs/1806.04015}{{\ttfamily
  1806.04015}}].

\bibitem{Lazeyras:2017hxw}
T.~Lazeyras and F.~Schmidt, \emph{{Beyond LIMD bias: a measurement of the
  complete set of third-order halo bias parameters}},
  \href{https://doi.org/10.1088/1475-7516/2018/09/008}{\emph{JCAP} {\bfseries
  1809} (2018) 008} [\href{https://arxiv.org/abs/1712.07531}{{\ttfamily
  1712.07531}}].

\bibitem{Baldauf:2013hka}
T.~Baldauf, U.~s. Seljak, R.~E. Smith, N.~Hamaus and V.~Desjacques, \emph{{Halo
  stochasticity from exclusion and nonlinear clustering}},
  \href{https://doi.org/10.1103/PhysRevD.88.083507}{\emph{Phys. Rev. D}
  {\bfseries 88} (2013) 083507}
  [\href{https://arxiv.org/abs/1305.2917}{{\ttfamily 1305.2917}}].

\bibitem{Hahn:2016kiy}
C.~Hahn, R.~Scoccimarro, M.~R. Blanton, J.~L. Tinker and S.~A.
  Rodríguez-Torres, \emph{{The effect of fibre collisions on the galaxy power
  spectrum multipoles}}, \href{https://doi.org/10.1093/mnras/stx185}{\emph{Mon.
  Not. Roy. Astron. Soc.} {\bfseries 467} (2017) 1940}
  [\href{https://arxiv.org/abs/1609.01714}{{\ttfamily 1609.01714}}].

\bibitem{Kitaura:2015uqa}
F.-S. Kitaura et~al., \emph{{The clustering of galaxies in the SDSS-III Baryon
  Oscillation Spectroscopic Survey: mock galaxy catalogues for the BOSS Final
  Data Release}}, \href{https://doi.org/10.1093/mnras/stv2826}{\emph{Mon. Not.
  Roy. Astron. Soc.} {\bfseries 456} (2016) 4156}
  [\href{https://arxiv.org/abs/1509.06400}{{\ttfamily 1509.06400}}].

\bibitem{Gil-Marin:2014sta}
H.~Gil-Marín, J.~Noreña, L.~Verde, W.~J. Percival, C.~Wagner, M.~Manera
  et~al., \emph{{The power spectrum and bispectrum of SDSS DR11 BOSS galaxies
  – I. Bias and gravity}},
  \href{https://doi.org/10.1093/mnras/stv961}{\emph{Mon. Not. Roy. Astron.
  Soc.} {\bfseries 451} (2015) 539}
  [\href{https://arxiv.org/abs/1407.5668}{{\ttfamily 1407.5668}}].

\bibitem{Gil-Marin:2014baa}
H.~Gil-Marín, L.~Verde, J.~Noreña, A.~J. Cuesta, L.~Samushia, W.~J. Percival
  et~al., \emph{{The power spectrum and bispectrum of SDSS DR11 BOSS galaxies
  – II. Cosmological interpretation}},
  \href{https://doi.org/10.1093/mnras/stv1359}{\emph{Mon. Not. Roy. Astron.
  Soc.} {\bfseries 452} (2015) 1914}
  [\href{https://arxiv.org/abs/1408.0027}{{\ttfamily 1408.0027}}].

\end{thebibliography}\endgroup

\end{document}